\documentclass[fleqn,usenatbib]{mnras}

\usepackage{newtxtext,newtxmath}
\usepackage{subfigure}

\usepackage[T1]{fontenc}
\usepackage{ae,aecompl}

\usepackage{graphicx}	
\usepackage{amsmath}	
\usepackage{amssymb}	

\usepackage{soul}
\usepackage[usenames,dvipsnames]{xcolor}

\defcitealias{Mingarelli2017}{M17}	 

\title[GWB Angular Power Spectra]{On the Amplitude and Stokes Parameters of a Stochastic Gravitational-Wave Background}

\author[Ciar\'{a}n Conneely et al.]{
Ciar\'{a}n Conneely,$^{1}$\thanks{E-mail: c.conneely14@imperial.ac.uk}
Andrew H. Jaffe,$^{1}$
Chiara M.~F.~Mingarelli$^{2}$
\\
$^{1}$Astrophysics, Blackett Laboratory, Imperial College, London SW7 2AZ, UK\\
$^{2}$Center for Computational Astrophysics, Flatiron Institute, 162 Fifth Ave, New York, NY 10010, USA
}

\date{Accepted XXX. Received YYY; in original form ZZZ}

\pubyear{2018}

\begin{document}
\label{firstpage}
\pagerange{\pageref{firstpage}--\pageref{lastpage}}
\maketitle

\begin{abstract}
The direct detection of gravitational waves has provided new opportunities for studying the universe, but also new challenges, such as the detection and characterisation of stochastic gravitational-wave backgrounds at different gravitational-wave frequencies. In this paper we examine two different methods for their description, one based on the amplitude of a gravitational-wave signal and one on its Stokes parameters. We find that the Stokes parameters are able to describe anisotropic and correlated backgrounds, whereas the usual power spectra of the amplitudes cannot -- i.e. the Stokes spectra are sensitive to properties such as the spatial distribution of the gravitational-wave sources in a realistic backgrounds.
\end{abstract}




\section{Introduction}

The observation of gravitational waves by the Laser Interferometer Gravitational-Wave Observatory (LIGO) detectors \citep{GW150914} was the result of several decades of work, both from the LIGO/Virgo collaborations and elsewhere. As of writing, there have been six confirmed (plus one probable) observations of black hole or neutron star binaries published, with varying degrees of localisation. These are single, individually separable sources but in the future it is predicted that detectors such as pulsar timing arrays \citep[PTAs; e.g.][]{Detweiler:1979, HellingsDowns:1983, Jaffe2003, Yardley:2011fk, Lentati:2015qwp, Arzoumanian:2018saf} and the Laser Interferometer Space Antenna \citep[LISA; e.g.][]{LISA:2017, Cornish:2001bb} will be able to observe a stochastic background -- i.e. one where there are multiple, nonseparable signals. Such a signal can have varying properties. It could be astrophysical \citep{Regimbau2011} or cosmological \citep{Caprini2015} in origin, monochromatic or polychromatic, isotropic or anisotropic \citep{msmv13, Ungarelli:2001xu, ThraneEtAl:2009}.
It may also be made up of many individual sources that are theoretically resolvable but are too numerous/low amplitude to do so at the current time (e.g. galactic white dwarf binaries) or one that is not due to the fact that it simply exists everywhere (e.g. inflationary gravitational waves, e.g. \citealt{lms+16}). 

How best to analyse this broad range of multi-frequency stochastic backgrounds is the subject of this paper. Here we consider two methods to construct power spectra for a gravitational-wave background, using analogies to Cosmic Microwave Background (CMB) -- one involving a decomposition of the $h_+$ and $h_\times$ amplitudes \citep{Gair2014} and the other one of the GW Stokes parameters \citep{Seto:2008sr,Gubitosi2016,KatoSoda:2016}. 

The power spectral methods we use here make no major assumptions about properties of the background, only that it is observable and there is some directional dependence. As such, they are very general and can be applied to any gravitational frequency. We introduce the mathematics of the formalisms in Section~\ref{Sec:Form}. In Section~\ref{Sec:Backgrounds} we apply the formalisms to models of various backgrounds -- both astrophysical and cosmological -- and consider which method is appropriate for the description of different backgrounds and what can be learned from the power spectra in each case. Finally in Section~\ref{Sec:conclusions} we conclude and give an outlook for future work using these techniques.\footnote{Readers interested in the code used to produce the simulations and analyses presented here should contact the authors.}

\section{Formalism} \label{Sec:Form}
For the different types of gravitational-wave backgrounds, we will consider fields on the sky in harmonic space. That is, for a scalar field $F(f,\hat{k})$, where $f$ is the temporal frequency of gravitational waves or photons, on which the field may depend, and $\hat{k}$ is a three-dimensional unit vector giving the direction of propagation\footnote{Note that this can be equivalently considered in terms of the direction of observation, i.e. $-\hat{k}$. Also, for PTAs, the notation $\hat{k}=\hat{\Omega}$ is commonly used.}, we define
\begin{align} \label{Eq:CMB-T}
a^{F}_{\ell m}(f) = \int \text{d}^2\hat{k} F(f, \hat{k}) Y_{\ell m}^*(\hat{k})\;,
\end{align}
where $Y_{\ell m}$ are the spherical harmonics. In this construction, the $\ell$ subscript represents the angular scale of a perturbation and $m$ its orientation. As in the case of the CMB, the majority of the background fields considered in this paper are statistically isotropic and so it is reasonable to consider the average over orientations, and hence the power spectrum. 
For isotropic fields, this is defined by the expectation
\begin{align}
 \langle a_{\ell m}^F a_{\ell' m'}^{F'*} \rangle = C^{FF'}_\ell \delta_{\ell\ell'}\delta_{mm'}\;,
  \label{Eq:SpectraDef}
\end{align}
 where $F'$ may be the same or a similarly constructed field and $\langle \cdots \rangle$ represents an ensemble average, i.e. one over infinite realisations of the background. This is an idealised result; a single (full-sky, noise-free) realisation motivates the definition of the estimator
\begin{align} \label{Eq:SpectraDef-Measured}
\hat{C}^{FF'}_{\ell} \equiv \frac{1}{2\ell + 1} \sum_{m = -\ell}^\ell a_{\ell m}^F a_{\ell m}^{F'*}\;,
\end{align}
with isotropic expectation
\begin{align}
\langle \hat{C}^{FF'}_\ell\rangle = \langle a_{\ell m}^F a_{\ell m}^{F'*} \rangle = C^{FF'}_\ell\;.
\end{align}
Even with the assumption of negligible noise, a particular instance of the spectrum, $\hat{C}^{FF'}_\ell$, will differ from the true value, $C^{FF'}_\ell$, because of sample variance. In this case, the standard deviation of the difference can be estimated as
\begin{align}
\Delta C_\ell^{FF'} = \sqrt{\frac{\left(\hat{C}^{FF'}_\ell\right)^2 + \hat{C}^{FF}_\ell \hat{C}^{F'F'}_\ell}{(2\ell +1)f_{\text{sky}}} } \label{Eq:Error} \, .
\end{align}
The $f_{\text{sky}}$ term represents the fraction of the sky considered for incomplete or masked background -- $f_{\text{sky}} = 1$ for a background including the whole sky.

These errors are derived assuming real, Gaussian fields and so, where necessary, we will discuss more appropriate errors. Note further that many of the backgrounds considered in this paper will be anisotropic and so, in these cases, we define the power spectra to be the appropriate squared spherical harmonic coefficients as in equation~\ref{Eq:SpectraDef-Measured}. 

A potential alternative to these power spectra is a (2-point) correlation function, which is defined as
\begin{align}
C(\theta) =& \langle F(\hat{k}) F(\hat{k}') \rangle_{\hat{k}\cdot \hat{k}' = \cos\theta} \, ,\label{Eq:CFDef}
\end{align} 
and is very closely related to the power spectrum by
\begin{align}
C(\theta) 
=&
\sum_{\ell} \frac{2\ell + 1}{2\pi} C_\ell P_{\ell}(\cos\theta) \, . \label{Eq:CFfromPS}
\end{align}
This can be difficult to compute analytically or numerically if, as in many examples below, we assume that we can localise sources to arbitrary degrees of accuracy. In this case, $C_\ell \not\rightarrow 0$ and so $C(\theta)$ does not converge. In measured examples this will not be an issue, because this assumption of arbitrary sensitivity will not be true and response and/or window functions will reduce sensitivity at higher $\ell$s. Because of the issues with convergence and because they provide no new information (just a re-representation of the information in the power spectrum), correlation functions will not be considered further here.

\subsection{CMB Stokes Parameters} 
The key idea behind the formalisms discussed and compared in this paper is that a gravitational wave field shares many properties with an electromagnetic one. In particular, we consider how the temperature field $T$, or, equivalently, the total power $I^\mathrm{EM}$, and linear polarisation terms $Q^\mathrm{EM}$ and $U^\mathrm{EM}$ are analysed for the CMB. The fourth Stokes parameter, $V^\mathrm{EM}$, is usually assumed to be zero because Compton scattering cannot produce any net circular polarisation. Temperature is a scalar field on the sky and so can be decomposed as in equation~\ref{Eq:CMB-T} to give coefficients $a^T_{\ell m}$.

The linear polarisation is more complicated -- neither $Q^\mathrm{EM}$ nor $U^\mathrm{EM}$ are scalar or spin-weighted fields on the sky. The combinations $Q^\mathrm{EM} \pm i U^\mathrm{EM}$ are, however, spin-$\pm 2$ and so can decomposed using $\pm 2$ spin-weighted spherical harmonics, derived for general spin $\pm s$ from the spin-0 spherical harmonics \citep[as in, e.g.][]{Newman1966,Goldberg1967} giving
\begin{align} \label{Eq:sYlm-def}
{}_{s}Y_{\ell m}(\hat{k}) =& 
\begin{cases}
\sqrt{\frac{(l-s)!}{(l+s)!}} \eth^s Y_{\ell m}(\theta, \phi), &\text{for } 0\leq s\leq \ell
\\
\sqrt{\frac{(l-s)!}{(l+s)!}} (-1)^s \bar{\eth}^{-s} Y_{\ell m}(\theta, \phi), &\text{for } -\ell\leq s\leq 0
\end{cases}
\notag\\
=& \sqrt{\frac{(\ell + m)!}{(\ell + s)!}\frac{(\ell - m)!}{(\ell - s)!}\frac{2\ell + 1}{4\pi}} \sin^{2\ell}(\theta/2) e^{im\phi}
\notag\\
&\times\sum_r   \binom{\ell - s}{r} \binom{\ell +s}{r+s-m}(-1)^{\ell - r-s} \cot^{2r+s-m}(\theta/2)  \, ,
\end{align}
and are zero for $|s|>|\ell|$. Here, $\theta$ and $\phi$ are the angular polar coordinates of the ${\hat k}$ unit vector. The $\eth$ and $\bar{\eth}$ derivatives are operators which raise and lower, respectively, the spin of a field and are defined as
\begin{align}
\eth \eta =& -(\sin\theta)^s \left(\frac{\partial}{\partial \theta} + i \csc\theta \frac{\partial}{\partial \phi} \right) (\sin\theta)^{-s} \eta \, ,
\notag\\
\bar{\eth} \eta =& -(\sin\theta)^{-s} \left(\frac{\partial}{\partial \theta} - i \csc\theta \frac{\partial}{\partial \phi} \right) (\sin\theta)^{s} \eta 
\end{align}
for an initially spin-s function $\eta$.
Using these, we obtain \citep[e.g.][]{Seljak1997} 
\begin{align} \label{Eq:CMB-pm2}
{}_{\pm 2}a_{\ell m} = \int \text{d}^2\hat{k} (Q^\mathrm{EM} \pm i U^\mathrm{EM})(\hat{k}) {}_{\pm 2} Y_{\ell m}^*(\hat{k})
\end{align}
which give scalar and pseudo-scalar combinations 
\begin{align} \label{Eq:CMB-EB}
a^{E}_{\ell m} =& -\frac{1}{2}\left({}_{+ 2}a_{\ell m} + {}_{- 2}a_{\ell m} \right) \, ,
\notag\\
a^{B}_{\ell m} =& +\frac{i}{2}\left({}_{+ 2}a_{\ell m} - {}_{- 2}a_{\ell m} \right)
\end{align}
that can be used to construct corresponding scalar and pseudo-scalar fields $E$ and $B$ by, for example, $E = \sum_{\ell m} a^{E}_{\ell m} Y_{\ell m}$.
The fact that spin-weighted spherical harmonics are zero for $|s|>|\ell|$ means that $a^{E,B}_{\ell m} = 0$ for $\ell = 0,1$.

Using these harmonics, we can construct power spectra as in equation~\ref{Eq:SpectraDef-Measured}. Measurement of these power spectra can be used to learn about various properties of the early universe. For example, we can use the strength and shape of the $\hat{C}_\ell^{TT}$ signal to infer the value of cosmological parameters in the early universe including the dimensionless matter and dark energy densities \citep[e.g.][]{PlanckParameters2015}, and a measurement of a non-zero $\hat{C}_\ell^{BB}$ would be taken by many to be a ``smoking gun'' of inflationary gravitational waves. Additionally, because $E$  and $T$ fields are scalar and $B$ is pseudo-scalar, we expect $C^{TB}_\ell$ and $C^{EB}_\ell$ to be zero by parity. As such, any statistically significant measurement of a non-zero value for either of these spectra could say a lot about the early universe -- for example, that primordial gravitational waves were parity violating \citep[e.g.][]{Contaldi2008}.

A further comment is that all of the CMB power spectra are, by construction, real. For the auto-spectra, this follows immediately from their definition. For the cross-spectra it is due to the fact that the  Stokes parameters are real, and so their harmonic coefficients can be shown to satisfy 
\begin{align} \label{Eq:m-mrelation}
a^{M*}_{\ell m} =& (-1)^m a^{M}_{\ell, - m}, \, M \in \{I, E, B \} \, .
\end{align}  
This combined with the fact that ${}_{s}Y_{\ell 0}$ is real (as can be seen by setting $m=0$ in equation~\ref{Eq:sYlm-def}), and so $a^F_{\ell m} \in \mathbb{R}$, gives
\begin{align} \label{Eq:RealCrossSpec}
\hat{C}^{FF'}_\ell =& a^{F}_{\ell 0} a^{F'*}_{\ell 0} + \sum_{m=1}^\ell \left(a^{F}_{\ell m} a^{F'*}_{\ell m} + a^{F}_{\ell,-m} a^{F'*}_{\ell,-m}\right)
\notag\\
=& a^{F}_{\ell 0} a^{F'*}_{\ell 0} + \sum_{m=1}^\ell \left(a^{F}_{\ell m} a^{F'*}_{\ell m} + a^{F*}_{\ell m} a^{F'}_{\ell m}\right) \in \mathbb{R} \, .
\end{align}

\subsection{Gravitational Wave Amplitude}
\label{Sec:GWamp}
A gravitational wave can be represented as a small perturbation, $h_{\mu\nu}(t, \vec{x})$, about a static metric. This can be calculated as a sum over the Fourier components  
\begin{align}
h_{\mu\nu}(t, \vec{x}) = \int_{-\infty}^{\infty} \text{d}f \int \text{d}^2\hat{k} \tilde{h}_{\mu\nu}(f, \hat{k}) e^{i2\pi f(t - \hat{k}\cdot \vec{x}/c)} \, ,
\end{align}
where the $\tilde{h}_{\mu\nu}(f, \hat{k})$ are the contributions to the perturbation propagating in direction $\hat{k}$ -- i.e a field on the 2-sphere -- and the reality condition on $h_{\mu\nu}(t, \vec{x})$ is satisfied by $\tilde{h}_{\mu\nu}(f, \hat{k}) = \tilde{h}_{\mu\nu}^*(-f, \hat{k})$. This field will form the basis of the methods in this paper. Given a set of coordinate axes, $h$ and $\tilde{h}$ can be decomposed in terms of two polarisations, $+$ and $\times$ -- for example, ${h}_+(f, \hat{k})$ and ${h}_\times(f, \hat{k})$. These terms are not themselves scalar fields on the 2-sphere but can used to create fields of varying spin-weight.

One such method is to directly use the $h_+$ and $h_\times$ amplitude signals. As the gravitational wave tensor is rank-2, it can be shown that combinations ${h}_+\pm i {h}_\times$ have spin-$\pm 2$ respectively -- consistent with the spin of the graviton. They can, therefore, be decomposed as
\begin{align}
{}_{\pm 2}a_{\ell m}(f) = \int \text{d}^2\hat{k} (h_+ \pm i h_\times)(f, \hat{k}) {}_{\pm 2}Y_{\ell m}^*(\hat{k}) \, .
\end{align}
These harmonics then can be combined into the more useful gradient (scalar), $G$, and curl (pseudo-scalar), $C$, combinations -- 
\begin{align} \label{Eq:GC-def}
a^G_{\ell m} =& +\frac{1}{\sqrt{2}}\left( {}_{+2}a_{\ell m} + {}_{-2}a_{\ell m} \right) \, ,
\notag\\
a^C_{\ell m} =& -\frac{i}{\sqrt{2}}\left( {}_{+2}a_{\ell m} - {}_{-2}a_{\ell m} \right) \, . 
\end{align}
Note the difference in coefficients between equations~\ref{Eq:CMB-EB} and \ref{Eq:GC-def}. The factor of $- \sqrt{2}$ is largely irrelevant for the discussion herein, what is important is the scalar and pseudo-scalar nature of the coefficients and their corresponding fields. 

Using these coefficients, the power spectra can be computed in an analogous way as equations~\ref{Eq:SpectraDef} and~\ref{Eq:SpectraDef-Measured}. 
We evaluate all amplitudes near some single frequency $f$ (see Section~\ref{Sec:Backgrounds} for some specific examples in which this is most useful), although a generic background may not be stationary in time and so can have correlations between different frequencies.
In the CMB example, the reality of the $I$, $Q$ and $U$ fields means that the $a^{I}_{\ell m}$, $a^{E}_{\ell m}$ and $a^{B}_{\ell m}$ harmonics satisfy equations~\ref{Eq:m-mrelation} and~\ref{Eq:RealCrossSpec}, but this is not true in this case. This is, of course, due entirely to the complex-valued Fourier transform of the original real fields $h_{\mu\nu}(t, \vec{x})$. 
This implies that the harmonics satisfy $a^{P}_{\ell m}(f) = (-1)^m a^{P*}_{\ell, - m}(-f)$ for $P = G, C$ and the angular spectra are Hermitian in terms of frequency, i.e. $C^{PP'}_\ell(f) = C^{PP'*}_\ell(-f)$. 

The auto-spectra, $C^{GG}_\ell$ and $C^{CC}_\ell$, are, as before, real, but in general the complex nature of the amplitudes will mean that equation~\ref{Eq:m-mrelation} will not always be true and the cross-spectrum can be complex. 
As $C^{GC}_\ell = C^{CG*}_\ell$, by construction, we will present the signal in terms of $\mathbb{R}[C^{GC}_\ell]$ and $\mathbb{I}[C^{GC}_\ell]$. The complex nature of the cross-spectrum is, in principle, moot as conservation of parity again implies that $C^{GC}_{\ell}(f) = 0$. Therefore, any measurement of a non-zero cross-spectrum will imply parity violation and so have implications for the physics of the background \citep[e.g.][]{Crowder:2012ik}.
In fact, we will see in sections~\ref{Sec:WhiteNoise} and~\ref{Sec:SinglePointSource} examples where backgrounds which explicitly violate parity symmetry can be purely imaginary.

An equivalent derivation for the harmonics and power spectra can be calculated using tensor spherical harmonics -- see \citet{Gair2014} in analogy to \citet{Kamionkowski1997Pol}.

As mentioned, the standard deviation between the true and observed spectra, as given in equation~\ref{Eq:SpectraDef-Measured}, is only valid when the initial fields are real. As that is not the case here the result is more complicated and is given in equations~\ref{Eq:Errors-for-Amplitude}.

\subsection{Gravitational Wave Stokes Parameters}
Alternatively, as is the case in CMB decomposition, we can combine the amplitude fields into Stokes parameters \citep[e.g.][]{Breuer1975,Lightman1979,Gubitosi2016}
\begin{align} \label{Eq:Stokes}
I^\mathrm{GW}(f, \hat{k}) =& \langle |h_+(f, \hat{k})|^2 \rangle + \langle |h_\times(f, \hat{k})|^2 \rangle \, ,
\notag\\
Q^\mathrm{GW}(f, \hat{k}) =&  \langle |h_+(f, \hat{k})|^2 \rangle - \langle |h_\times(f, \hat{k})|^2 \rangle \, ,
\notag\\
U^\mathrm{GW}(f, \hat{k}) =& - 2 \langle  \mathbb{R}[h_+(f, \hat{k}) h_\times^*(f, \hat{k})]\rangle \, ,
\notag\\
V^\mathrm{GW}(f, \hat{k}) =& - 2 \langle  \mathbb{I}[h_+(f, \hat{k}) h_\times^*(f, \hat{k})]\rangle \, .
\end{align}
Each parameter has the same meaning as its electromagnetic counterpart \citep{Jackson1975}:
\begin{itemize}
\item $I^\mathrm{GW}$ is the total power
\item $Q^\mathrm{GW}$ and $U^\mathrm{GW}$ are the linear polarisations
\item $V^\mathrm{GW}$ is the circular polarisation
\end{itemize}
The $\langle \cdots \rangle$ average is often \citep[e.g.][]{Seto:2008sr, Gubitosi2016} taken to be the ensemble average over all possible realisations, though in the electromagnetic case it is more usually assumed to be a temporal average. Neither average is completely well defined, especially for the case of gravitational waves.
The temporal average works in practice for the CMB because the observation times are much longer than both the period of the waves ($\sim 3.5 \times 10^{-13}$s) and the timescale for relevant changes to the signal, but LISA is sensitive to gravitational waves with hour periods, and nanohertz gravitational waves detectable by PTAs have periods of years to decades or longer. Because of this, it will not always be possible to measure the signal for long enough to compute such an average.
The ensemble averages have the issue that we do not always have a well-defined ensemble (or, equivalently, a well-defined probability distribution for the gravitational wave amplitudes). In this paper, we are mainly concerned with power spectra and their expected values, rather than the power spectra of the expected Stokes parameters. This is particularly an issue for the polarisation terms as, for many cases, the ensemble average of $Q^\mathrm{GW}$, $U^\mathrm{GW}$ and $V^\mathrm{GW}$ will be zero. This means that any spectra constructed from these will also be zero. However, any given realisation may be non-zero and so will have associated non-zero power spectra with non-zero expected values.

Because of these issues, it is mathematically convenient, and sometimes necessary, to ignore the averages and calculate the Stokes parameters at a single time or frequency.
This assumption is equivalent to removing some level of statistical freedom from the distribution. For example, assuming the signal to be monochromatic -- which we do in many cases below -- will have the same effect as ignoring a temporal average.
Specifically, this assumption characterises waves that are 100\% polarised, which is what one would expect from certain monochromatic sources such as binaries.

A significant consequence is that there are fewer degrees of freedom for the Stokes parameters than when considering the amplitude formalism in the above calculations -- four Stoke parameters combine to only give three amplitudes, satisfying $I^2 = Q^2 + U^2 + V^2$ for completely polarised waves. More generally the Stokes parameters for partially-polarised fields satisfy the inequality $I^2 \geq Q^2 + U^2 + V^2$ and their detailed analysis will be left to future work.

As is the case for electromagnetic Stokes parameters, $I^\mathrm{GW}$ is a scalar field and $V^\mathrm{GW}$ is pseudo-scalar and so we can construct $b^{I/V}_{\ell m}$ coefficients as in equation~\ref{Eq:CMB-T}. The combination $Q^\mathrm{GW} \pm i U^\mathrm{GW}$ are spin-$\pm 4$, rather than spin-$\pm 2$ as in electromagnetism, but we can construct ${}_{\pm 4}b_{\ell m}$ in the same way as equation~\ref{Eq:CMB-pm2} -- using the spin-$\pm 4$ spherical harmonics, ${}_{\pm 4}Y_{\ell m}$, instead of spin-$\pm 2$. The scalar and pseudo-scalar combinations of ${}_{\pm 4}b_{\ell m}$ are respectively \citep[e.g.][]{Gubitosi2016}
\begin{align}
b^{E}_{\ell m}(f) =& -\frac{1}{2}\left( {}_{+4}b_{\ell m} + {}_{- 4}b_{\ell m} \right) \, ,
\notag\\
b^{B}_{\ell m}(f) =& +\frac{i}{2}\left( {}_{+4}b_{\ell m} - {}_{- 4}b_{\ell m} \right) \, .
\end{align}
From these harmonics, we can construct 4 auto-power spectra and 6 cross-power spectra, of which $C^{IV}_{\ell}$, $C^{IB}_{\ell}$, $C^{VE}_{\ell}$ and $C^{EB}_{\ell}$ are expected to be zero by parity conservation. 
Even in the case of parity violation, the reality of the Stokes parameters leads to power spectra (including the cross-spectra) that are also real -- by the same logic as in the CMB. This is in contrast to the $C^{GC}_\ell$ spectrum in the amplitude formalism.
Additionally, the spin-$\pm 4$ nature of the linear polarisation terms means any of the spectra involving $E$ or $B$ will be identically zero for $\ell = 0,1,2,3$, regardless of parity conservation.
For all cases below, the $\mathrm{GW}$ superscript will be dropped.

The reality condition on $h_{\mu\nu}(t, \vec{x})$ can be used to show that $I(-f, \hat{k}) = I(f, \hat{k})$, $Q(-f, \hat{k}) = Q(f, \hat{k})$ and $U(-f, \hat{k}) = U(f, \hat{k})$ but $V(-f, \hat{k}) = -V(f, \hat{k})$. The harmonics therefore satisfy $a^{N}_{\ell m}(-f) = a^{N}_{\ell m}(f)$ for $N \in \{I, E, B\}$ and $a^{V}_{\ell m}(-f) = -a^{V}_{\ell m}(f)$. All of the auto-spectra are therefore symmetric in frequency, as are the cross-spectra that do not include $V$ but are antisymmetric for those that do.

For any given background defined from the amplitudes (i.e. a pair of fields on the sphere, $h_+(f, \hat{k})$ and $h_\times(f, \hat{k})$, satisfying the constraints), it is clear how to obtain the Stokes parameters from their definition. For the inverse, in our case of 100\% polarisation, it is almost possible to generate the amplitude signal from a Stokes parameter background. If $h_{+/\times} = |h_{+/\times}| e^{i\phi_{+/\times}}$ then
\begin{align} \label{Eq:AmpFromStoke}
|h_+| =& \sqrt{\frac{1}{2}(I+Q)} \, ,
\notag\\
|h_\times| =& \sqrt{\frac{1}{2}(I-Q)} \, ,
\notag\\
\phi_+ - \phi_\times =& \arctan\left( \frac{V}{U} \right) \, 
\end{align}
-- i.e. $h_+$ and $h_\times$ can be obtained up to a phase factor. This phase factor can be defined arbitrarily (and is assumed to be $\phi_\times \sim \text{U}(0, 2\pi)$ where relevant) and does not affect the power spectra as it will cancel in each correlation function ($h_{A} h_{A'}^*$) that appear in the calculation of the power spectra.

For many (but not all) examples below, the monopole signal of $I$ will be very large and so $C^{II}_0$ will be many orders of magnitude larger than any other spectral term. In such cases, we will instead present $C^{II}_\ell$ with $C^{II}_0 = 0$, which is equivalent to removing the monopole. The same idea is used in CMB calculations, where only the fluctuations about the monopole $T_0 = 2.73 K$ are considered for the temperature spectrum and not the monopole itself.

\section{Gravitational-Wave Backgrounds} \label{Sec:Backgrounds}
\subsection{White noise} \label{Sec:WhiteNoise}
The simplest stochastic background that could be considered is an isotropic, uncorrelated, unpolarised and stationary background -- such as that expected from inflation or a phase transition -- satisfying
\begin{align} \label{Eq:IsoUncUnp}
\langle h_{A}(f, \hat{k}) h_{A'}^*(f', \hat{k}') \rangle = \frac{1}{2}H(f) \delta(f-f') \delta^2(\hat{k}, \hat{k}') \delta_{AA'} \, , 
\notag \\
\,\,\, A,A' \in \{+,\times \} \, .
\end{align}
Here the $\delta^2(\hat{k}, \hat{k}')$ is a delta function in direction space, i.e. $\delta^2(\hat{k}, \hat{k}') = 0$ for $\hat{k} \neq \hat{k}'$ and $\int \text{d}^2\hat{k} \delta^2(\hat{k}, \hat{k}') = 1$. The average represents a sample from e.g. a Gaussian distribution, with different frequencies and directions being uncorrelated. As in \citet{Gair2014}, this will lead to two white power spectra, $C^{GG}_\ell (f) = C_\ell^{GG}(f) = C(f)$ with no correlation between modes $C_\ell^{GC}(f) = C_\ell ^{CG}(f) = 0$. 

More generally, an amplitude background that is allowed to be anisotropic and polarised but still spatially uncorrelated and stationary -- for example in a cosmology with a preferred direction or a background with significant foreground sources, such a binaries within the Milky Way. Such a background will have amplitude signals obeying 
\begin{align} \label{Eq:AniWN-def}
\langle h_{A}(f, \hat{k}) h_{A'}^*(f', \hat{k}') \rangle = \frac{1}{2}g_{AA'}(f, \hat{k}) \delta(f-f') \delta^2(\hat{k}, \hat{k}') \, ,
\notag \\
\,\,\, A,A' \in \{+,\times \} .
\end{align}
Using the summation properties of spin-weighted spherical harmonics in equations~\ref{Eq:Sum-sYlm1} and \ref{Eq:Sum-sYlm2}, it can be shown that, for $\ell \geq  2$, 
\begin{subequations} \label{Eq:GGCC-aniwn}
\begin{align} 
C_\ell ^{GG}(f) =& C_\ell ^{CC}(f) = \frac{1}{4\pi}   \int \text{d}^2 \hat{k}  \left( g_{++}(f, \hat{k}) + g_{\times\times}(f, \hat{k})\right) \label{Eq:GGCC-aniwn-a}
\notag\\
=& \frac{1}{8\pi}  \int \text{d}^2 \hat{k} \langle I(f, \hat{k}) \rangle \, ,
\\
C_\ell ^{GC}(f) =& -C_\ell ^{CG}(f) = \frac{1}{4\pi}  \int \text{d}^2 \hat{k}  \left( g_{+\times}(f, \hat{k}) - g_{\times+}(f, \hat{k})\right) \label{Eq:GGCC-aniwn-b}
\notag\\
=& -\frac{i}{8\pi}  \int \text{d}^2 \hat{k} \langle V(f, \hat{k}) \rangle \, .
\end{align}
\end{subequations}
Other than $\ell = 0, 1$, this is independent of $\ell$ and so, unless the amplitudes have correlations between different directions, the power spectra are always white. 
We also satisfy the parity constraints if there is no correlation between the two polarisations or $h_+$ is statistically identical to $h_\times$ -- in either case this means $g_{+\times} = g_{\times +}$. 
Further, the power spectra in equations~\ref{Eq:GGCC-aniwn} are proportional to an average over all direction of the ensemble average of the $I$ and $V$ Stokes parameters, respectively. This implies, that the power spectra here are insensitive to linear polarisation ($Q$ and $U$), and are only sensitive to the monopole (i.e. direction averaged value) of $\langle I \rangle$ and $\langle V \rangle$, whose power it spreads over $\ell \geq 2$.  In this case, as mentioned in Section~\ref{Sec:GWamp}, the $C^{GC}_\ell$ power spectrum is purely imaginary. However, as $V$ is pseudo-scalar under a change of parity (i.e. $V \rightarrow -V$), we will still have $C^{GC}_\ell = 0$ in a parity-symmetric background,.

The observation that the $G$ and $C$ power spectra are independent of $\ell$ (for $\ell \geq 2$) will be seen several times. It arises because of the $\delta^2(\hat{k}, \hat{k}')$ term in equations~\ref{Eq:IsoUncUnp} and \ref{Eq:AniWN-def} which is, in turn, due to the following -- if the phase and magnitude of a $h_{A}$ amplitude signal are statistically independent quantities, then
\begin{align}
\langle h_{A}(f, \hat{k}) h_{A'}^*(f', \hat{k}') \rangle =& \langle |h_{A}(f, \hat{k})|| h_{A'}^*(f', \hat{k}') |\rangle  \langle e^{i \phi(\hat{k})} e^{-i \phi(\hat{k}')} \rangle \, . 
\end{align}
However, the phases of two gravitational wave amplitudes will be uncorrelated in many common scenarios. For example, to have two binary black hole systems in different directions with correlated phases requires the orbits of the black holes to have been correlated to a significant degree -- which is unlikely due to a large number of local factors that will affect the orbits. We therefore expect $\langle e^{i \phi(\hat{k})} e^{-i \phi(\hat{k}')} \rangle \propto \delta^2(\hat{k}, \hat{k}')$. 

While non-standard cosmological events, such as those of formation of primordial black hole, can, in principle, cause such correlations (i.e.  $\langle h_{A}(f, \hat{k}) h_{A'}^*(f', \hat{k}') \rangle \allowbreak \not{\propto} \allowbreak \delta^2(\hat{k}, \hat{k}')$), it is not expected and so will indicate something significant about the background that will require explanation \citep[e.g.][]{Gair2014}. One potential background that could have correlated phases is that of  standing modes in squeezed gravitational waves \citep[e.g.][]{Grishchuk1990}. 
These, in theory, can have correlated phases, particularly between $\hat{k}$ and $-\hat{k}$. As the $h_+ \pm i h_\times$ fields considered are spin-$\pm 2$, their power spectra will not be sensitive to the dipole term -- though they would be sensitive to similar correlations on smaller scales. Such cosmological signals will not be considered further here, however.

The issue of uncorrelated phases also causes problems with simulation of the backgrounds. We want to analyse the same white noise background in terms of both gravitational wave amplitude and Stokes parameters. It is, however, not possible to generate the full range of possible Stokes parameters correlators (e.g. $\langle I(f, \hat{k}) I(f', \hat{k}') \rangle$) from a background generated purely from the statistical distributions given in equations~\ref{Eq:IsoUncUnp} or \ref{Eq:AniWN-def}. This is down to the idea of what we assume is Gaussian. As in the CMB case, we assume that it is the Stokes parameters that are Gaussian distributed and, using this, we generate the amplitude field statistics accordingly. 
To do this, we generate linear and circular polarisations from assumed power spectra and, removing the averages in the definition of the Stokes parameters, set the overall power using $I^2 = Q^2 + U^2 + V^2$. For the cross-spectra, we justify assuming that $C^{EB}_\ell = C^{VE}_\ell = 0$ by conservation of parity and $C^{VB}_\ell = 0$ by assuming that the $\phi_+$ and $\phi_\times$ are uncorrelated with each other and $|h_+|$ and $|h_\times|$ -- see equation~\ref{Eq:VB-corr}. This assumption also leads to the white input $C^{VV}_\ell$ spectrum, as can be seen using equation~\ref{Eq:WN-VVcorr} and the same methods as for equations~\ref{Eq:GGCC-aniwn}. It does not, however, immediately mean that the input $C^{EE}_\ell$ and $C^{BB}_\ell$ spectra are white and so here they are given some, albeit identical, shape. Using all four Stokes parameters, we can then use equations~\ref{Eq:AmpFromStoke} to calculate the amplitudes.

\begin{figure*} 
\centering  
\subfigure[Stokes auto-power spectra]{
\includegraphics[width = 0.45\textwidth]{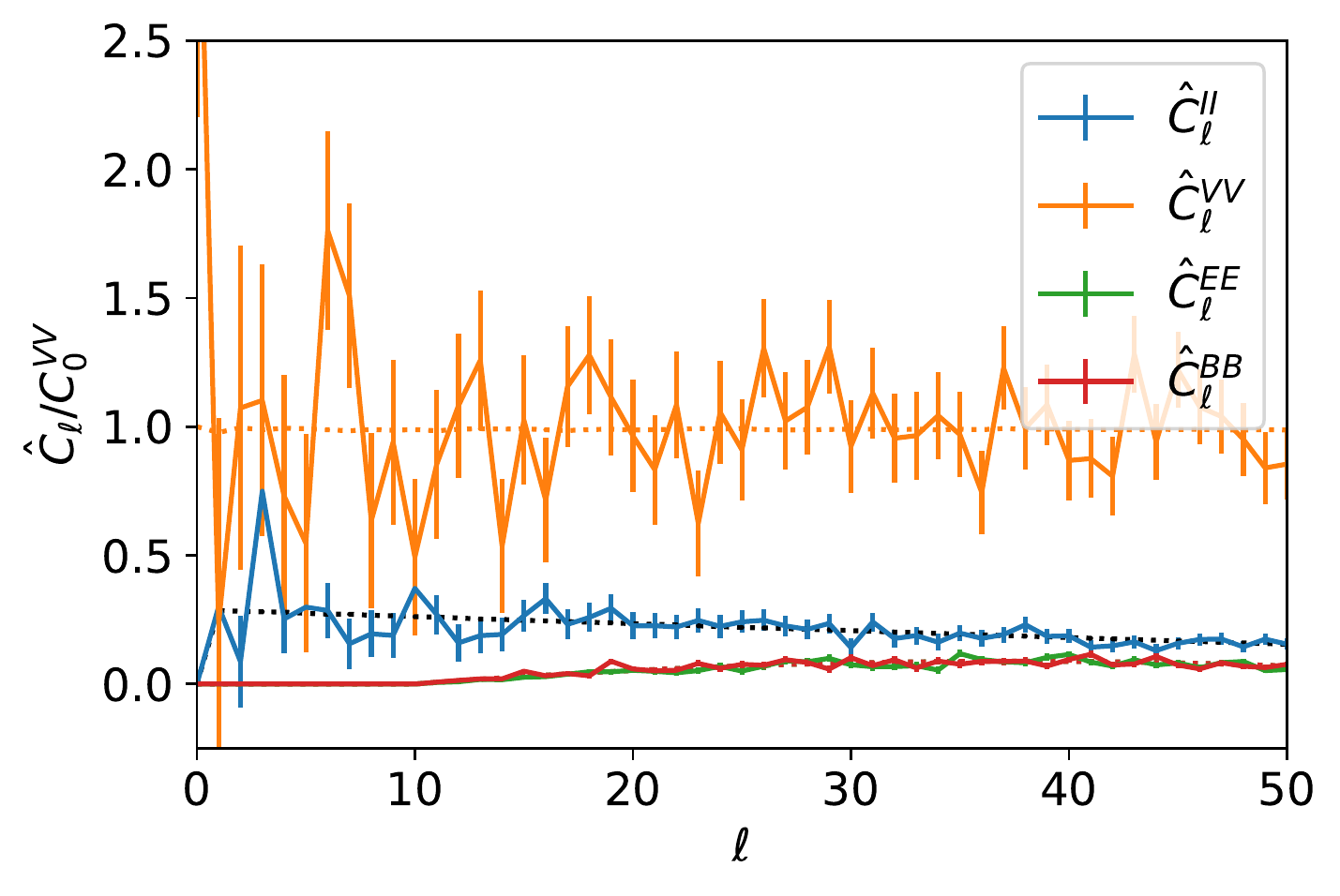}
\label{fig:c0c0VV}}
\subfigure[Stokes cross-power spectra]{
\includegraphics[width = 0.45\textwidth]{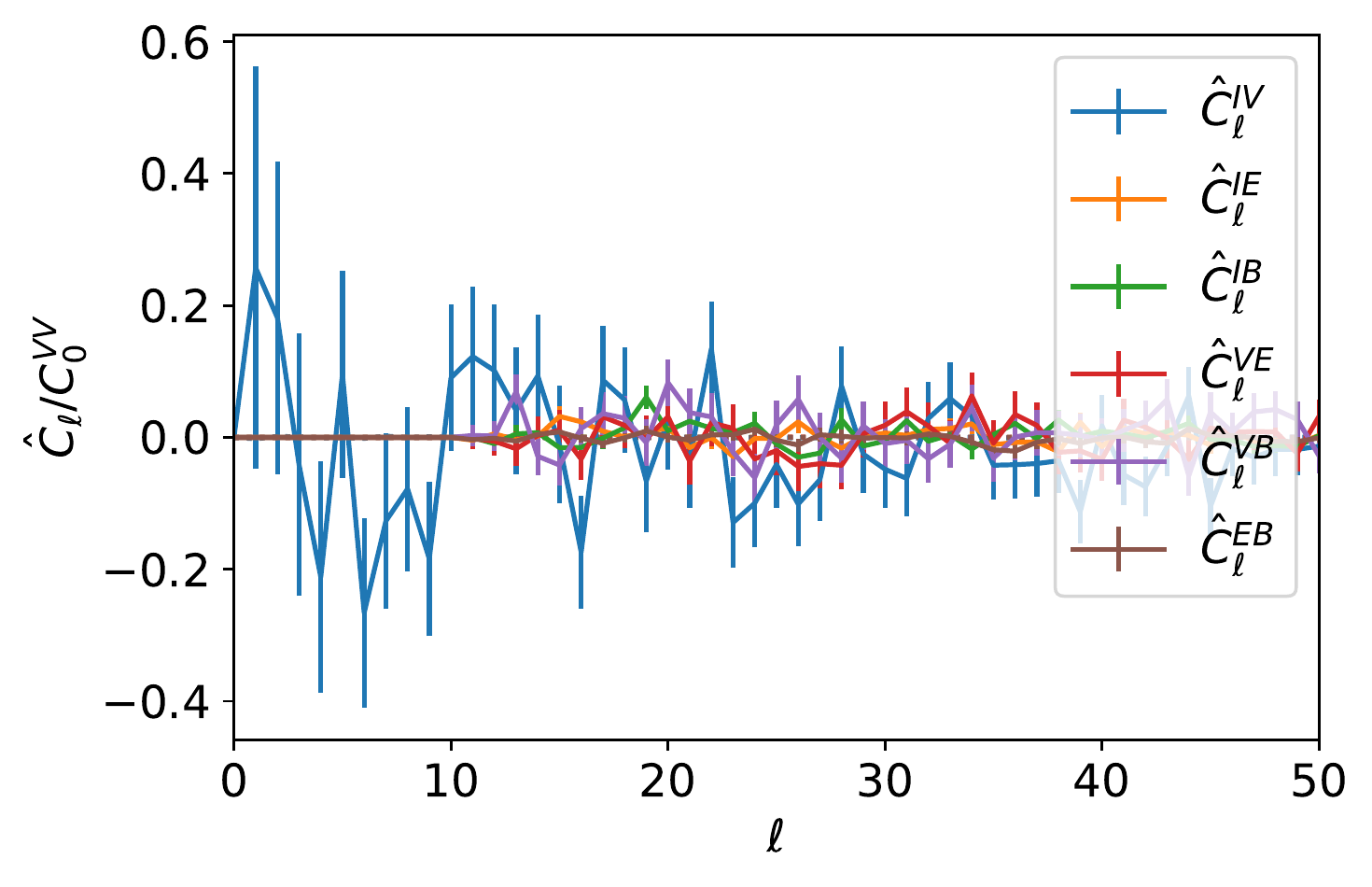}}
\subfigure[Amplitude auto-power spectra]{
\includegraphics[width = 0.45\textwidth]{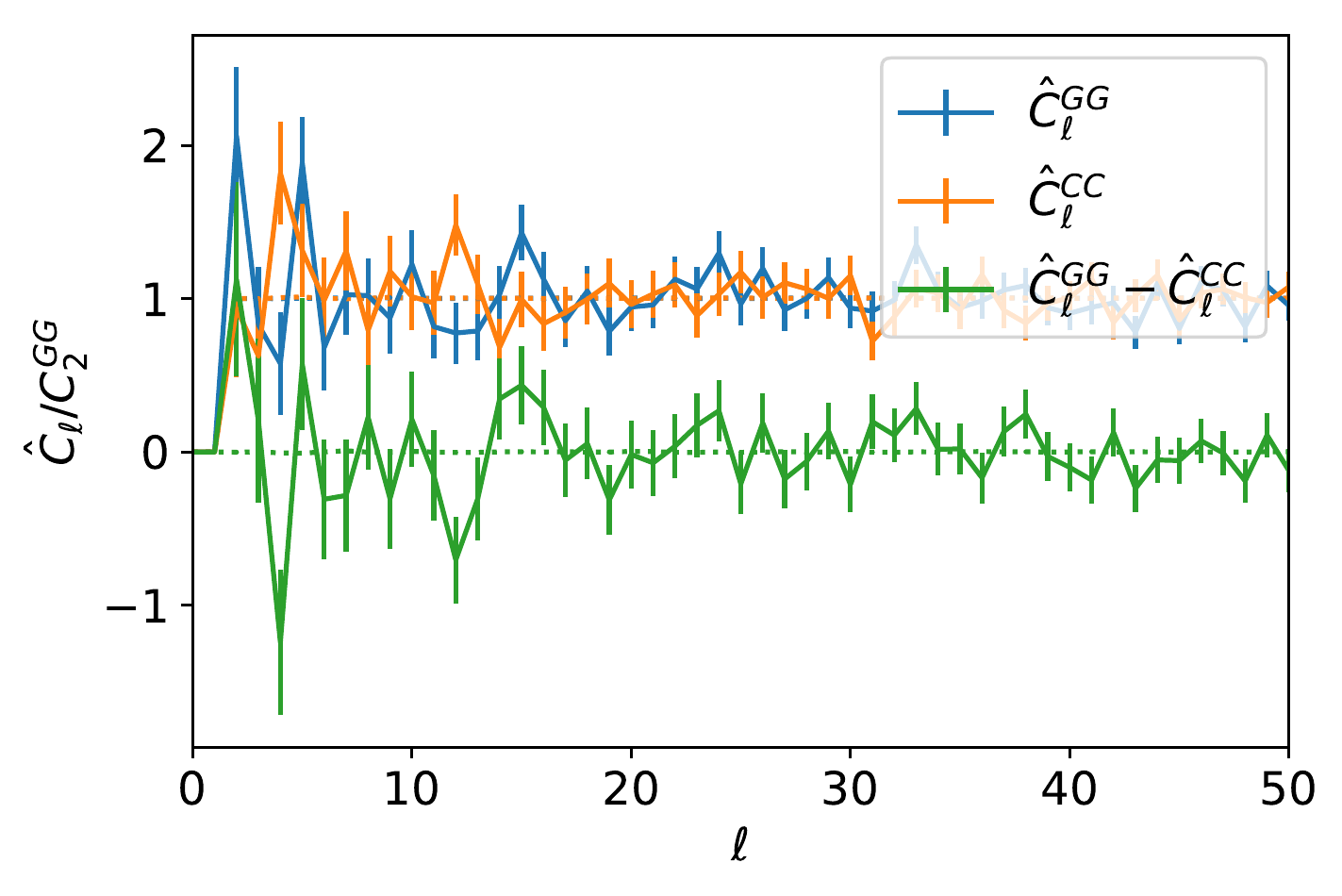}}
\subfigure[Amplitude cross-power spectrum]{
\includegraphics[width = 0.45\textwidth]{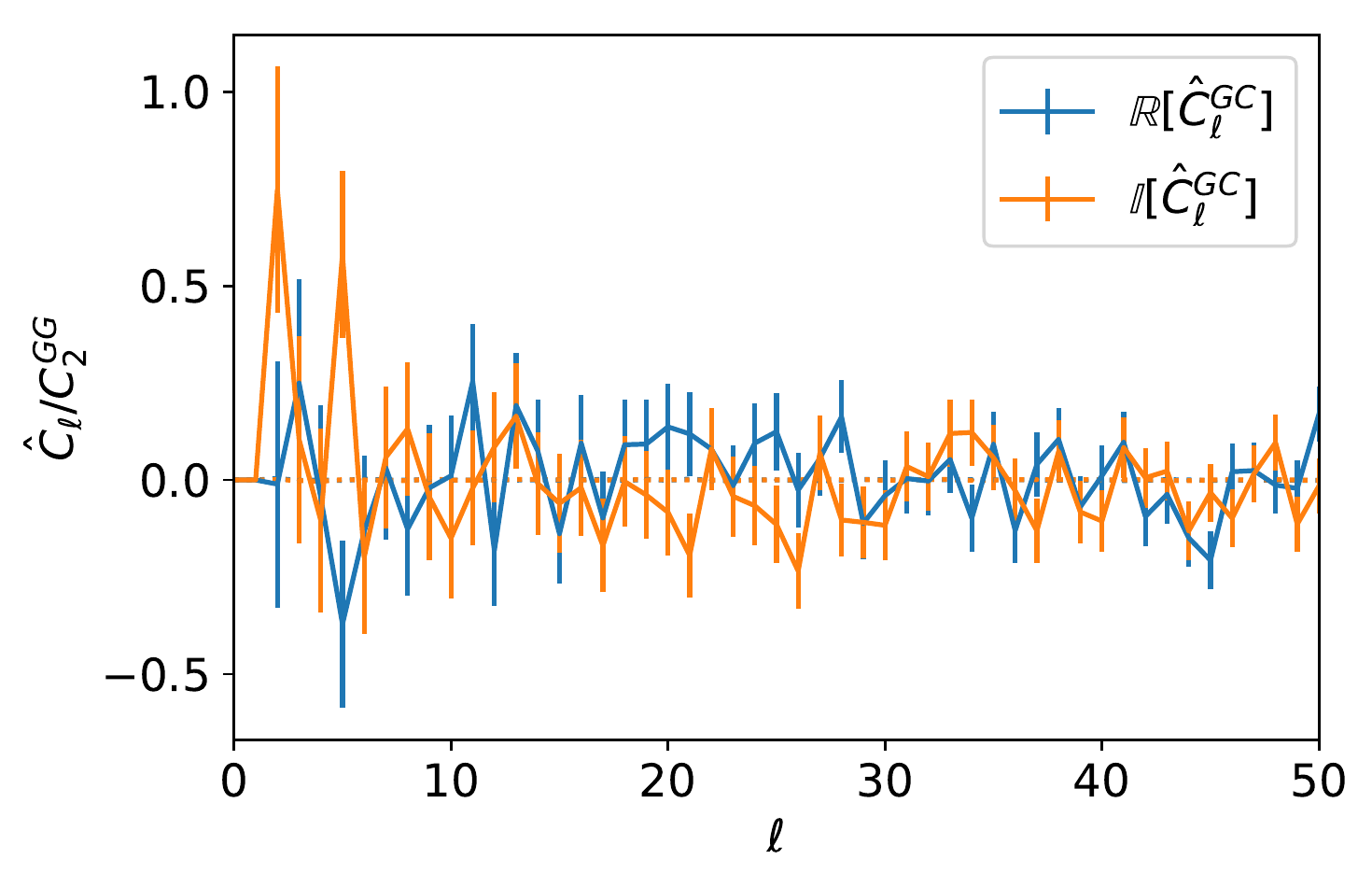}}
\caption{Power spectra for a  white noise background generated by known $C^{VV}_\ell$, $C^{EE}_\ell$ and $C^{BB}_\ell$ spectra. Solid lines are the spectra for a particular instance of the background, dashed lines indicate the power spectra averaged over 10,000 simulations, except in the cases of $C^{VV}_\ell$, $C^{EE}_\ell$ and $C^{BB}_\ell$ where the input spectra are used. Stokes parameter power spectra are normalised with respect to $C_0^{VV}$ (as the $C^{II}_0$ monopole has been removed) and amplitude power spectra with respect to  $C_2^{GG}$. } \label{Fig:WNPS}
\end{figure*}

The resultant power spectra from such a simulation are given in Fig.~\ref{Fig:WNPS}. Here we plot a single instance of the $\hat{C}_\ell$ power spectra in addition to the input value of the $C^{VV}_\ell$, $C^{EE}_\ell$ and $C^{BB}_\ell$ spectra and use an average over the 10,000 simulations as an estimate for the values of the remaining ensemble averaged spectra.
The errors bars used are defined by equation~\ref{Eq:SpectraDef-Measured} for the Stokes parameters (with $f_\text{sky} = 1$) and equations~\ref{Eq:Errors-for-Amplitude} for the $G$ and $C$ spectra. Note that, for simplicity, we have implicitly assumed that the amplitude fields, as well as the Stokes parameters, are Gaussian, as the effect on the error bars is negligible in this case.

We see that, as predicted, the $C_\ell^{GG}$ and $C^{CC}_\ell$ spectra are white and equal and both the real and imaginary parts of the $C^{GC}_\ell$ spectrum are zero -- as predicted by parity conservation. In contrast, the $C_\ell^{II}$ spectrum is not white -- and is in fact a slowly decreasing function of $\ell$. Here, unlike in other examples we will see, the $C_\ell^{II}$ spectrum is less than $C_\ell^{VV}$ for $\ell > 0$. This is a consequence of the removal of the monopole and Parseval's theorem, which can be shown to imply that
\begin{align}
\sum_\ell (2\ell + 1) C^{II}_\ell = \sum_\ell (2\ell + 1) \left( C^{VV}_\ell + C^{EE}_\ell + C^{BB}_\ell \right)
\end{align}
but the majority of the $I$ signal lies in $C^{II}_0$ and so there is less power distributed amongst other $\ell$.

As mentioned previously, the fact that we do not include any averages in our definition of the Stokes parameters has implied that the gravitational waves are monochromatic and 100\% polarised. This is one extreme. The other extreme is that the gravitational waves are completely unpolarised,  i.e. $Q = U = V = 0$ -- as is the case if we had assumed ensemble averages in our definition. In this case all of the Stokes parameter power spectra (auto- and cross-) would be zero except for $C^{II}_\ell$, and there would not be enough information to calculate any of amplitude power spectra.

\subsection{Galactic White dwarf Binaries}
One source of stochastic gravitational waves are galactic white dwarf binaries. It is predicted that there are around 100--300 million  white dwarf binary systems \citep[e.g.][]{Marsh2011, Nelemans:2013yg} of which only a fraction will be resolvable by LISA with a signal-to-noise ratio of more than five \citep[e.g.][]{Ruiter:2007xx,Timpano2006}.

A single binary system in a circular orbit\footnote{Circular orbits are chosen for simplicity and are justified by the fact that gravitational wave emission decreases the eccentricity of an orbit.}
is predicted to emit gravitational waves with amplitudes \citep[e.g.][]{Peters1963, Maggiore:1999vm} 
\begin{align} \label{Eq:WD-amplitude}
h_+(t) =& + A_+ \cos(2\psi) \cos(2\Omega t + \phi_0) + A_\times \sin(2\psi) \sin(2\Omega t + \phi_0) \, ,
\notag\\
h_\times(t) =& - A_+ \sin(2\psi) \cos(2\Omega t + \phi_0) + A_\times \cos(2\psi) \sin(2\Omega t + \phi_0) \, ,
\end{align}
where
\begin{align} \label{Eq:WD-coeff}
A_+ =& + \frac{2G^2 M_1 M_2}{c^4 r} \left( \frac{\Omega^2}{G(M_1 + M_2)} \right)^{1/3} (1+ \cos^2\varphi) \, ,
\notag\\
A_\times =& - \frac{4G^2 M_1 M_2}{c^4 r} \left( \frac{\Omega^2}{G(M_1 + M_2)} \right)^{1/3} \cos\varphi \, ,
\end{align}
$M_{1,2}$ are the masses of the white dwarfs, $\Omega$ the orbital angular frequency of the binary, $\phi_0$ the initial phases, $r$ the distance from the observer at Earth to the binary and $\psi$ and $\varphi$ are the 
principal polarisation and inclination, respectively, which describe the binary orientation as viewed by said observer. The inclination angle is the angle between the angular momentum direction of the binary, $\vec{L}$, and the direction to the binary from an observer at the solar barycentre, $-\hat{k}$. The principal polarisation angle is the angle of the semi-major axis of the binary with respect to the coordinates of the observer \citep{Rubbo2004,Timpano2006}. 

\begin{figure} 
\centering 
\subfigure[Probability of having a white dwarf binary]{
\includegraphics[width = 0.48\textwidth]{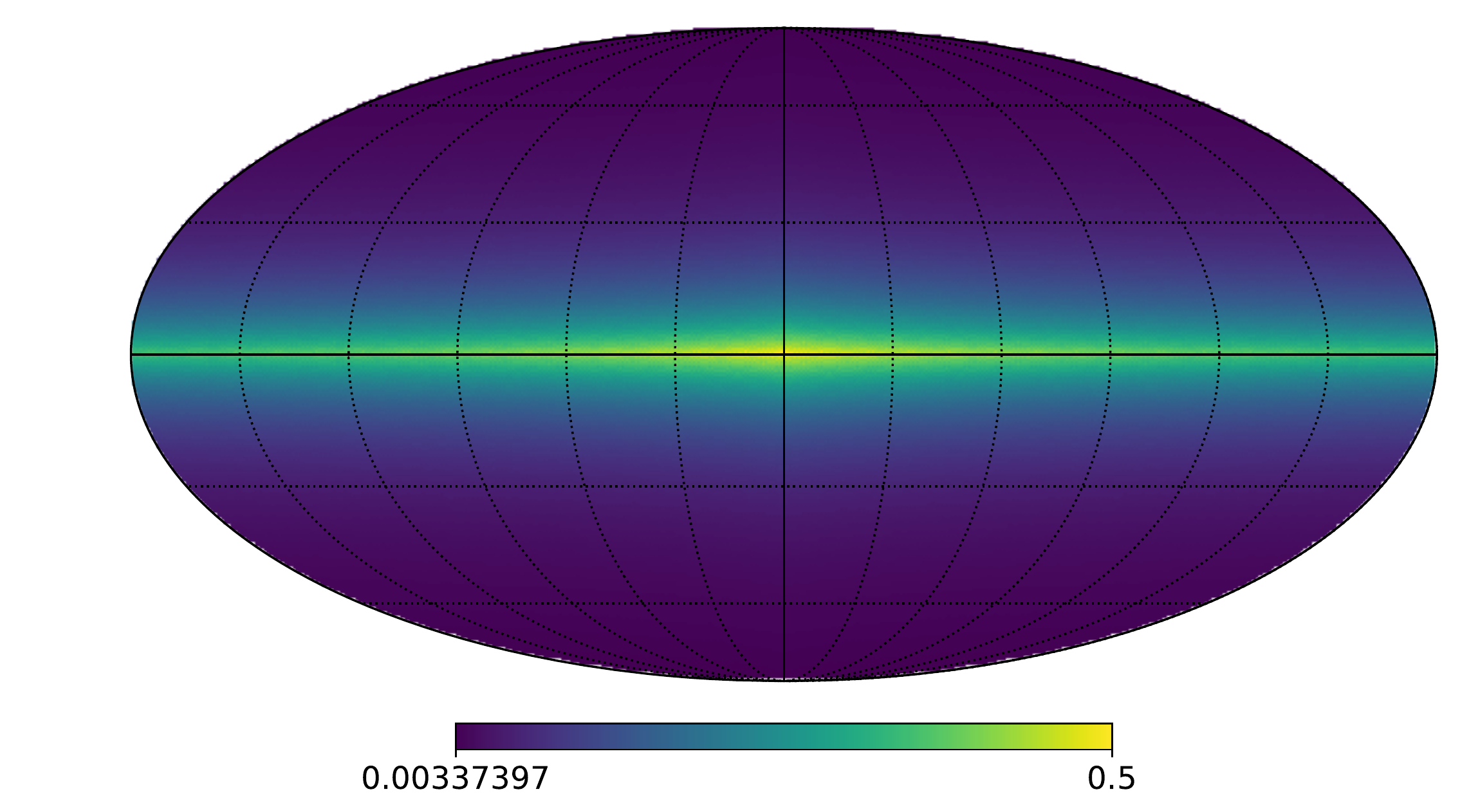}}
\subfigure[Positions of white dwarf binaries]{
\includegraphics[width = 0.48\textwidth]{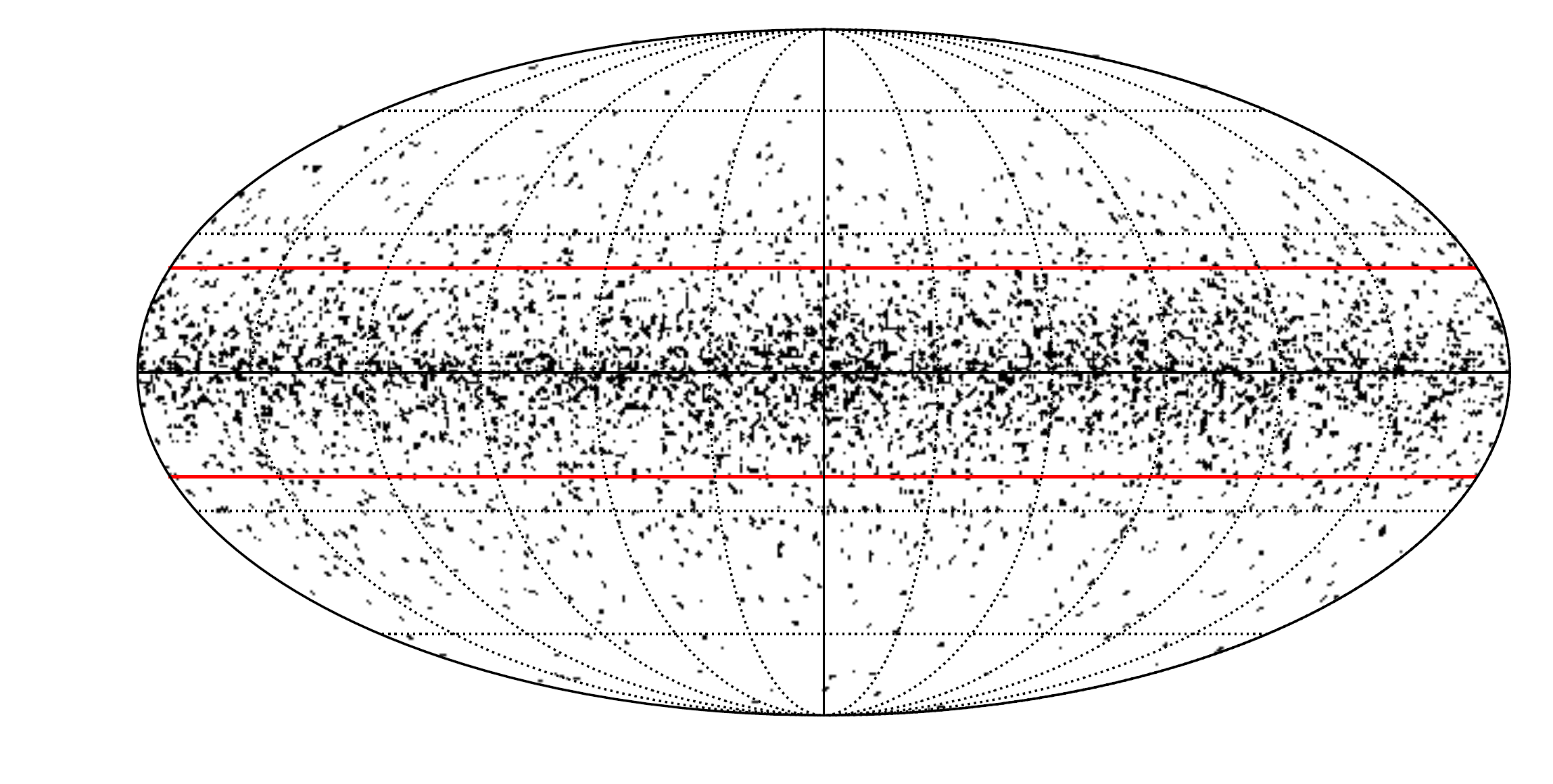}}
\caption{Probability distribution and positions of 4448 white dwarf binaries in a particular simulation. The red lines enclose the region containing 75\% of the binaries and so approximate our mask.} \label{Fig:WDPos}
\end{figure}

In frequency space and considering only frequency $f^* = \Omega/\pi$ this is 
\begin{align}
{h}_+(f^*) =& +\frac{1}{2} \left( A_+ \cos(2\psi) e^{i\phi_0} - i A_\times \sin(2\psi) e^{i\phi_0} \right) \, ,
\notag\\
{h}_\times(f^*) =& -\frac{1}{2} \left( A_+ \sin(2\psi) e^{i\phi_0} + i A_\times \cos(2\psi) e^{i\phi_0} \right) \, .
\end{align}
The probability distribution for the white dwarfs is modelled to be something similar to the galactic shape -- as shown in Fig.~\ref{Fig:WDPos}. Specifically, each direction (i.e. each pixel) is given an independent number $t \sim \text{U}(0,1)$ and if this is less than that pixel's value in the probability distribution (in Fig.~\ref{Fig:WDPos}) then the pixel is defined to have a binary. The distributions of $\varphi$, $\psi$ and $\phi_0$ are taken from \citet{Timpano2006} -- i.e. $\cos(\varphi) = - \hat{k} \cdot \vec{L}/|\vec{L}| \sim \text{U}(-1,1)$, $\psi \sim \text{U}(0,\pi)$ and $\phi_0 \sim \text{U}(0,2\pi)$. 

The distribution of white dwarf masses is in reality complicated, and for simplicity is assumed to be uniform -- $M_1, M_2 \sim \text{U}(0.4, 1.4) M_\odot$. The upper bound is due the Chandrasekhar limit and the lower is due to low mass stars not having had time to evolve. 
The distance to the binary, $r$, is modelled in such a way as to capture the Earth's location in the galaxy.
The frequency of the emitted radiation, $f^*$, is assumed to be constant for all sources in a background and is set to be 1 mHz, which lies in the LISA frequency band for unresolved sources  \citep{Nelemans:2013yg}. 

The probability distribution in this case is clearly anisotropic
and can be approximated as a mask on an isotropic distribution. As this is not actually a mask with a hard cut off we approximate the effect by choosing the region containing 75\% of the white dwarf binaries. This is  $(\theta,\phi) \in [\pi/2-0.39, \pi/2+0.39] \times [0, 2\pi) $ and corresponds to $f_\text{sky} \approx 0.249$. This will affect the error bars in accordance with equation~\ref{Eq:Error}. However, as the underlying distribution is highly non-Gaussian (in both the position of sources and signal strength), the errors bars plotted are calculated as one standard deviation over 1024 simulations, rather than using equations~\ref{Eq:Error} and~\ref{Eq:Errors-for-Amplitude}.

\begin{figure*} 
\centering 
\subfigure[Stokes auto-power spectra]{
\includegraphics[width = 0.45\textwidth]{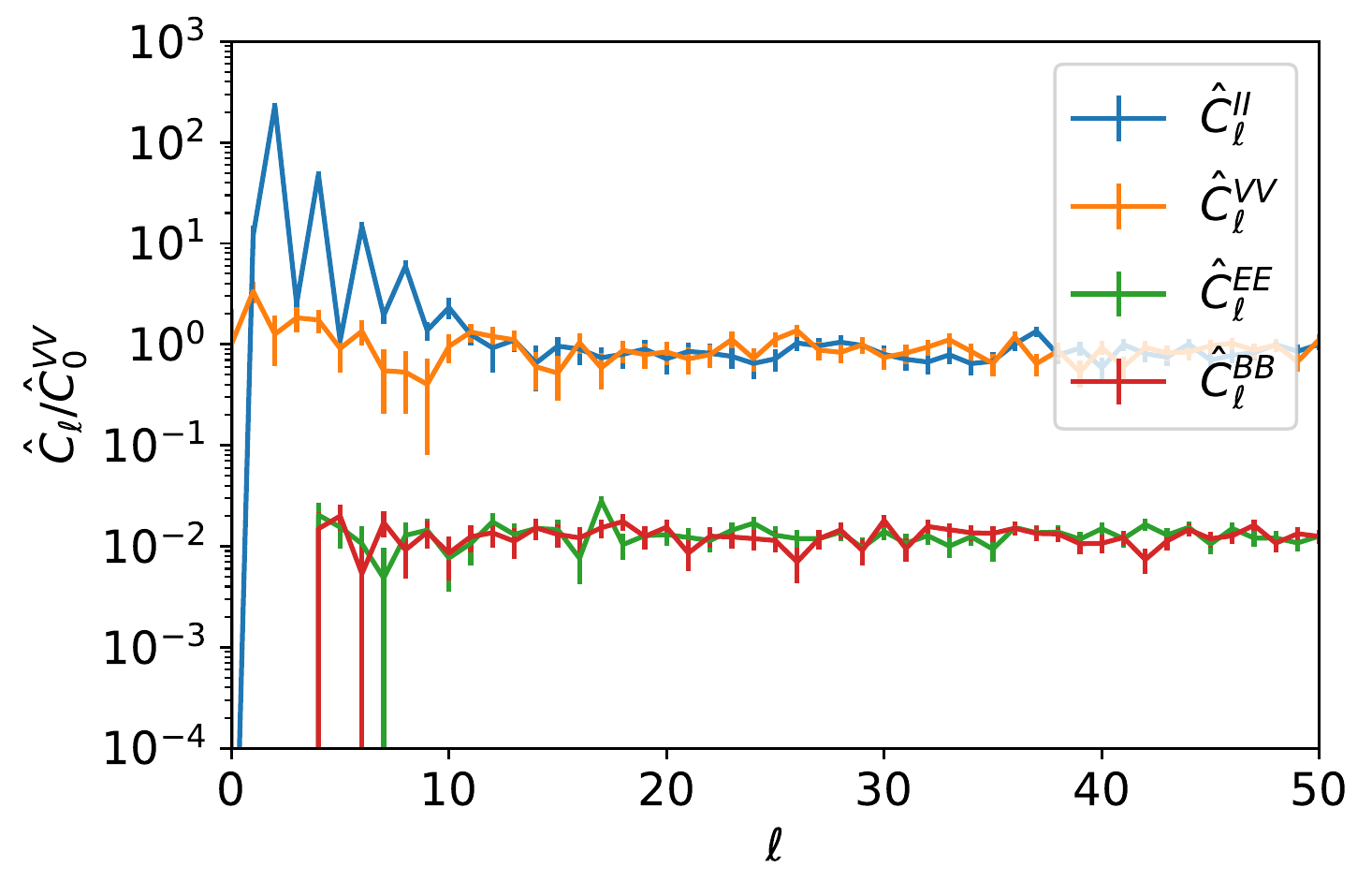}}
\subfigure[Stokes cross-power spectra]{
\includegraphics[width = 0.45\textwidth]{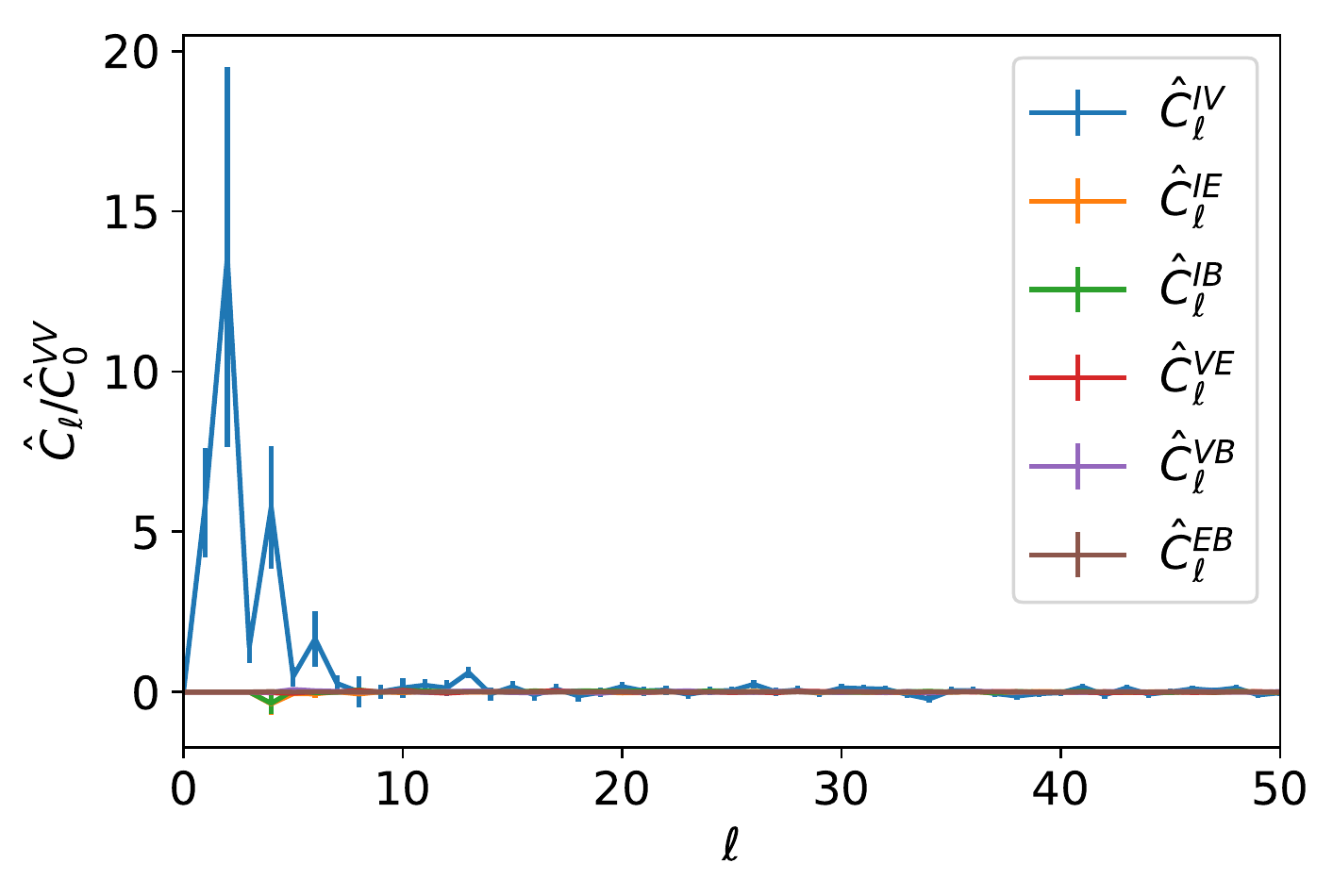}}
\subfigure[Amplitude auto-power spectra]{
\includegraphics[width = 0.45\textwidth]{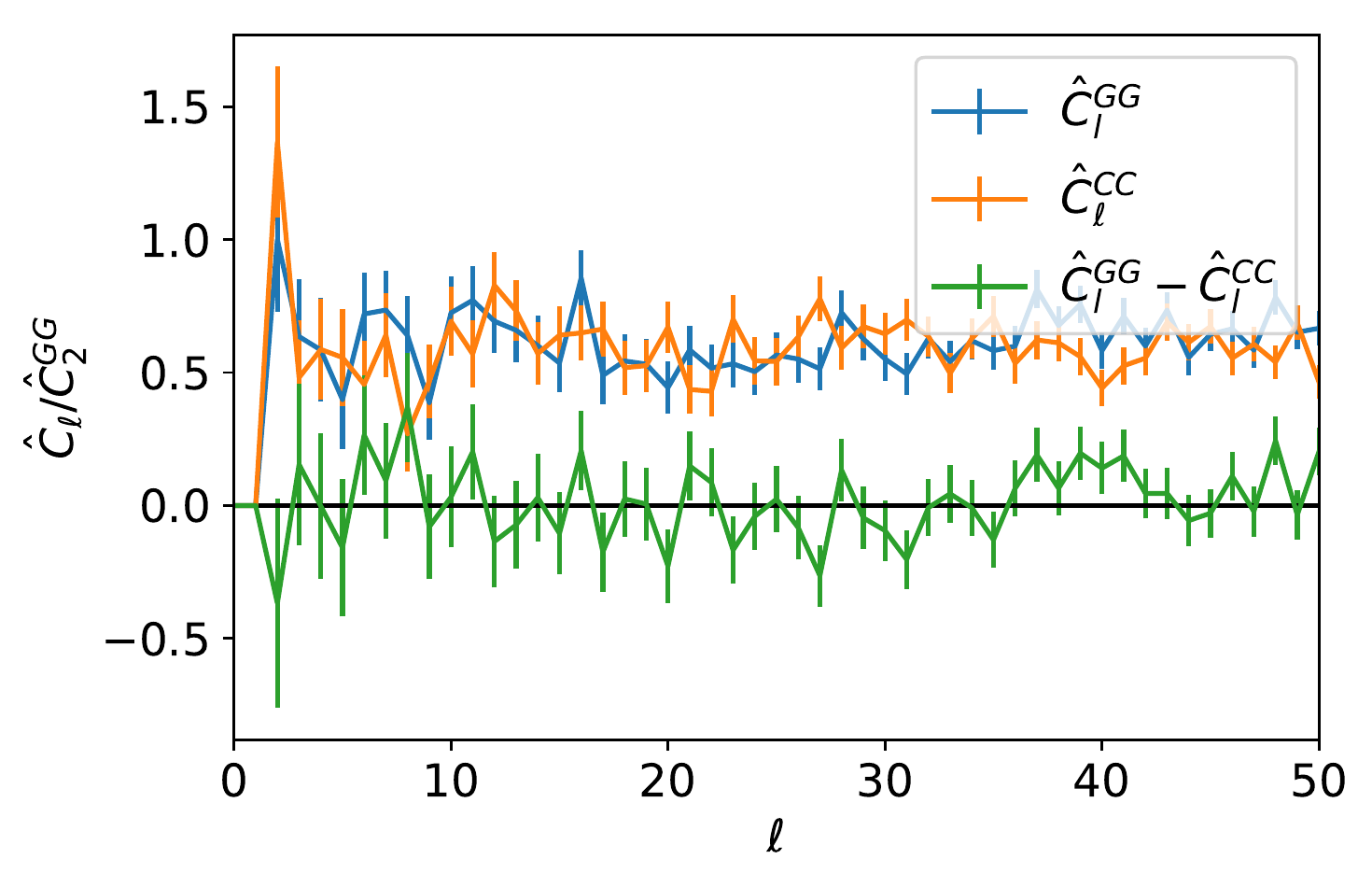}}
\subfigure[Amplitude cross-power spectrum]{
\includegraphics[width = 0.45\textwidth]{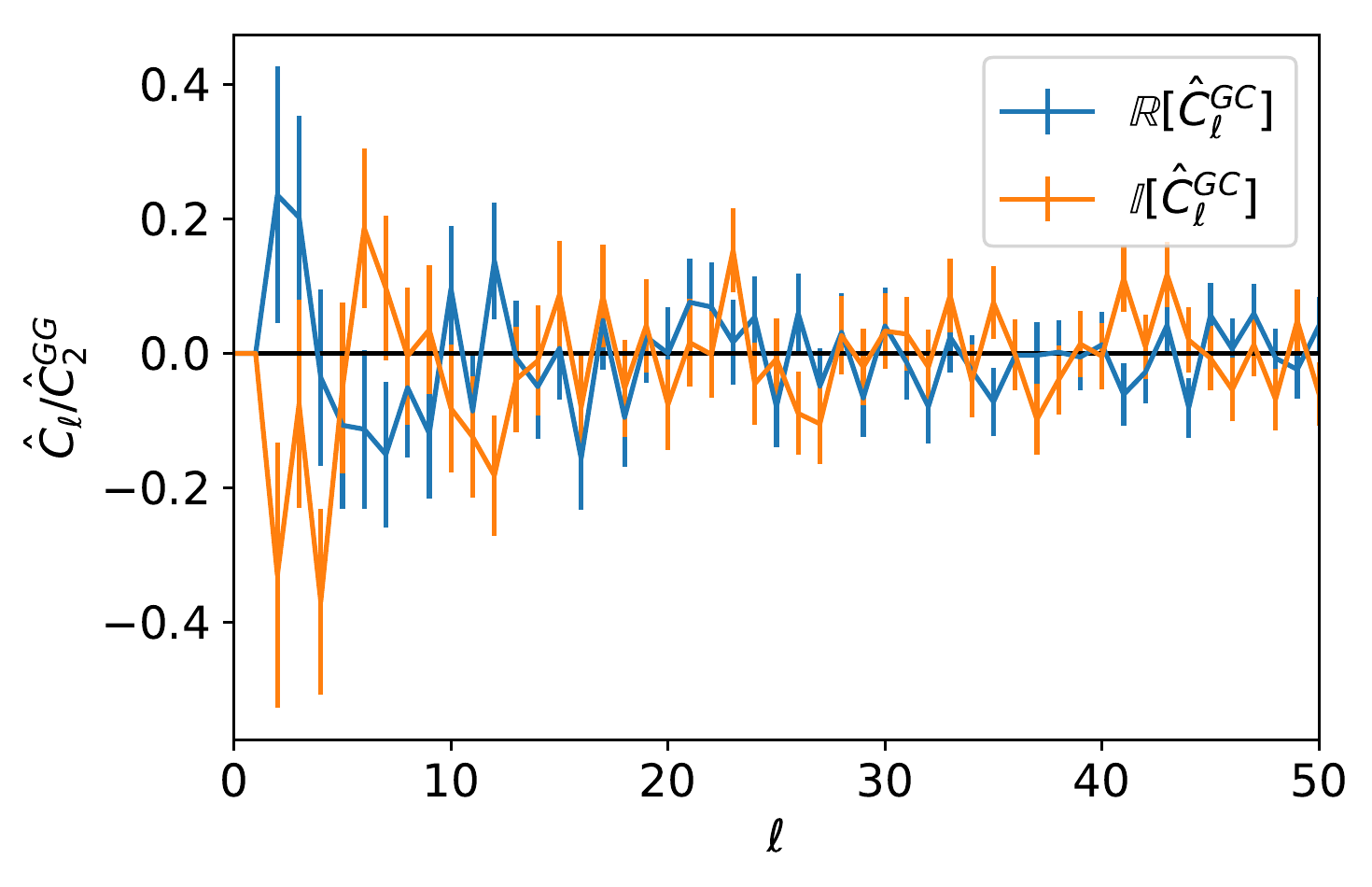}}
\caption{Power spectra for a white dwarf background. Amplitude power spectra are normalised with respect to $\hat{C}^{GG}_2$ and Stokes parameters with respect to $\hat{C}^{VV}_0$. For clarity, a horizontal black line shows $C_\ell=0$ where appropriate.}
\label{Fig:WDPS}
\end{figure*}

Each binary is given a set of parameters and the amplitudes and subsequently the Stokes parameters are calculated. The resultant power spectra are given in Fig.~\ref{Fig:WDPS}. 
It can be seen that the majority of the power spectra are approximately white. In particular we can see that $\hat{C}^{GG}_\ell$ and $\hat{C}_\ell^{CC}$ are consistent with both each other and a white power spectrum -- as are $\hat{C}^{EE}_\ell$ and $\hat{C}_\ell^{BB}$.

\begin{figure*} 
\centering 
\subfigure[$\log(I/I_\mathrm{max})$ for a white dwarf background]{
\includegraphics[width = 0.48\textwidth]{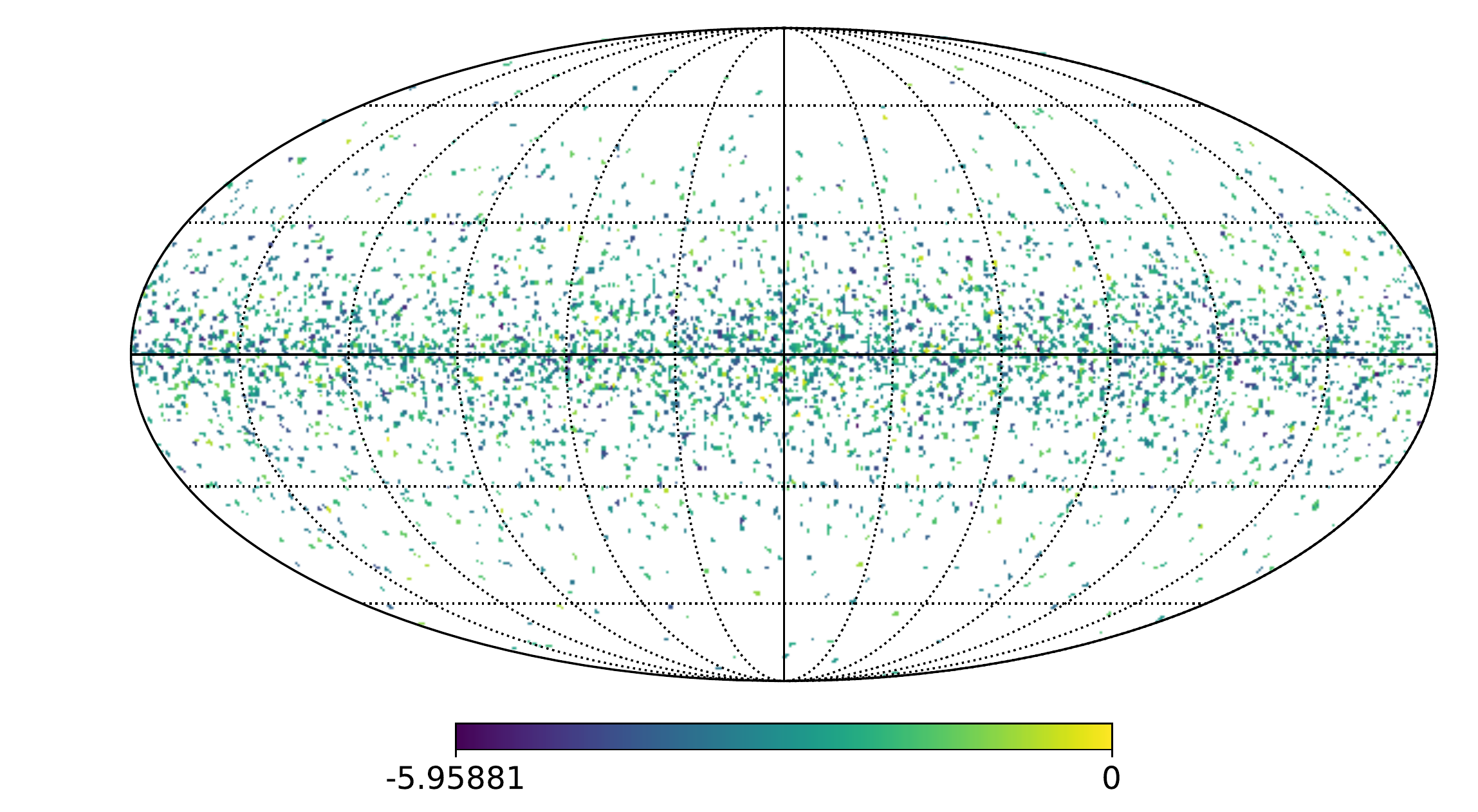}}
\subfigure[Effect of the mask on $C^{II}_\ell$]{
\includegraphics[width = 0.48\textwidth]{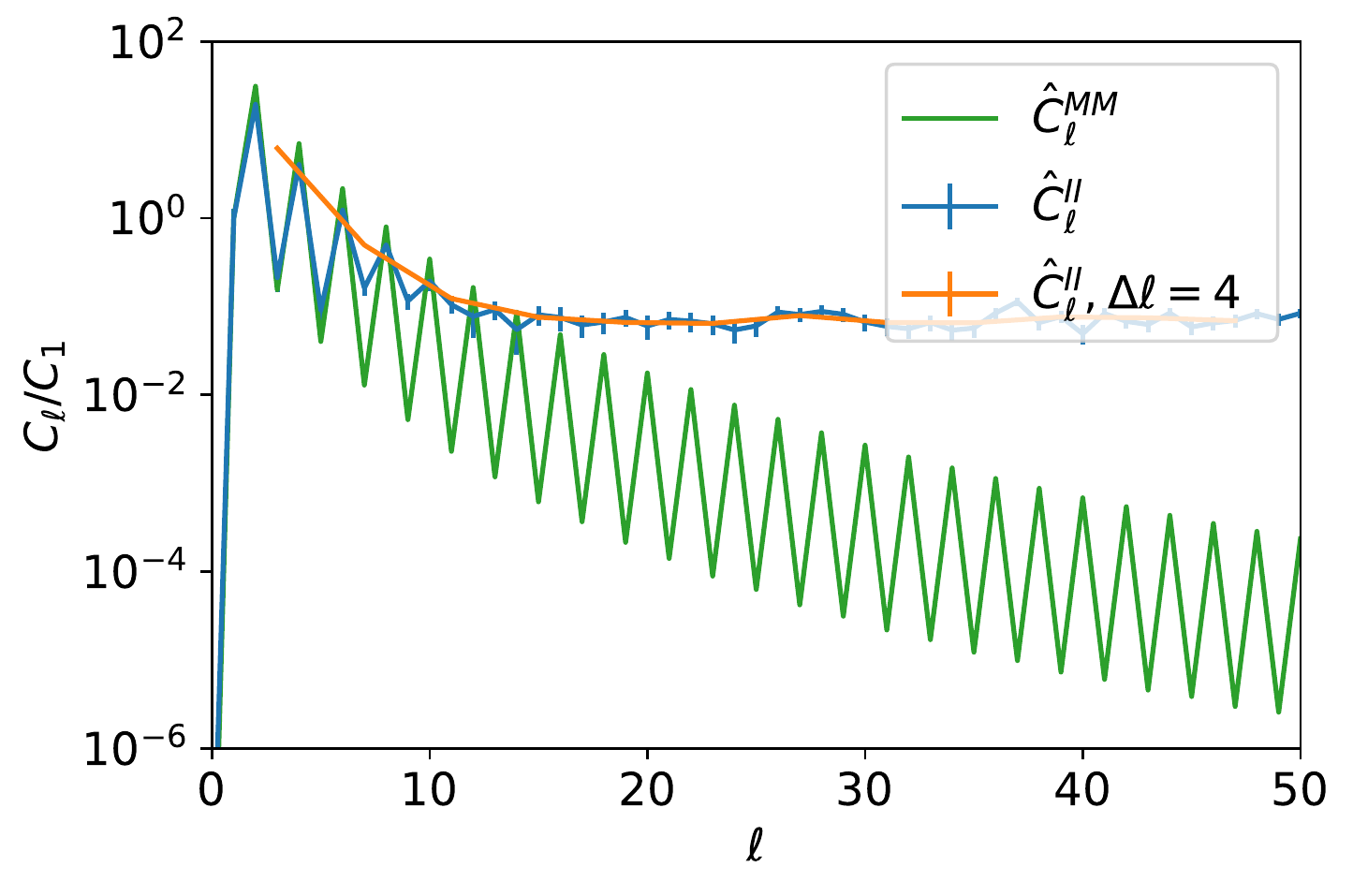}
\label{Fig:Mask-vs-I-b}}
\caption{Mollweide plot of the natural log of the $I$ Stokes parameter, normalised with respect to $\text{max}(I)$, for a white dwarf background, and the power spectra for the $I$ Stokes parameter for all $\ell \leq 50$, for $I$  averaged over $\Delta \ell = 4$ and for the probability distribution from Fig.~\ref{Fig:WDPos} ($M$). The power spectra for the $I$ field are normalised with respect to $\hat{C}^{II}_1$ and $\hat{C}^{MM}_\ell$ with respect to $\hat{C}^{MM}_1$. }
\label{Fig:Mask-vs-I}
\end{figure*}

The main interesting features of these power spectra lie in the relative amplitude of the $\hat{C}^{II}_\ell$ and $\hat{C}_\ell^{VV}$, and $\hat{C}^{EE}_\ell$ and $\hat{C}_\ell^{BB}$ spectra as well as the structure in the low $\ell$ $\hat{C}_\ell^{II}$ spectrum which can be viewed as a measure of the shape of the galaxy itself. Indeed, in Fig.~\ref{Fig:Mask-vs-I} we compute the power spectrum for the probability distribution by assuming it to be a scalar field and applying equations~\ref{Eq:CMB-T} and \ref{Eq:SpectraDef-Measured}. Using this, we can see that, for large scales, the $I$ power spectrum closely matches the probability distribution, only changing for small scales (large $\ell$) -- which follows from the approximately point source nature of the binaries. 

Also plotted in Fig.~\ref{Fig:Mask-vs-I} is the effect of the finite mask size on the independence of the harmonics. If we assume that there is some correlations of the scale of $\Delta \ell \sim 1/f_{\text{sky}} \sim 4$, as plotted, then the majority of the structure on large scale is averaged out. The broader shape -- i.e. a decrease is strength from $\ell = 1$ to $\ell = 15$ -- is still present and so this method could, in principle, still be used for measurement of the shape of the galaxy but, for the sake of generality, will not considered further here.

The cross-spectra that are predicted to be zero by parity conservation ($\hat{C}^{GC}_\ell$, $\hat{C}^{EB}_\ell$, $\hat{C}^{IV}_\ell$, $\hat{C}^{IB}_\ell$ and $\hat{C}^{VE}_\ell$) are consistent with zero, as expected, as are several which are not required to be ($\hat{C}^{IE}_\ell$ and $\hat{C}^{VB}_\ell$). The latter observation follows from Appendix~\ref{App:CorrSP} and because in this simulation the phases are independent and $h_+$ and $h_\times$ are statistically identical. Despite being consistent with zero, there appears to be structure in the $\hat{C}^{IV}_\ell$ spectrum, but will be mostly due the fact that the $I$ and $V$ harmonics have a larger magnitude than the $E$ and $B$ and so the $\hat{C}^{IV}_\ell$ cross-spectrum will be larger for any given background.

It is worth noting that, as we have restricted to only a single binary in any given direction, we have explicitly assumed monochromaticity of the source in each direction. As such, the considered background is 100\% polarised and so this presents an upper limit on the power spectra of the polarisation terms (e.g. $C^{VV}_\ell$, $C^{EE}_\ell$ and $C^{BB}_\ell$).

In a future work (Brevik et al., 2018, in preparation) we will consider a more realistic simulation of the white dwarf background using full population synthesis methods to characterise the distribution of sources. The characterisation of these simulations in terms of their angular and frequency spectra will enable experiments such as LISA to distinguish evolutionary models for the white dwarf population.

\subsection{Single point source} \label{Sec:SinglePointSource}
At the time of writing, a stochastic gravitational-wave background is yet to be detected. Individual signals have, however, been observed. In this section we consider the effect of such a source on the power spectra. 
We approximate the spatial distribution of a single point source by delta function $\delta^2(\hat{k}, \hat{k}_0)$, for some arbitrary direction $\hat{k}_0$.  The fact that is not possible to identify the direction of a gravitational wave point source with the same degree of accuracy as an electromagnetic signal will only effect the sensitivity of a detector, not the source power spectra.

The predicted power spectra can computed be analytically. For the general example, $h_{+/\times}(f, \hat{k}) = h_{+/\times}(f) \delta^2(\hat{k}, \hat{k}_0)$ and $N(f, \hat{k}) = N(f) \delta^2(\hat{k}, \hat{k}_0), N \in \{I, Q, U, V \}$ it can be shown, using equations~\ref{Eq:Sum-sYlm1} and \ref{Eq:Sum-sYlm2}, that
\begin{subequations} \label{Eq:DeltaFuncPred}
\begin{align} 
C^{GG}_\ell =& C^{CC}_\ell(f) = \frac{1}{4\pi}(|h_+|^2 + |h_\times|^2) = \frac{1}{4\pi} I(f) \, ,
\\
C^{GC}_\ell =& \frac{1}{4\pi}(h_+ h_\times^* - h_+^* h_\times) = -\frac{i}{4\pi} V(f) \, , \label{Eq:DeltaFuncPred-b}
\\
C^{II}_\ell =& \frac{I^2(f)}{4\pi} \, ,
\\
C^{VV}_\ell =& \frac{V^2(f)}{4\pi} \, ,
\\
C^{EE}_\ell =& C^{BB}_\ell = \frac{1}{8\pi}(Q^2(f) + U^2(f)) \, ,
\\
C^{IV}_\ell =& \frac{I(f)V(f)}{4\pi} \, ,
\\
C^{IE}_\ell =& C^{IB}_\ell = C^{VE}_\ell = C^{VB}_\ell = C^{EB}_\ell = 0 \, .
\end{align}
\end{subequations}
This means that all of the spectra are expected to be white (excluding where they are zero by construction) and, as in the white noise examples, the $C^{GC}_\ell$ is imaginary and not necessarily zero by construction but measures the level of circular polarisation of the background.  That the spectra are predicted to be white follows from the fact there is a single source in this model and so there is nothing for the source to correlate with.

\begin{figure*} 
\centering 
\subfigure[Stokes auto-power spectra]{
\includegraphics[width = 0.48\textwidth]{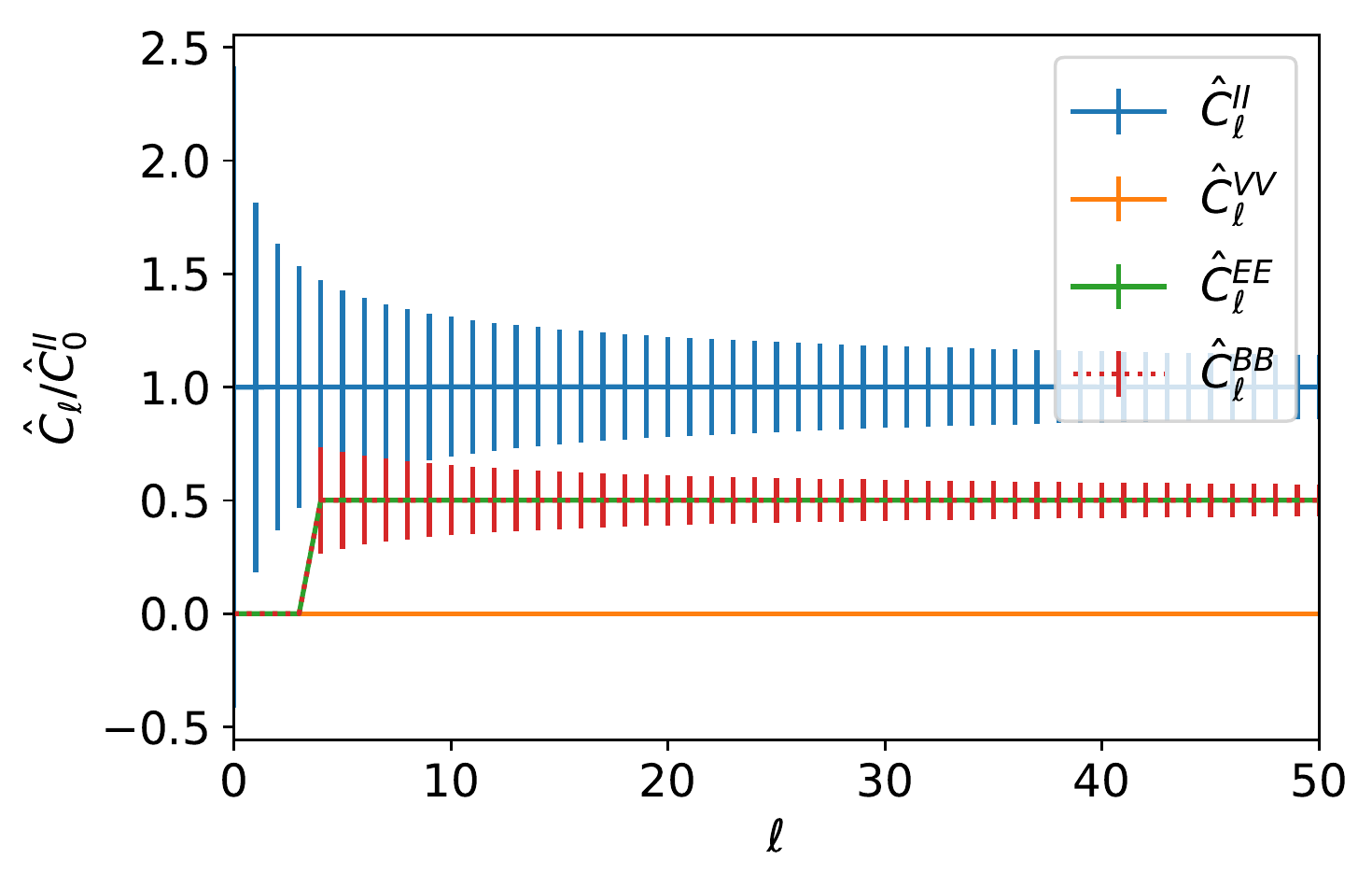}}
\subfigure[Stokes cross-power spectra]{
\includegraphics[width = 0.48\textwidth]{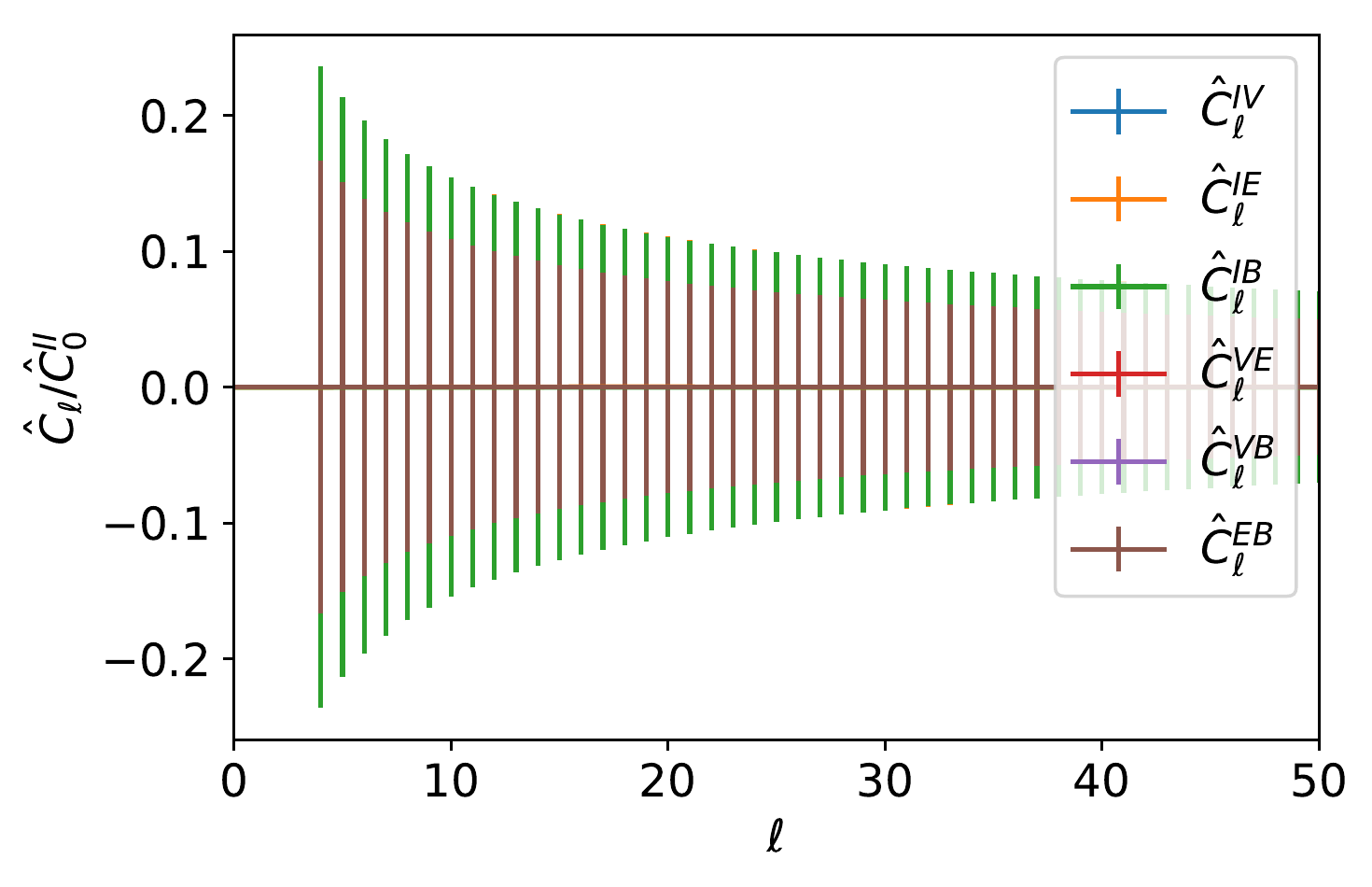}}
\subfigure[Amplitude auto-power spectra]{
\includegraphics[width = 0.48\textwidth]{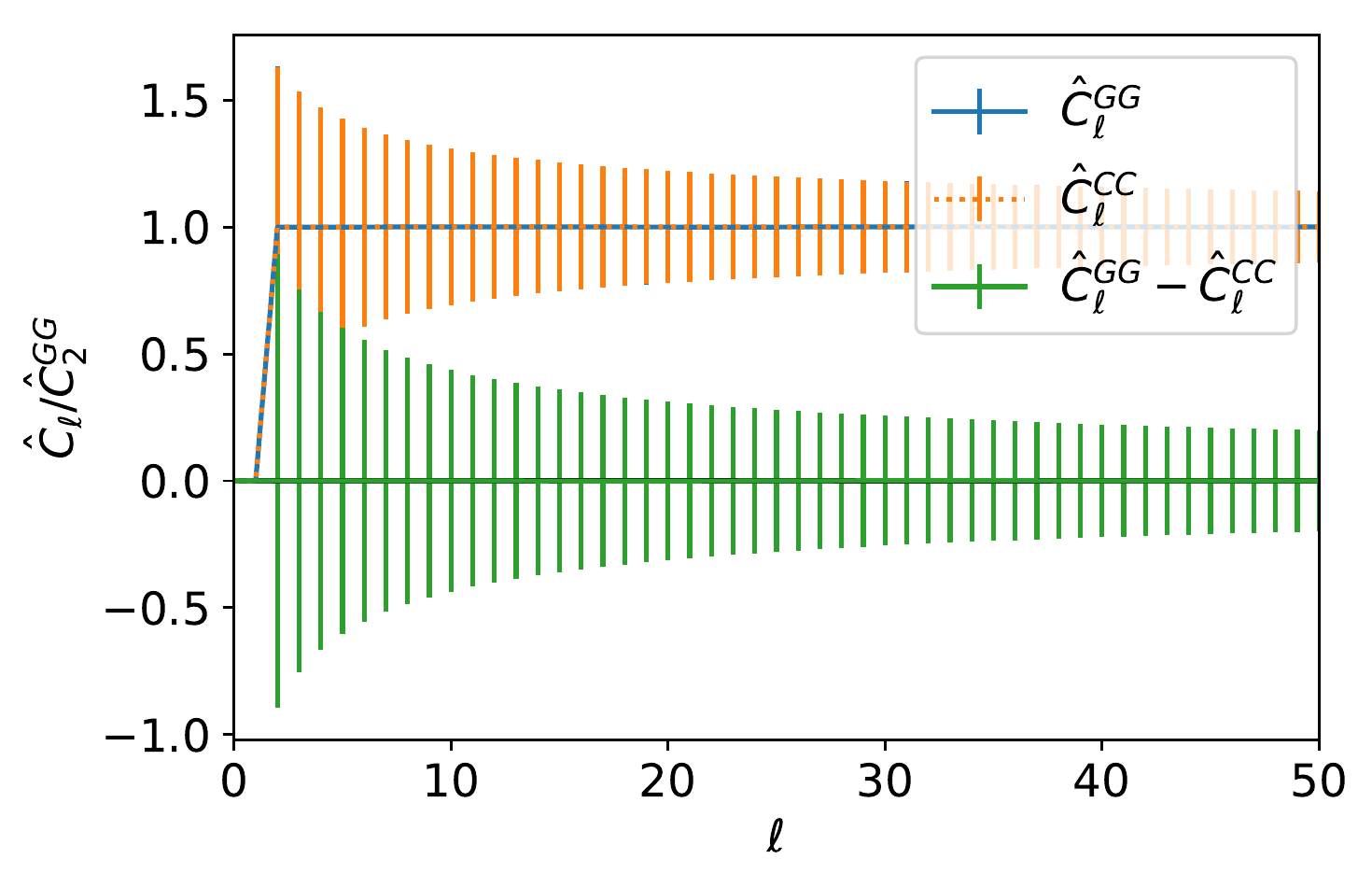}}
\subfigure[Amplitude cross-power spectrum]{
\includegraphics[width = 0.48\textwidth]{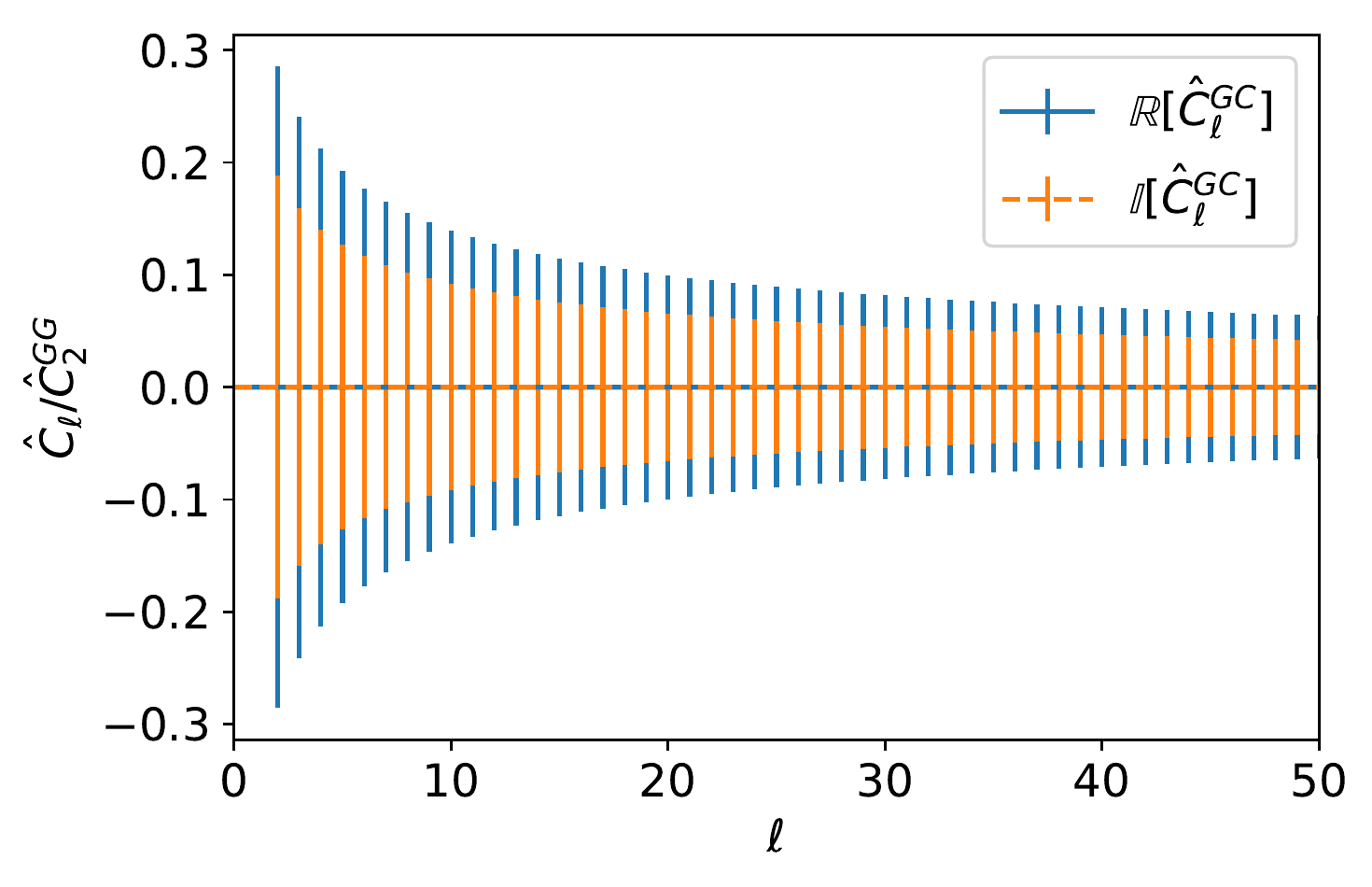}}
\caption{Power spectra for a single source ($h_+ = e^{4.2122 i}$, $h_\times = 0$) on the sky. Stokes parameter power spectra are normalised with respect to $\hat{C}_0^{II}$ and amplitude power spectra  with respect to $\hat{C}_2^{GG}$.}
\label{Fig:DeltaFunction}
\end{figure*}

The power spectra for a single polarised ($h_+$) signal on the sky are calculated numerically and plotted in  Fig.~\ref{Fig:DeltaFunction}. This single source gives Stokes parameters of $I = Q = |h_+|^2$ in the same direction and $U = V = 0$. Here we can see that for this simple example, there are no significant features in any of the auto-power spectra -- they are constant for all $\ell$, except where they are zero by the definition of the spin-weighted spherical harmonics. This is consistent with the predicted power spectra from equations~\ref{Eq:DeltaFuncPred} -- noting that $\hat{C}^{GC}_\ell = \hat{C}^{VV}_\ell = 0$ and $\hat{C}^{EE}_\ell = \hat{C}^{BB}_\ell = \hat{C}^{II}_\ell/2$ because $V = 0$ and $U = 0$, respectively.

While the source distribution is not Gaussian (we choose $|h_+| = \delta^2(\hat{k}, \hat{k}_0)$ and $\phi_+ \sim U(0, 2\pi)$), the error bars presented in Fig.~\ref{Fig:DeltaFunction} are those given by equations~\ref{Eq:Error} and~\ref{Eq:Errors-for-Amplitude} as a representative example. Note that because the real and imaginary parts of $h_+$ are not equal, neither are the error bars for $\mathbb{R}[C^{GC}_\ell]$ and $\mathbb{I}[C^{GC}_\ell]$.

\subsection{Multiple Sources}
\citet{Mingarelli2017} (hereafter \citetalias{Mingarelli2017}) assembled a galaxy catalogue of massive galaxies from the 2 Micron All Sky Survey (2MASS; \citealt{2MASS2006}) and computed the probability of each galaxy hosting a SMBHB emitting gravitational waves in the PTA band. Moreover, they computed the expected contribution of these nearby SMBHBs to the isotropic gravitational-wave background, and using methods from \cite{msmv13} and \cite{TaylorEtAl:2015}, found that these local sources contribute to anisotropy in the background at a level of $\sim 15-20\%$ of the monopole.

Here we build on this work, making heavy use of \citetalias{Mingarelli2017}'s open access code \citep{nano_gw}.
One notable difference between the method used in \citetalias{Mingarelli2017} and the version here is that we consider both the amplitude and Stokes parameters of the background.
\citetalias{Mingarelli2017} instead consider the characteristic strain, $h_c$, which is an inclination and principal polarisation averaged term. 
As we will see, the square of the characteristic strain is closely related to the power $I$. 

Assuming circular orbits, equations~\ref{Eq:WD-amplitude} and \ref{Eq:WD-coeff} apply, with a few modifications. Because the sources considered are extragalactic, the redshift of the emitting galaxy, $z$, is factored into the equations -- though the fact that we only consider galaxies out to 225 Mpc means that its numerical effect will be small.  
In this case, the distance $r$ is ambiguous and is replaced by luminosity distance $D_L$. Similarly, the orbital angular frequency, $\Omega$, is not directly observable and the observed gravitational wave frequency $f_o = \Omega/[\pi(1+z)]$ is used instead. Equations~\ref{Eq:WD-amplitude} and \ref{Eq:WD-coeff} therefore become
\begin{align} \label{Eq:Redshifted-amplitude}
h_+(t) =& A_+ \cos(2\psi) \cos(2\pi f_o t + \phi_0) + A_\times \sin(2\psi) \sin(2\pi f_o t + \phi_0) \, ,
\notag\\
h_\times(t) =& - A_+ \sin(2\psi) \cos(2\pi f_o t + \phi_0) 
\notag\\
&+ A_\times \cos(2\psi) \sin(2\pi f_o t + \phi_0)
\end{align}
and  
\begin{align} \label{Eq:Redshifted-coeff}
A_+ =& +\frac{2G^2 M_1 M_2}{c^4 D_{L}} \left( \frac{[\pi f_{o}(1+z)]^2}{G(M_1 + M_2)} \right)^{1/3} (1+ \cos^2\varphi) \, ,
\notag\\
A_\times =& - \frac{4G^2 M_1 M_2}{c^4 D_{L}} \left( \frac{[\pi f_{o}(1+z)]^2}{G(M_1 + M_2)} \right)^{1/3} \cos\varphi \, ,
\end{align}
where $\varphi$ and $\psi$ are defined as before.
Using this, it can be shown that
\begin{subequations} \label{Eq:IQUV-bin}
\begin{align} 
I =& \frac{5}{16} \langle I \rangle_{\psi\varphi}  (1 + 6 \cos^2\varphi + \cos^4\varphi) \, , \label{Eq:IQUV-bin-a}
\\
Q =& \frac{5}{16} \langle I \rangle_{\psi\varphi}  (1 - 2 \cos^2\varphi + \cos^4\varphi) \cos(4\psi) \, , \label{Eq:IQUV-bin-b}
\\
U =& \frac{5}{16} \langle I \rangle_{\psi\varphi}  (1 - 2 \cos^2\varphi + \cos^4\varphi) \sin(4\psi) \, , \label{Eq:IQUV-bin-c}
\\
V =& \frac{5}{4} \langle I \rangle_{\psi\varphi}  (\cos\varphi + \cos^3\varphi) \label{Eq:IQUV-bin-d} \, ,
\end{align}
\end{subequations}
where the average $\langle \cdots \rangle_{\psi\varphi}$ is over all possible inclinations and principal polarisation axes of the binary,
\begin{align}
\langle I \rangle_{\psi\varphi}  =&  \frac{1}{8} \frac{32}{5} \left( \frac{2 G^2 M_1 M_2}{c^4 D_L} \left( \frac{[\pi f_0(1+z)]^2}{G(M_1 + M_2)} \right)^{1/3} \right) ^2 
\notag\\
=& \frac{1}{2} \frac{32}{5}
\left( \frac{G^{5/3} }{c^4} \frac{\mathcal{M}_c^{5/3}}{D_L} \left[ \pi f_0(1+z) \right]^{2/3} \right)^2 \, ,
\end{align}
and $\mathcal{M}_c = (M_1 M_2)^{3/5}/(M_1 + M_2)^{1/5}$ is the chirp mass. Comparing this to equation~2 from \citetalias{Mingarelli2017} we see that $\langle I \rangle_{\psi\varphi} = h^2/2$ where $h$ is defined to be the inclination and principal polarisation averaged strain.  As we consider each source to be emitting from a single pixel, $h^2_c$  will be (\citetalias{Mingarelli2017}; \citealt{Finn:2000sy}) 
\begin{align}
h^2_c =&  \sum_i \frac{h^2 f_i}{\Delta f}
= 2\sum_i \frac{\langle I_i \rangle_{\psi\varphi} f_i}{\Delta f} \, ,
\end{align}
where $\langle I_i \rangle_{\psi\varphi}$ is the inclination and principal polarisation averaged power of the i-th source, $f_i$ its frequency and $\Delta f$ the inverse of the observation time. Firstly, given a particular instance of the $h_c$ background, and a simulated set of $\varphi$ and $\psi$, all four Stokes parameters can be computed and so can their corresponding power spectra. However, as the exact distribution of $\varphi$ and $\psi$ are known, it is possible to calculate the expected power spectra for the distribution of orientations. It is worth noting that this is the expected value of the power spectra of a field, not the power spectra of the expected field -- which is important as $\langle Q \rangle_{\psi\varphi}  = \langle U \rangle_{\psi\varphi}  = \langle V \rangle_{\psi\varphi}  = 0$.

The power spectra for such a background of point sources is an easy extension of the single-source version. Restricting to overall power $I$, it can be shown that
\begin{align}
\hat{C}_\ell^{II} 
=&
\frac{1}{4\pi}\Big(\sum_i I^2_i + \sum_{ij, i\neq j} I_i I_j P_{\ell}(\cos\beta_{ij}) \Big) \, ,
\end{align}
where $I_i$ are the individual powers and $\beta_{ij}$ the angle between sources $i$ and $j$. 

This has some interesting properties. Firstly, while it is not always white, for a single source, we recover the delta function source as considered in the previous section. If each source has statistically independent power and sky-location, then for a very large number of sources, 
\begin{align}
\hat{C}_\ell^{II} 
\rightarrow& 
\begin{cases}
\frac{1}{4\pi} \left(\sum_{i}  I_i^2 + \sum_{ij, i\neq j}  I_i    I_j  \right), &\ell=0 \\
\frac{1}{4\pi} \sum_{i}  I_i^2, &\ell\neq 0 \, .
\end{cases}
\end{align}
This is due to the fact that each Legendre polynomial with $\ell >0$ has average value zero over the range $[-1,1]$. Because of this, $C^{II}_0$ dominates in the limit of a large number of sources and the rest of the spectrum is white. Of course, this has similar properties to  the ensemble average -- that is, one over sky locations, power of individual sources and number of sources --
\begin{align}
{C}_\ell^{II} =& \frac{1}{4\pi} \bar{N} \left(\langle I^2 \rangle + \langle I \rangle^2 \delta_{\ell 0} \right) \, ,
\end{align}
where $\bar{N}$ is the average number of sources and $\langle I \rangle$ is the ensemble averaged power for a source -- i.e. averaged over all properties of the binary and galaxy.

This effect can be seen to some extent in Fig.~\ref{Fig:WDPS}. As there is a large number of white dwarfs, we had to remove the large $\ell =0$ value and can observe that the $\hat{C}^{II}_\ell$ spectrum is approximately white for $\ell \gg 0$. The observation that the spectrum is not white for low $\ell$ implies that there is correlation of power/sky location on large scales -- i.e. the shape of the galaxy. For a lower number of sources, there can be interesting features -- as will be seen below.

\begin{figure*} 
\centering 
\subfigure[Sky location of sources]{
\includegraphics[width = 0.3\textwidth]{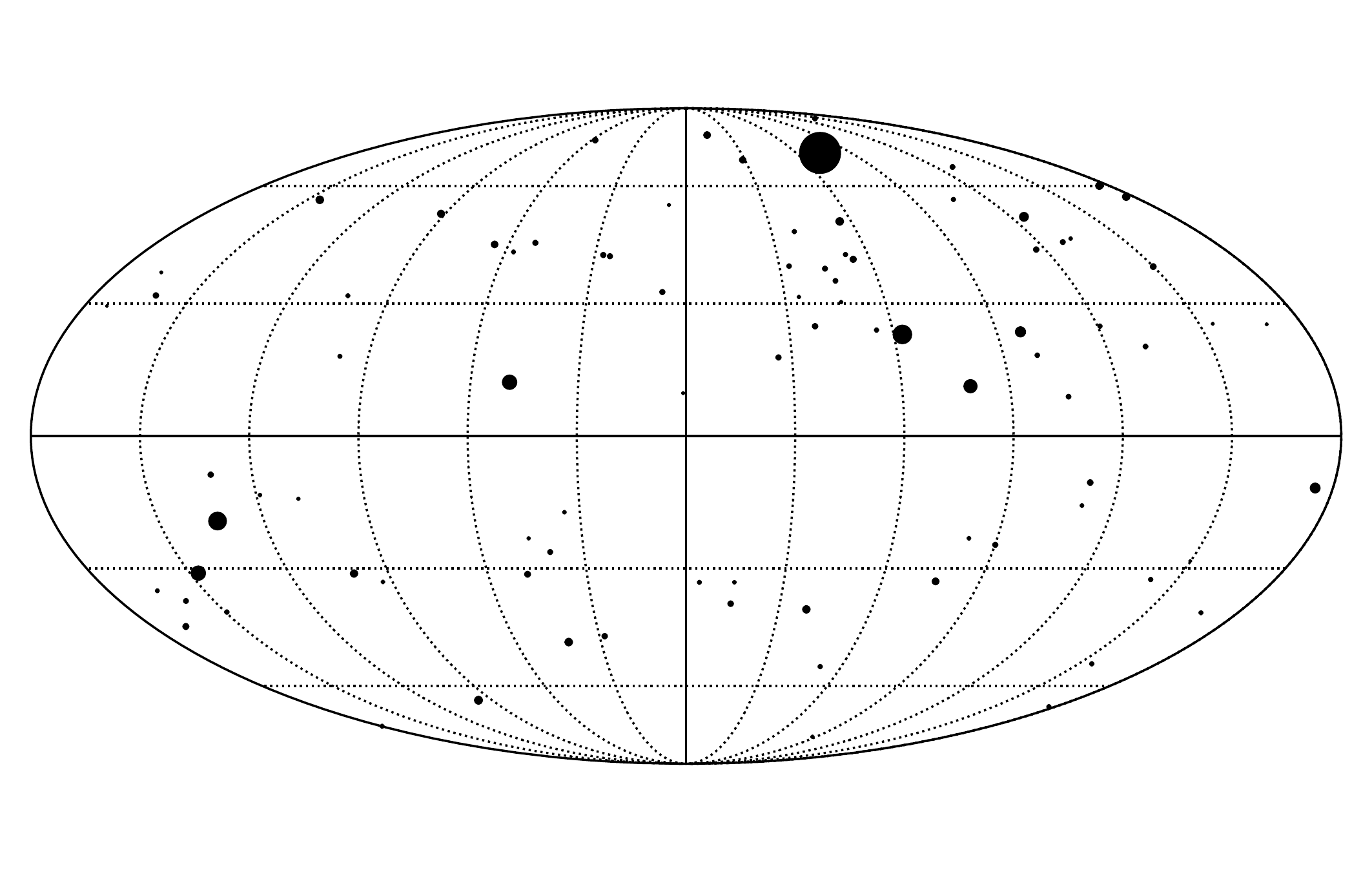}}
\subfigure[``All-sky'' background]{
\includegraphics[width = 0.3\textwidth]{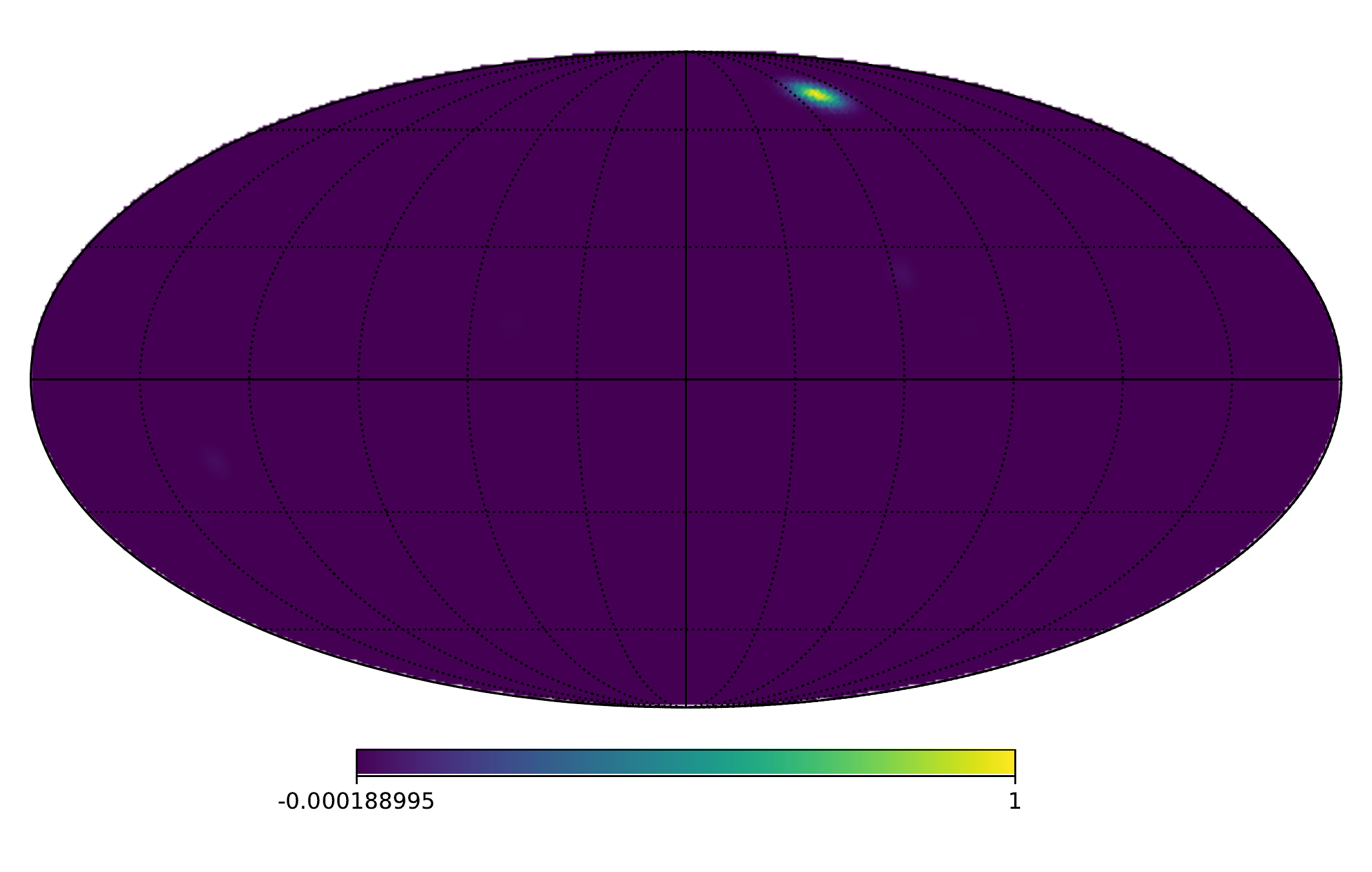}}
\subfigure[``Noise'' background]{
\includegraphics[width = 0.3\textwidth]{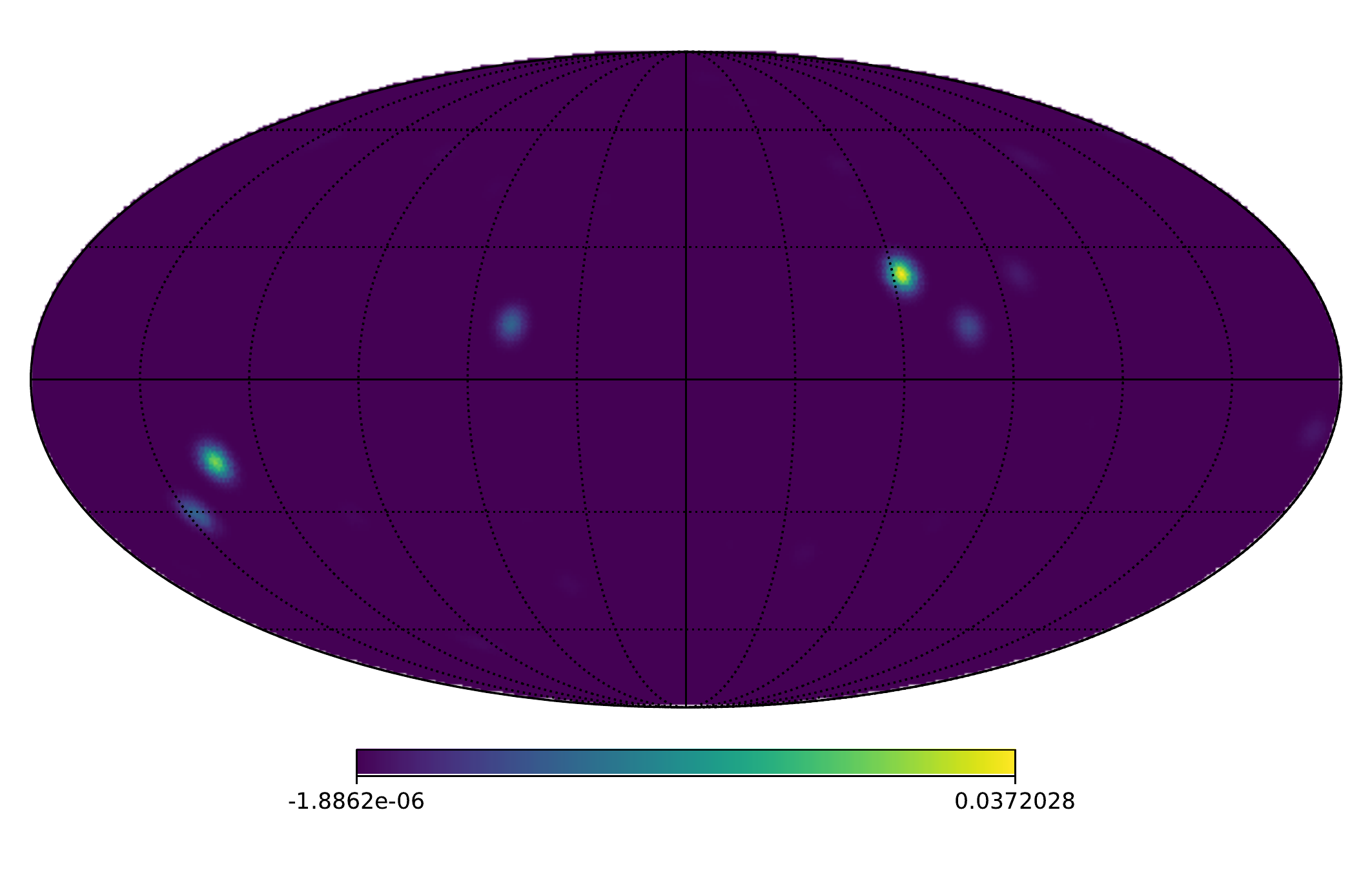}}
\caption{Going from continuous gravitational-wave sources to gravitational-wave backgrounds. (a) A sample Monte Carlo realisation from \citetalias{Mingarelli2017}: 87 local nanohertz gravitational-wave sources from 2MASS, marker sizes scaled to show the relative strain of the sources. (b) $\langle I \rangle_{\psi\varphi}$ for the gravitational-wave background generated from the sources on the left hand side, which we call the ``all-sky'' map. This includes a loud source which is visible in the top centre-right of the figure. (c) The background without this loud source,  which we call the ``noise'' background. Both $\langle I \rangle_{\psi\varphi}$ backgrounds normalised with respect to $\textrm{max}(\langle I \rangle_{\psi\varphi})$ from the ``all-sky'' background -- i.e. from the dominant source. } \label{Fig:Moll-All-Noise}
\end{figure*}

\begin{figure*} 
\centering 
\subfigure[$C^{II}_\ell$]{
\includegraphics[width = 0.32\textwidth]{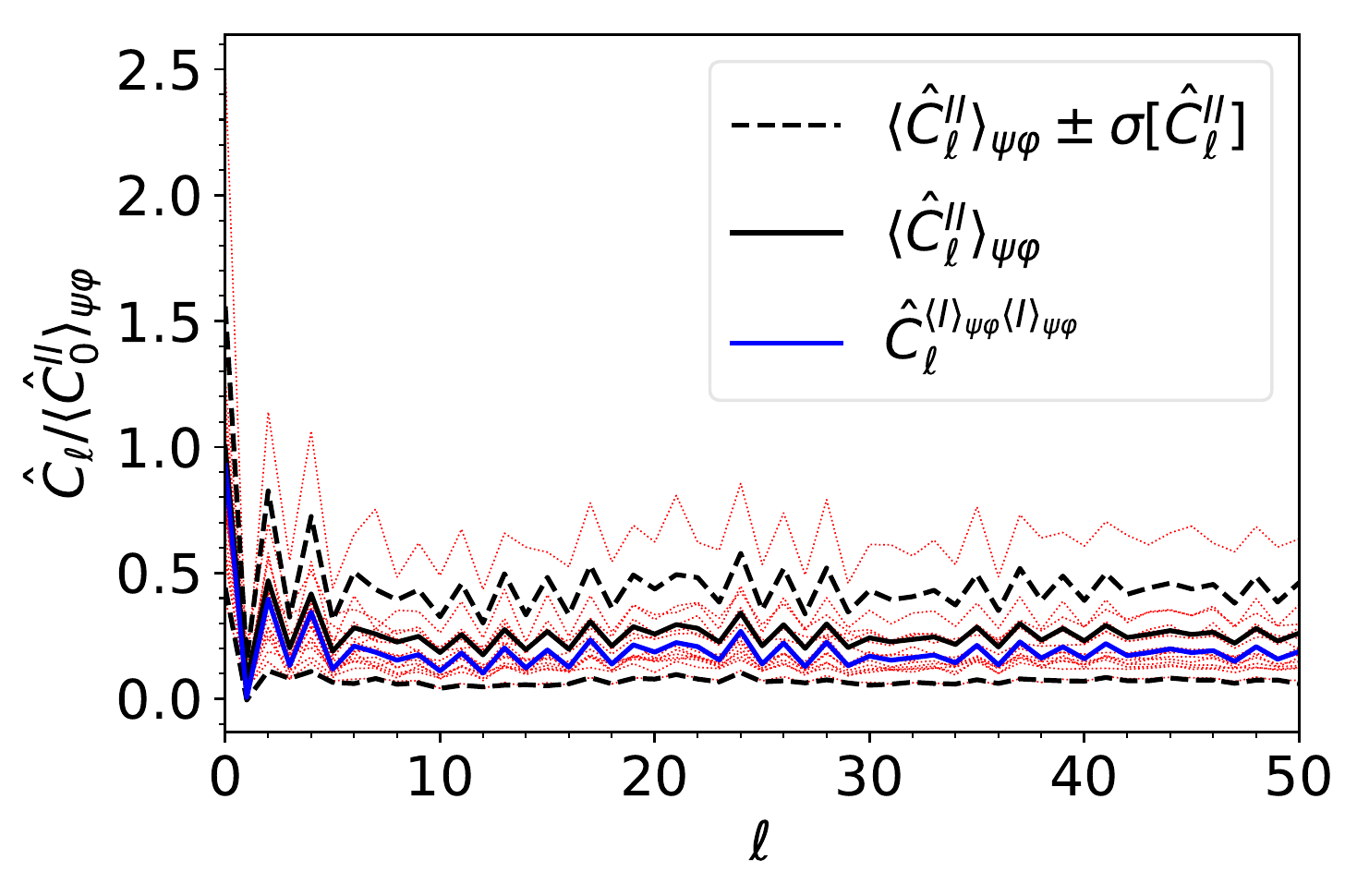}
\label{Fig:Multiple-Noise-a}}
\subfigure[$C^{VV}_\ell$]{
\includegraphics[width = 0.32\textwidth]{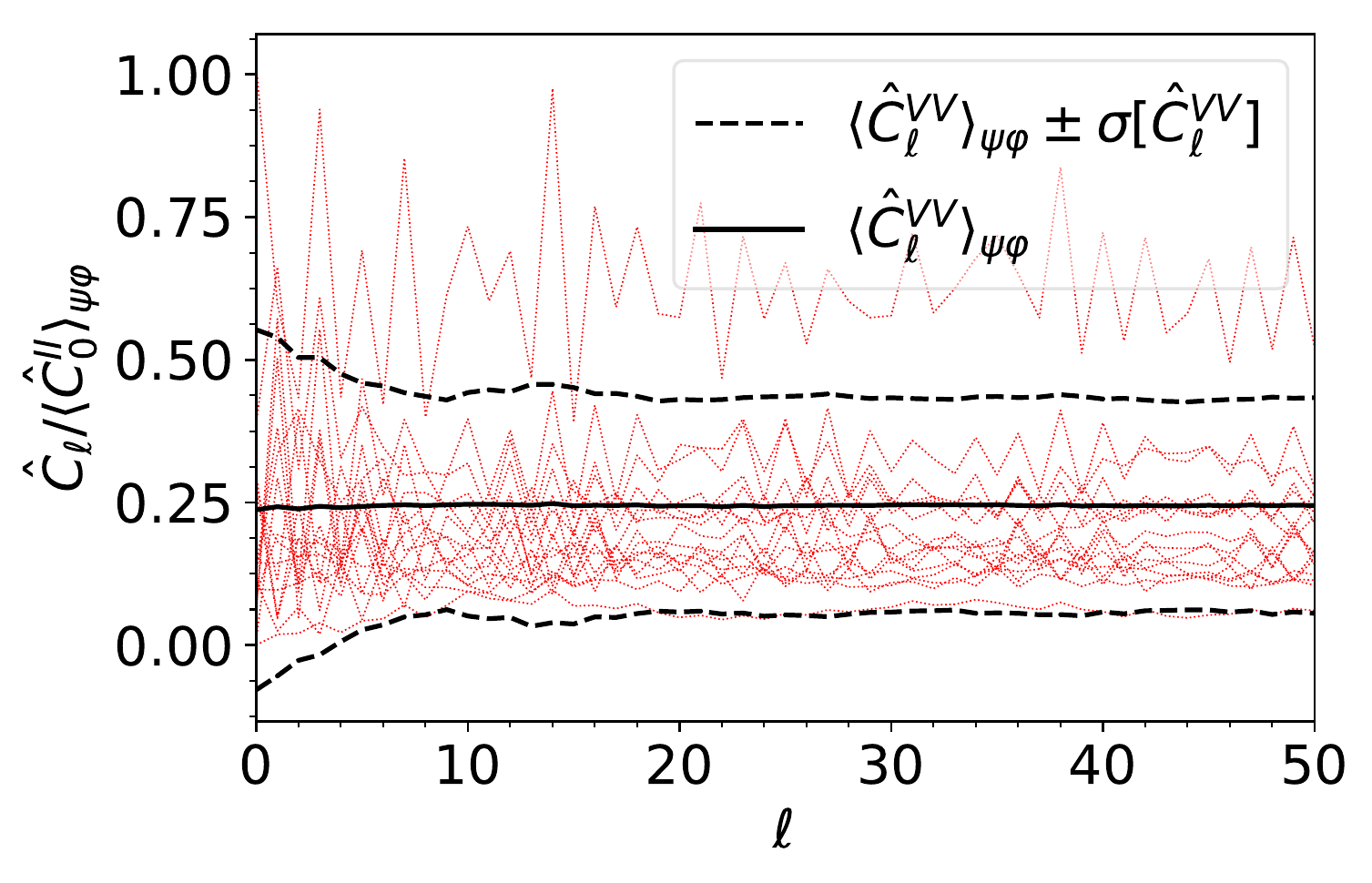}}
\subfigure[$C^{II}_\ell - C^{VV}_\ell$]{
\includegraphics[width = 0.32\textwidth]{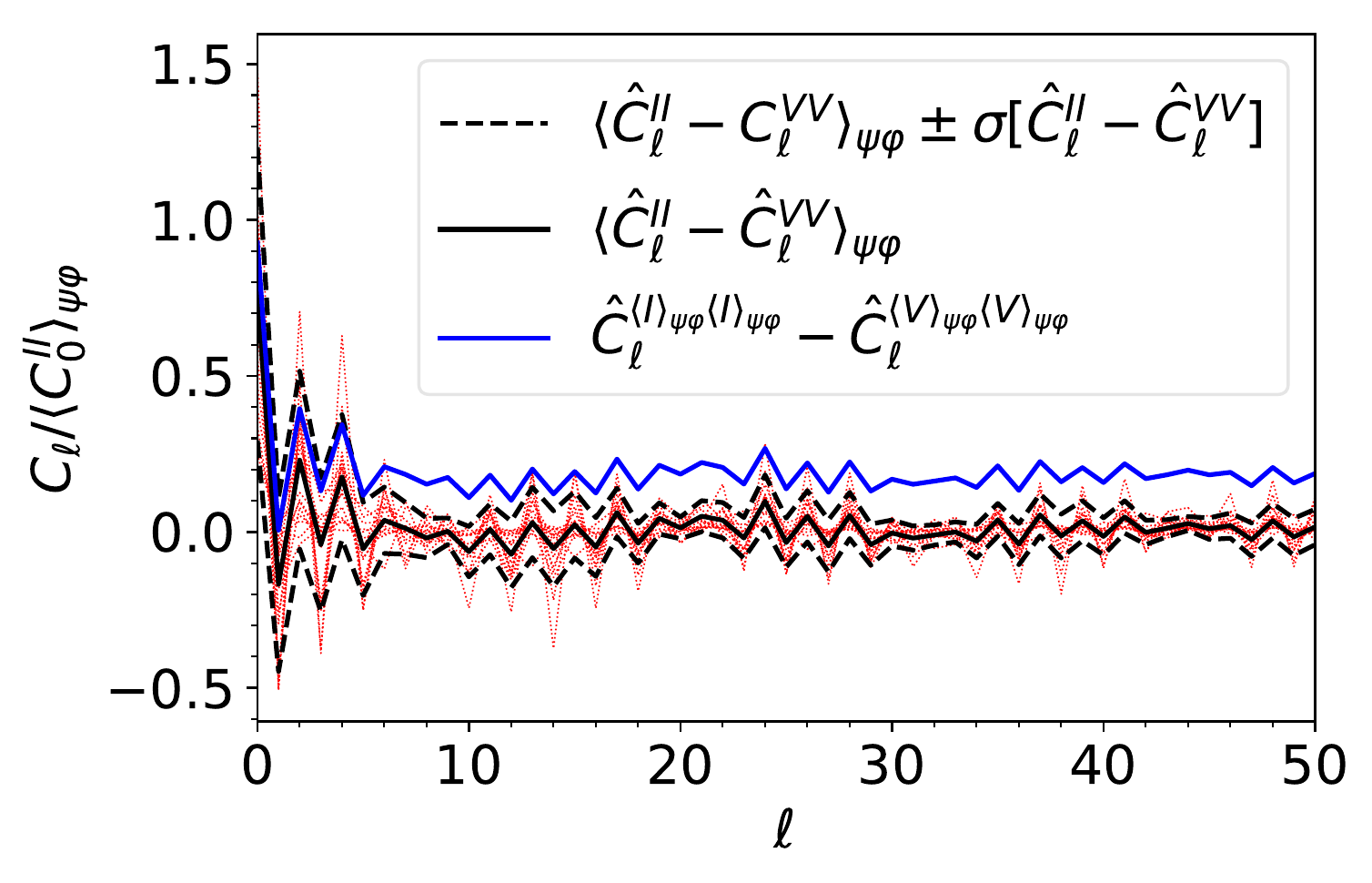}}
\subfigure[$C^{EE}_\ell$]{
\includegraphics[width = 0.32\textwidth]{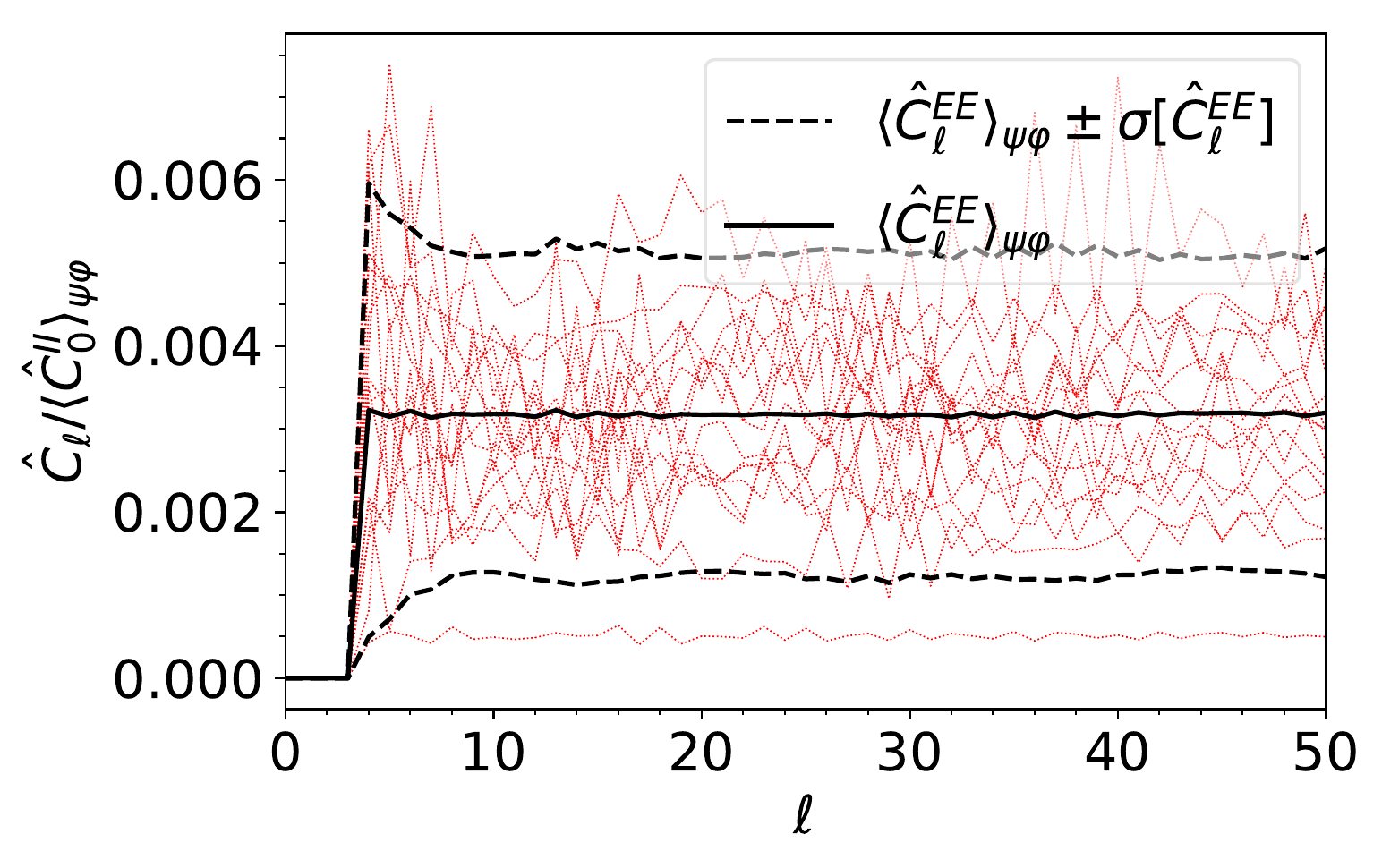}}
\subfigure[$C^{BB}_\ell$]{
\includegraphics[width = 0.32\textwidth]{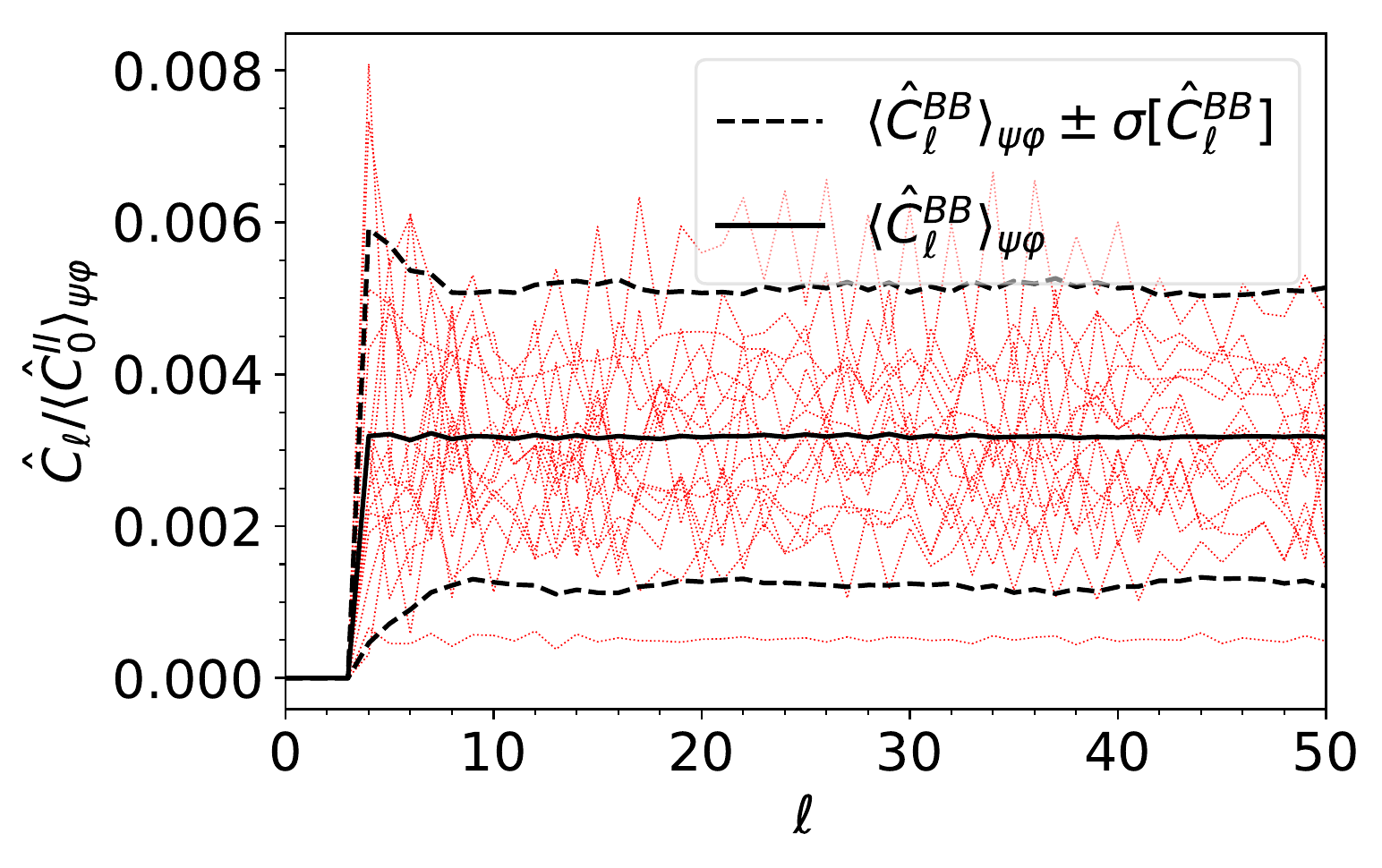}}
\subfigure[$C^{EE}_\ell - C^{BB}_\ell$]{
\includegraphics[width = 0.32\textwidth]{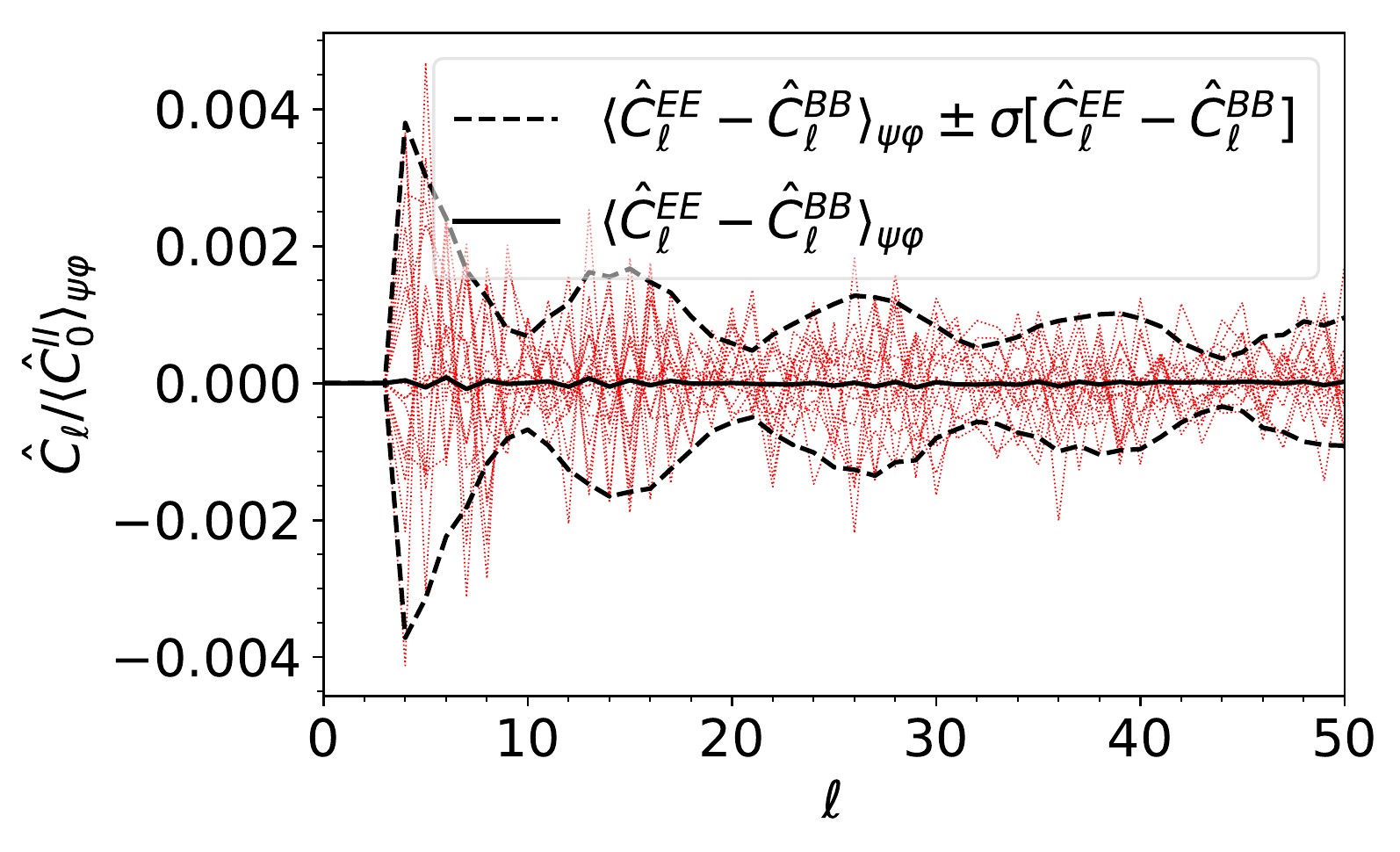} \label{Fig:Multiple-Noise-f}}
\subfigure[$C^{GG}_\ell$]{
\includegraphics[width = 0.32\textwidth]{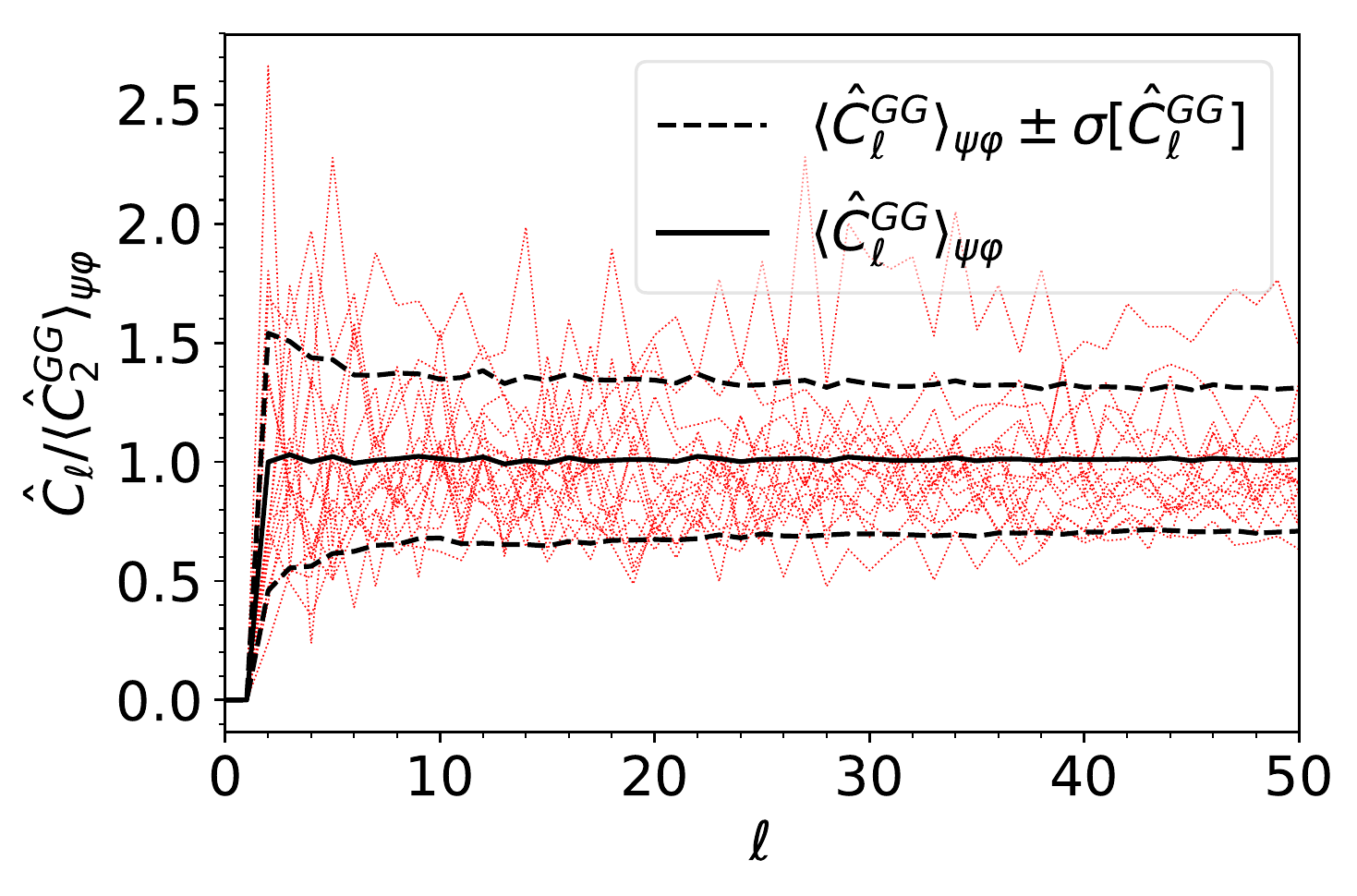}}
\subfigure[$C^{CC}_\ell$]{
\includegraphics[width = 0.32\textwidth]{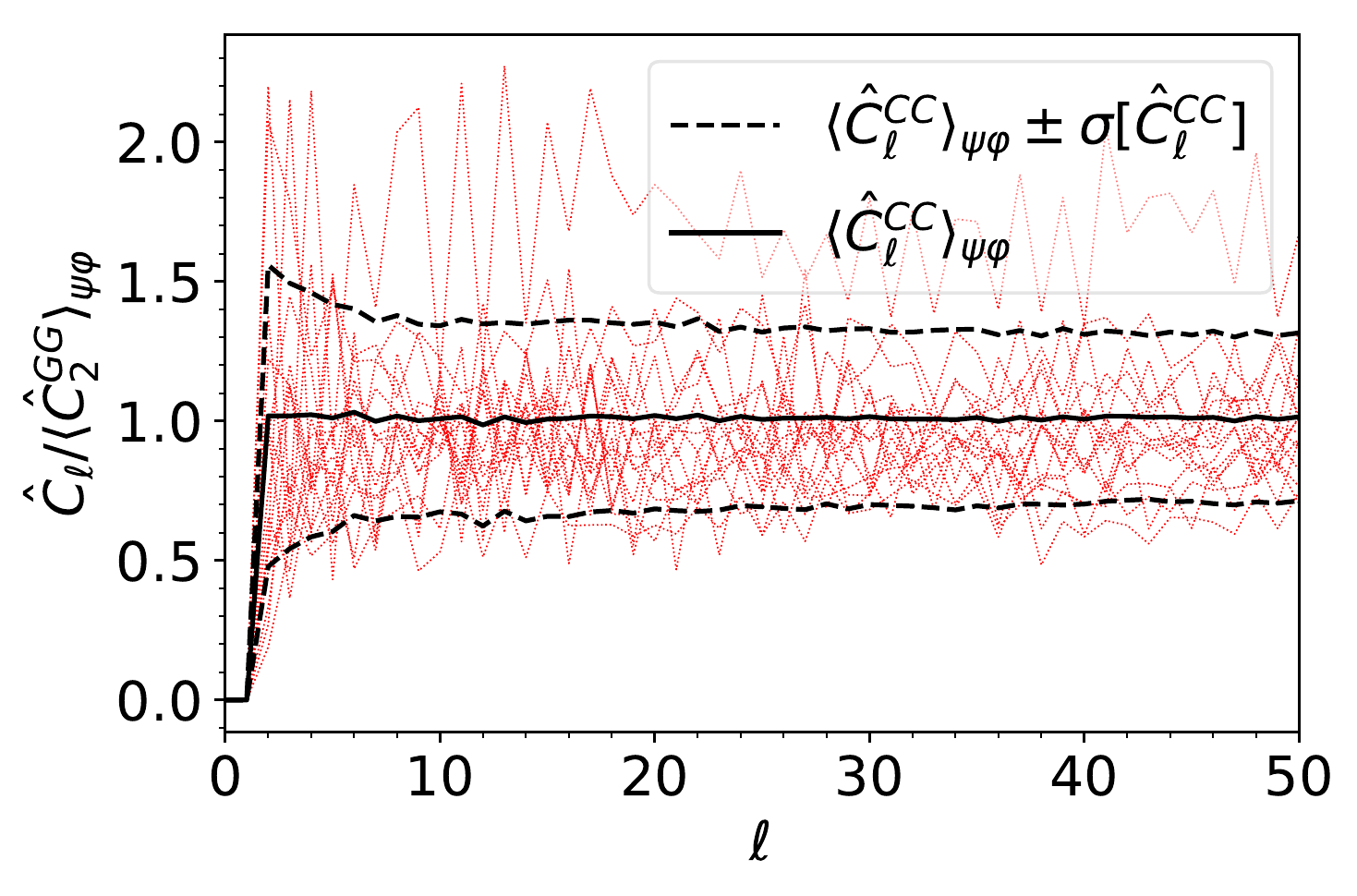}}
\subfigure[$C^{GG}_\ell - C^{GG}_\ell$]{
\includegraphics[width = 0.32\textwidth]{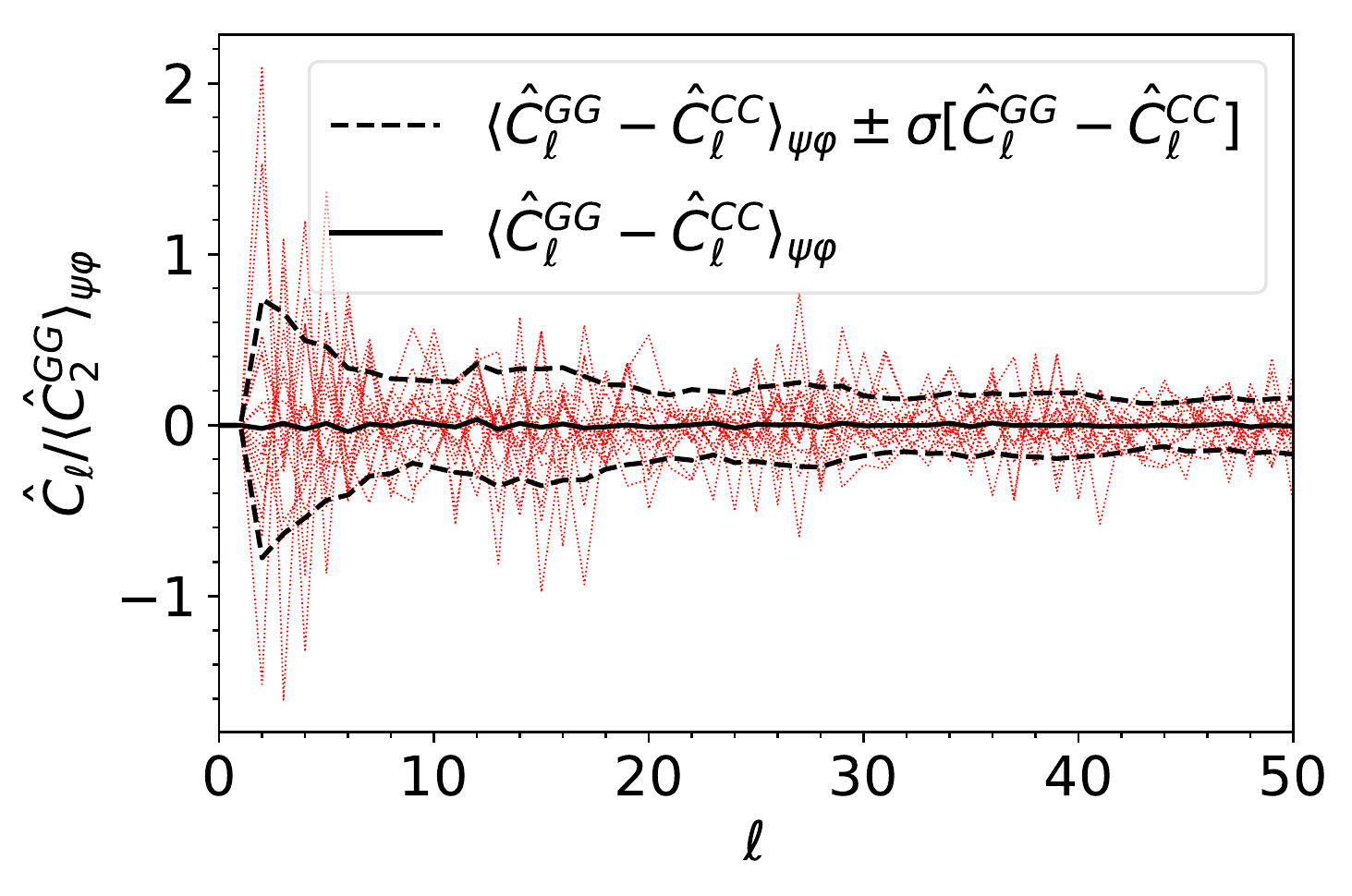}}
\subfigure[{$\mathbb{R}[C^{GC}_\ell]$}]{
\includegraphics[width = 0.32\textwidth]{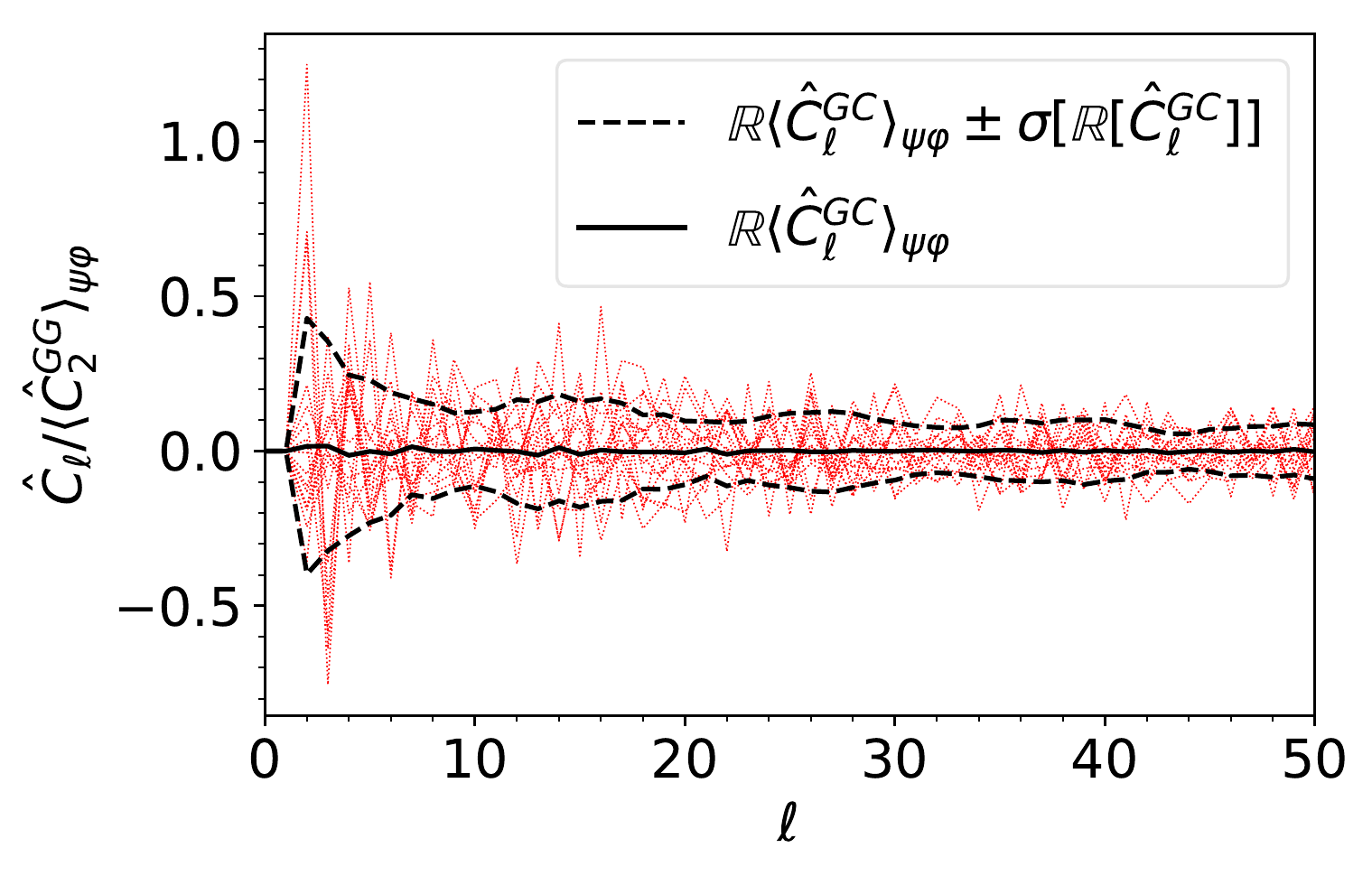}}
\subfigure[{$\mathbb{I}[C^{GC}_\ell]$}]{
\includegraphics[width = 0.32\textwidth]{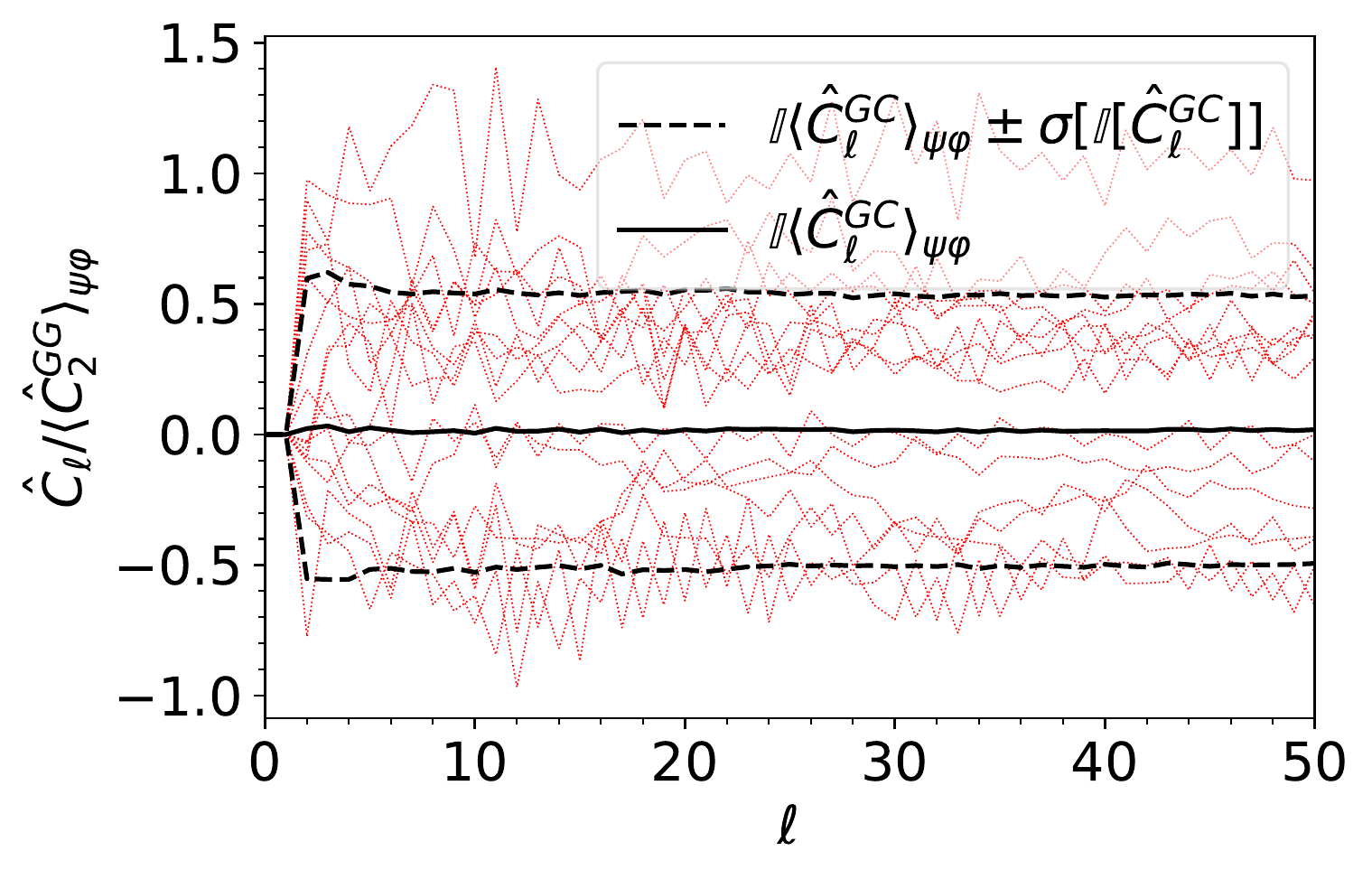}}
\caption{ Plots of power spectra of a sample of 16 of 512 simulations of different principal polarisations and inclinations for binaries taken from the ``noise'' background in \citetalias{Mingarelli2017} (red dashed lines) -- i.e. nearby unresolvable SMBHBs emitting gravitational waves with nHz frequencies.  The figure also shows the average over the power spectra from each of the simulations (black solid line) and the average $\pm$ one standard deviation (black dash lines). Power spectra of the averaged Stokes parameters are  shown where relevant (blue solid line). 
The power spectra constructed from the Stokes parameters are normalised with respect to $\langle C^{II}_0 \rangle_{\psi\varphi}$, and those constructed from the amplitudes with respect to $\langle C^{GG}_2 \rangle_{\psi\varphi}$ -- both from the ``noise'' background.} \label{Fig:Multiple-Noise}
\end{figure*}

\begin{figure*} 
\centering 
\subfigure[$C^{II}_\ell$]{
\includegraphics[width = 0.32\textwidth]{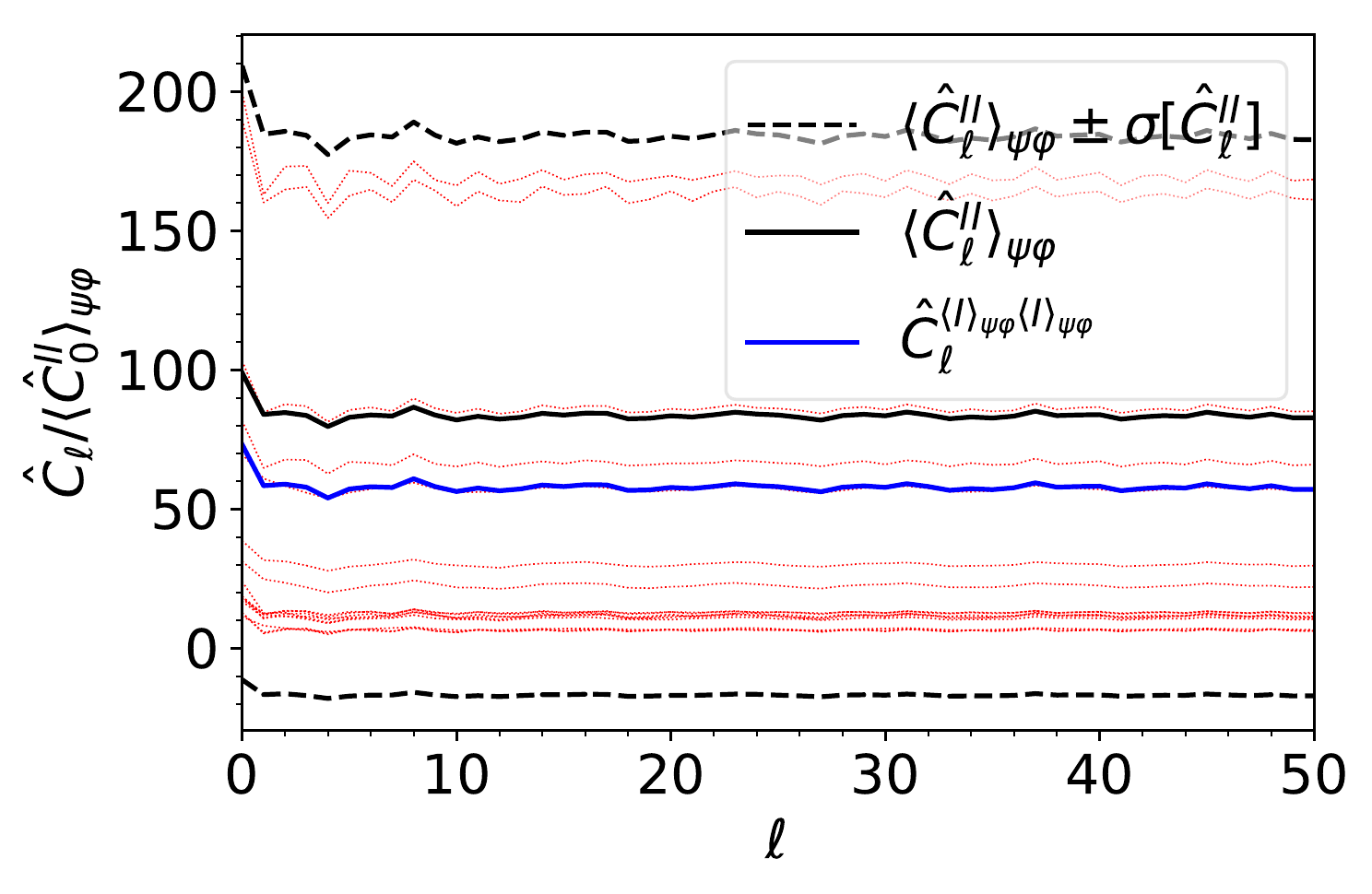}}
\subfigure[$C^{VV}_\ell$]{
\includegraphics[width = 0.32\textwidth]{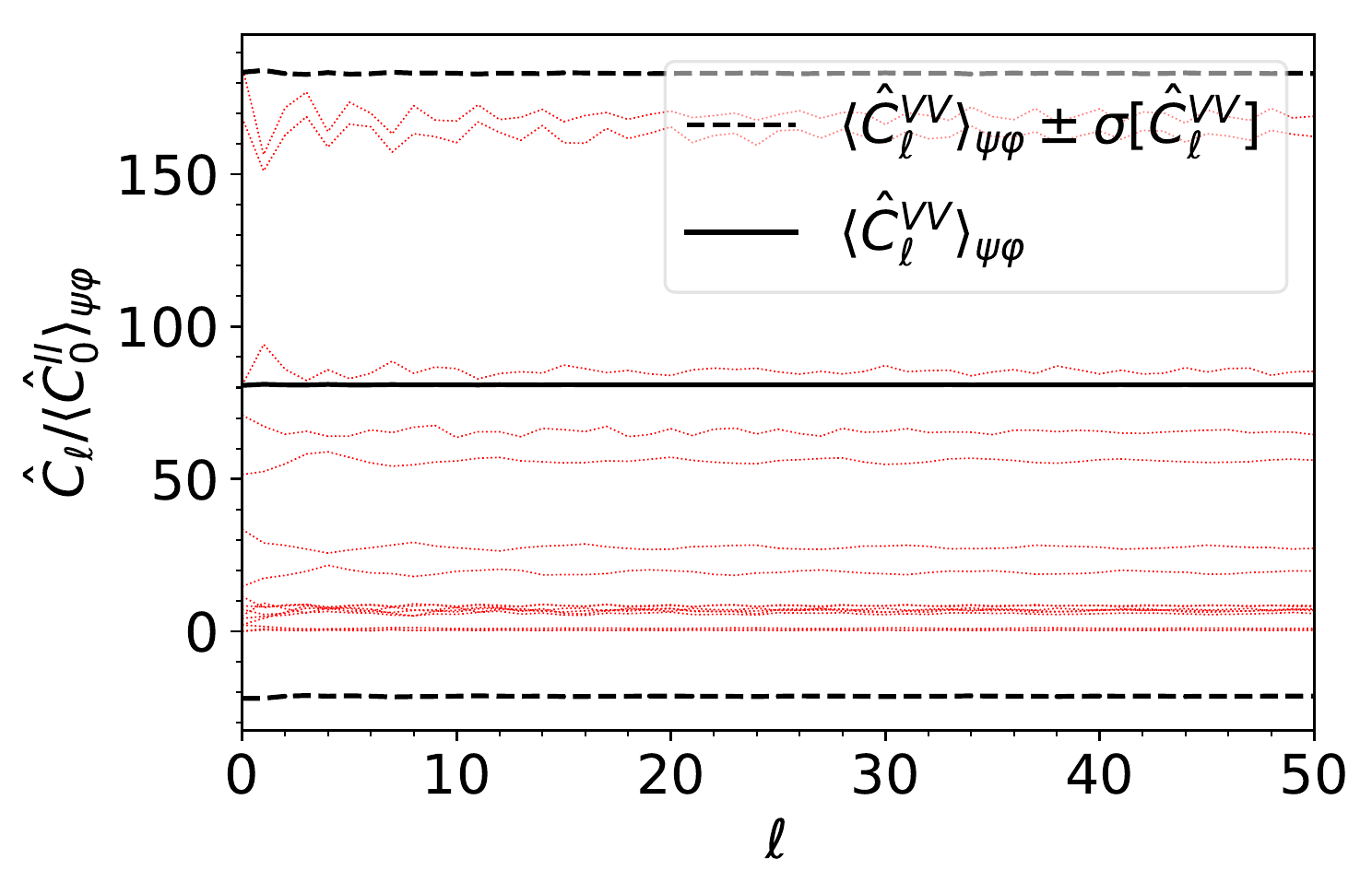}}
\subfigure[$C^{II}_\ell - C^{VV}_\ell$]{
\includegraphics[width = 0.32\textwidth]{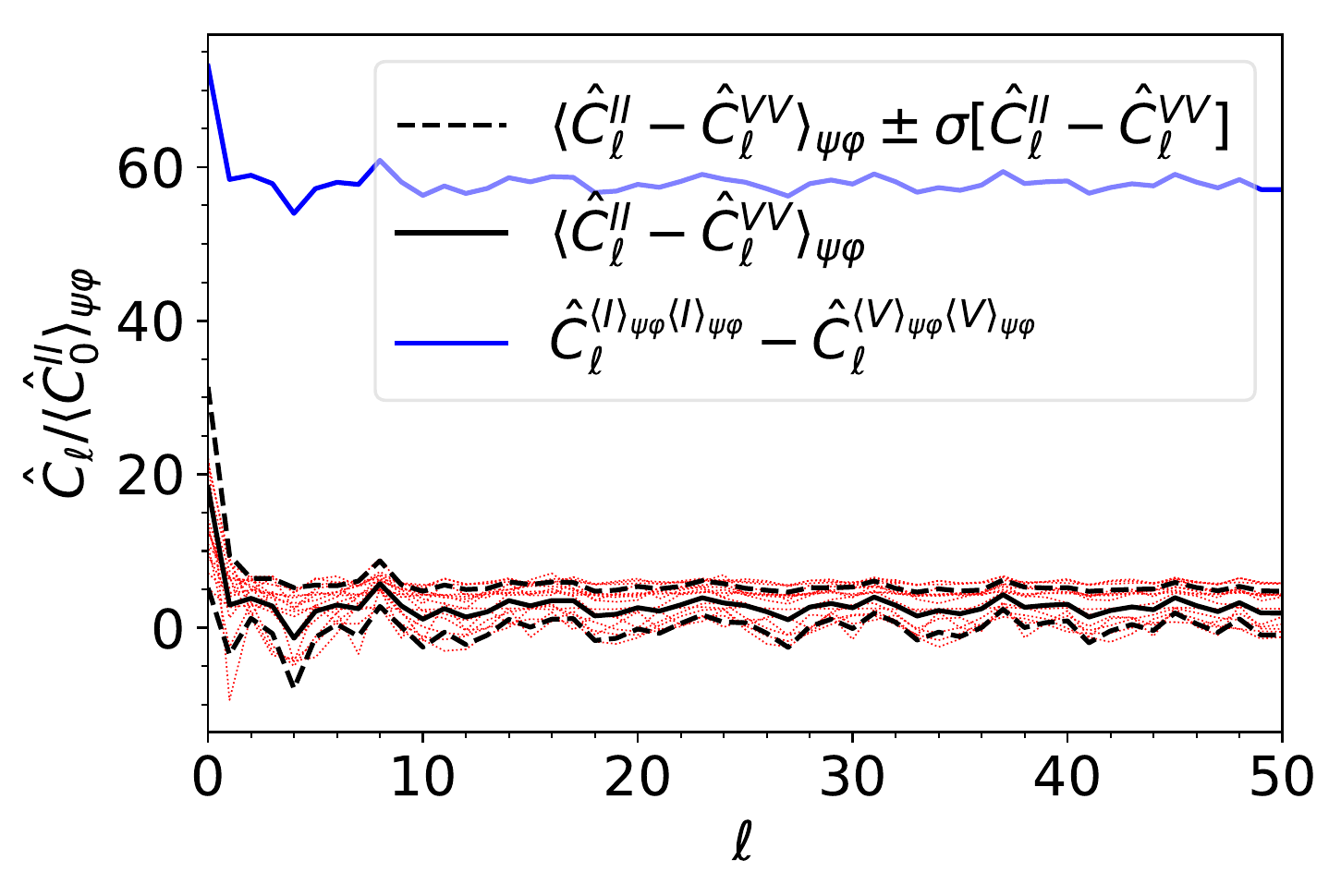}}
\subfigure[$C^{EE}_\ell$]{
\includegraphics[width = 0.32\textwidth]{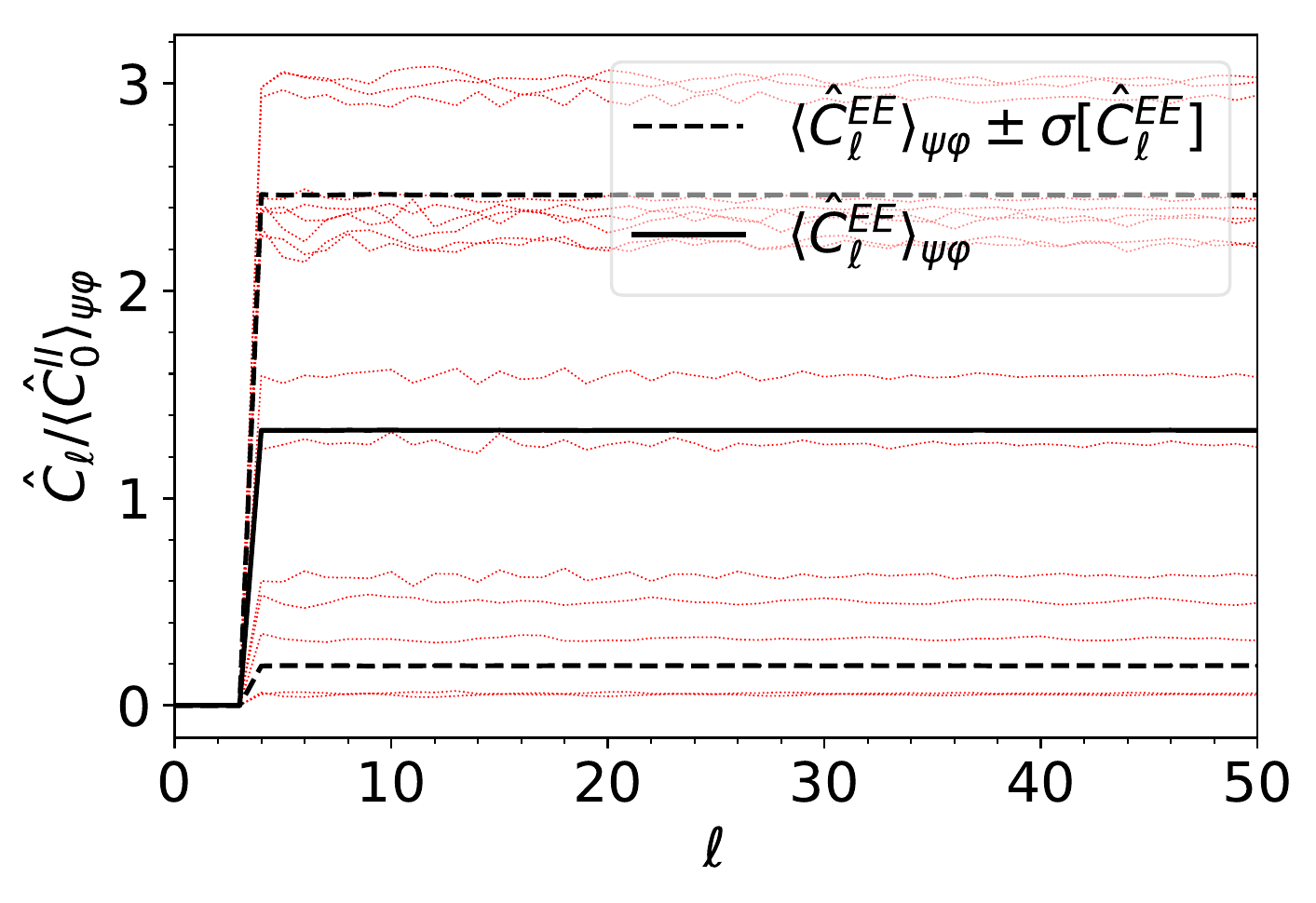}}
\subfigure[$C^{BB}_\ell$]{
\includegraphics[width = 0.32\textwidth]{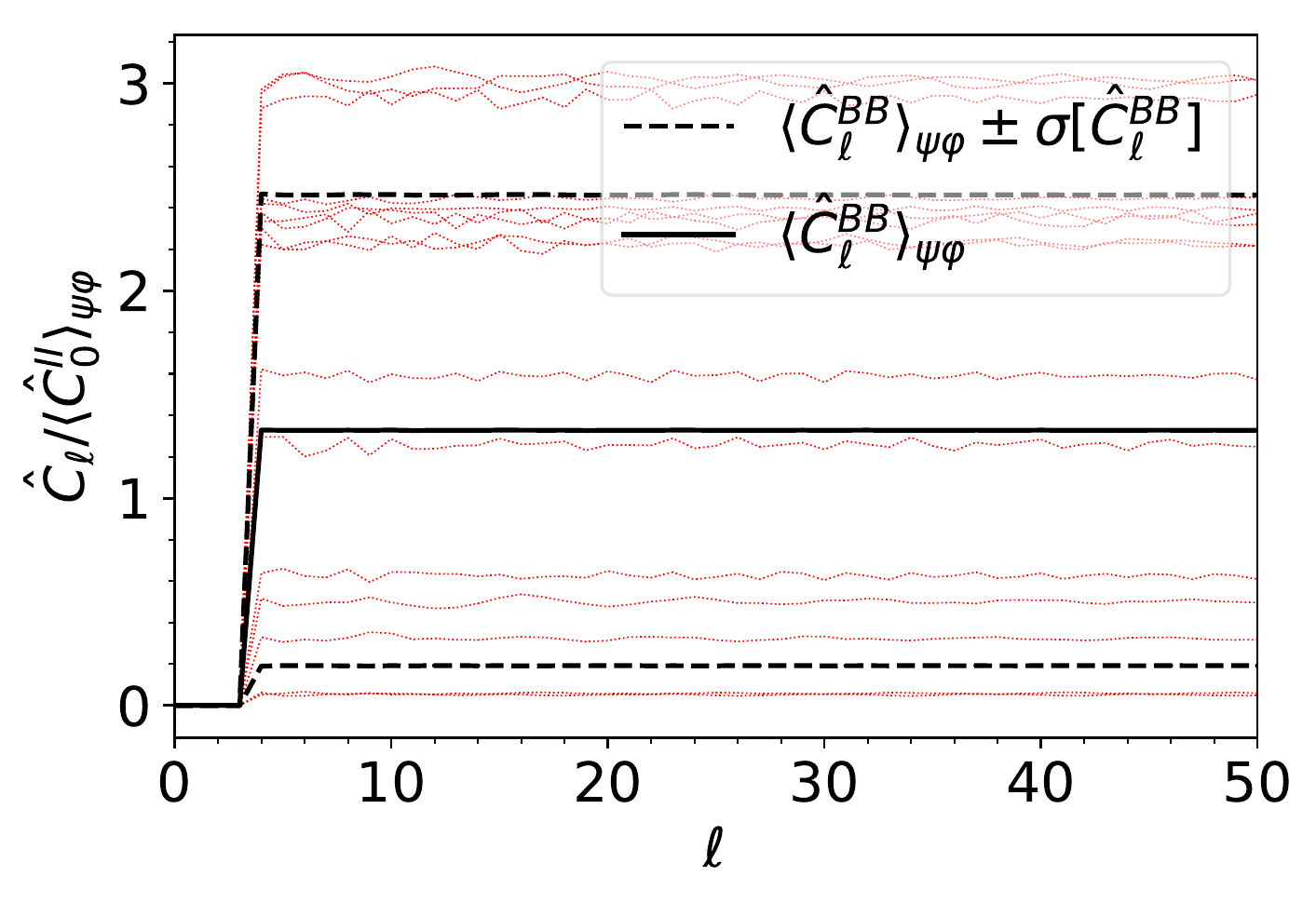}}
\subfigure[$C^{EE}_\ell - C^{BB}_\ell$]{
\includegraphics[width = 0.32\textwidth]{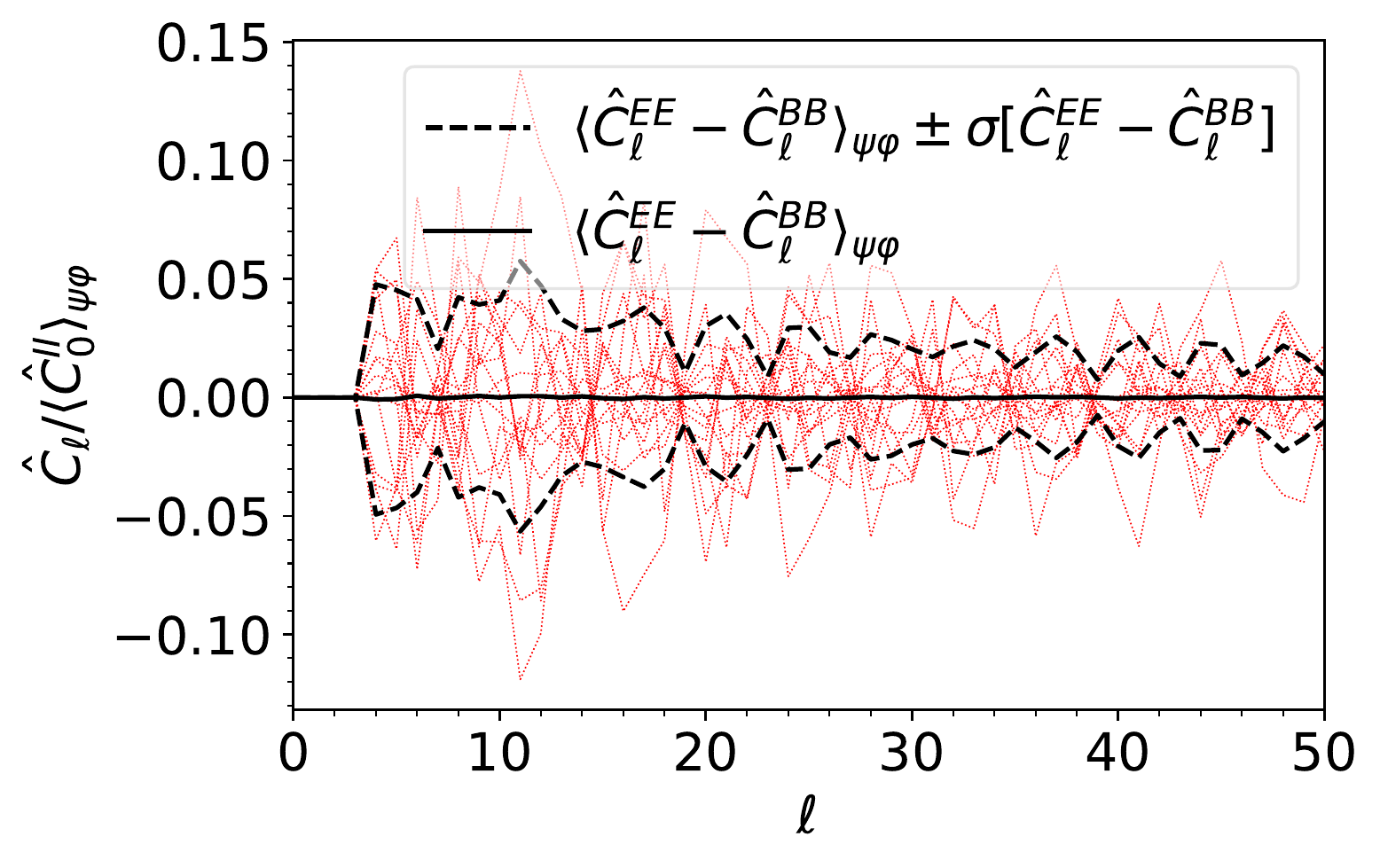} \label{Fig:Multiple-All-f}}
\subfigure[$C^{GG}_\ell$]{
\includegraphics[width = 0.32\textwidth]{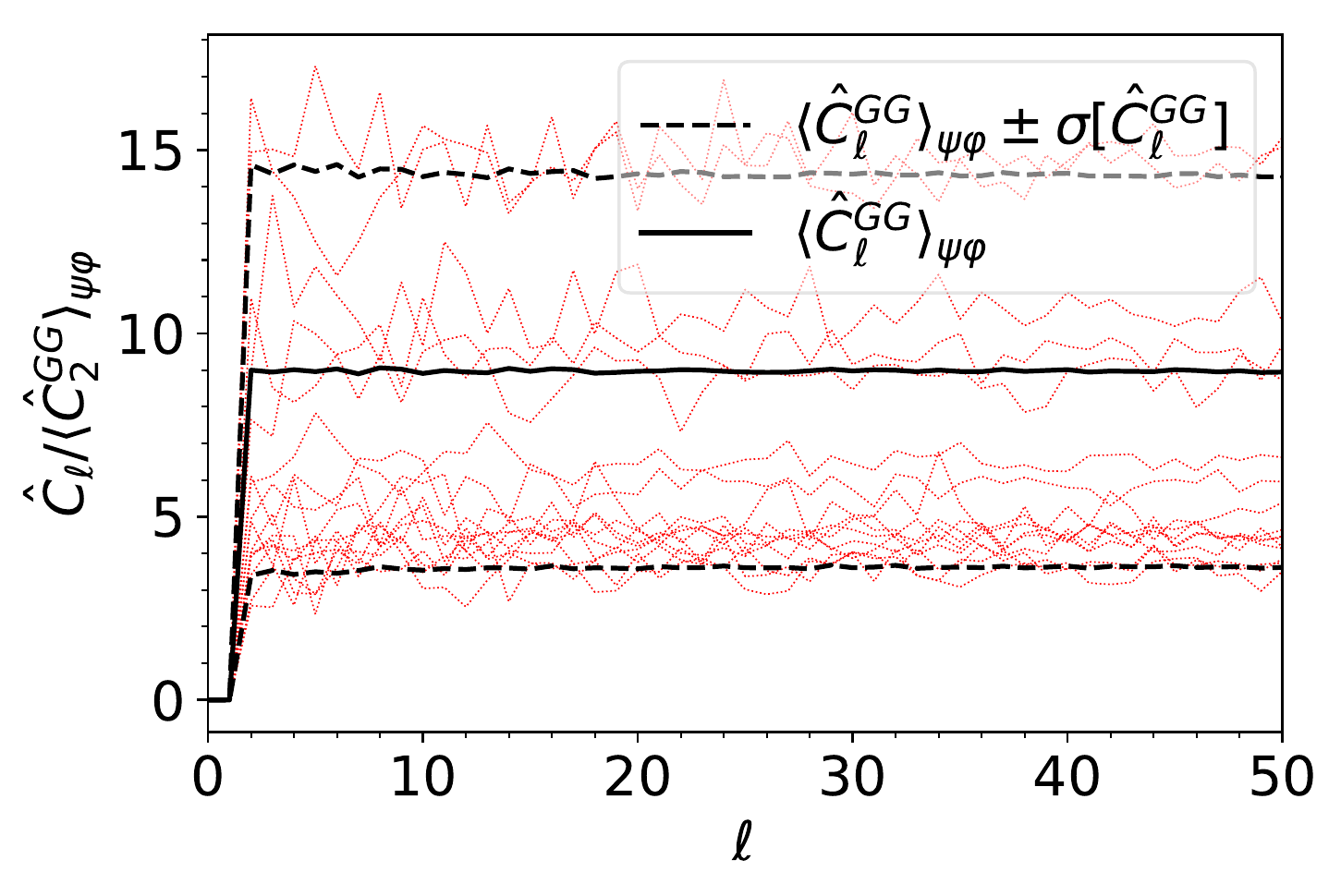}}
\subfigure[$C^{CC}_\ell$]{
\includegraphics[width = 0.32\textwidth]{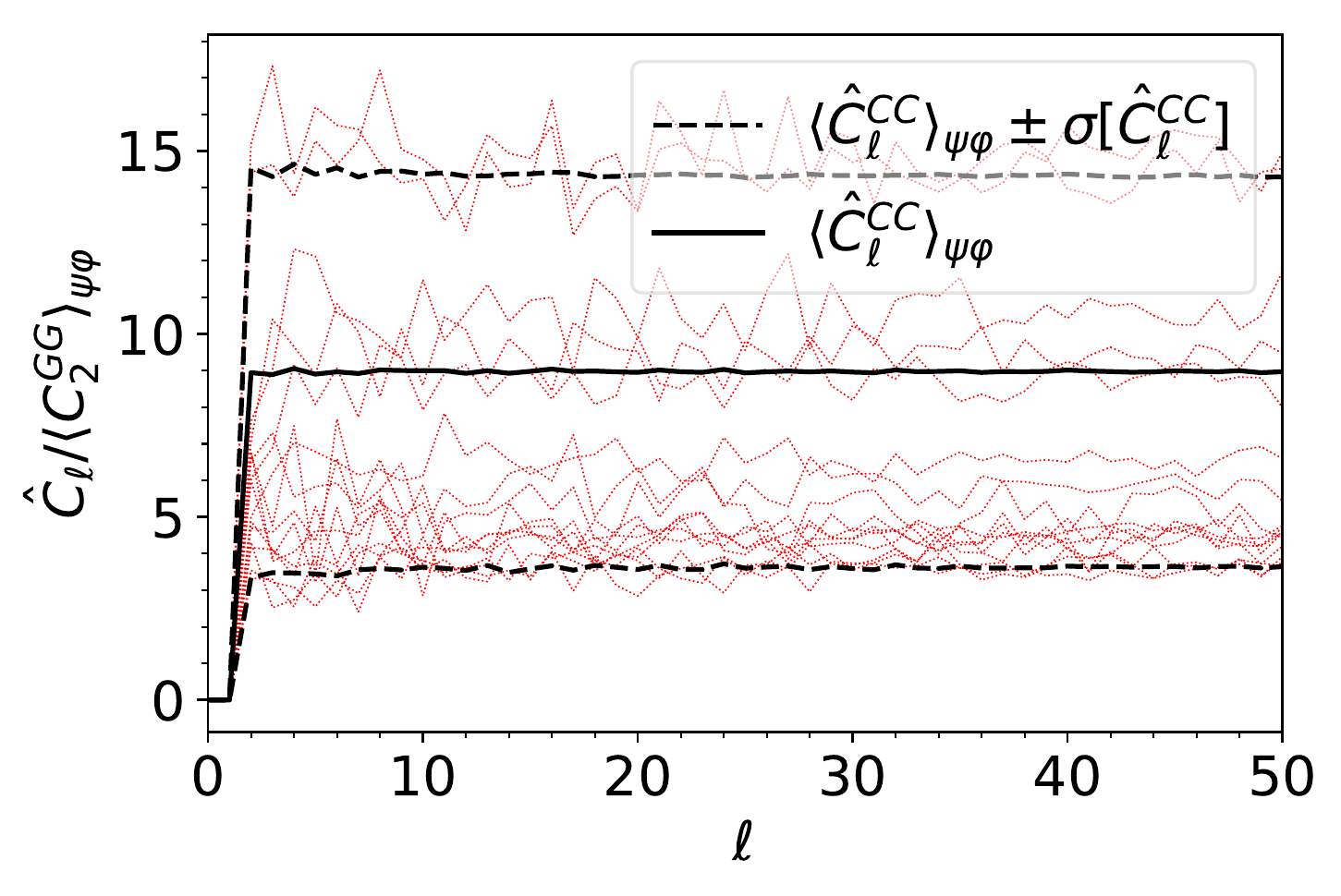}}
\subfigure[$C^{GG}_\ell - C^{GG}_\ell$]{
\includegraphics[width = 0.32\textwidth]{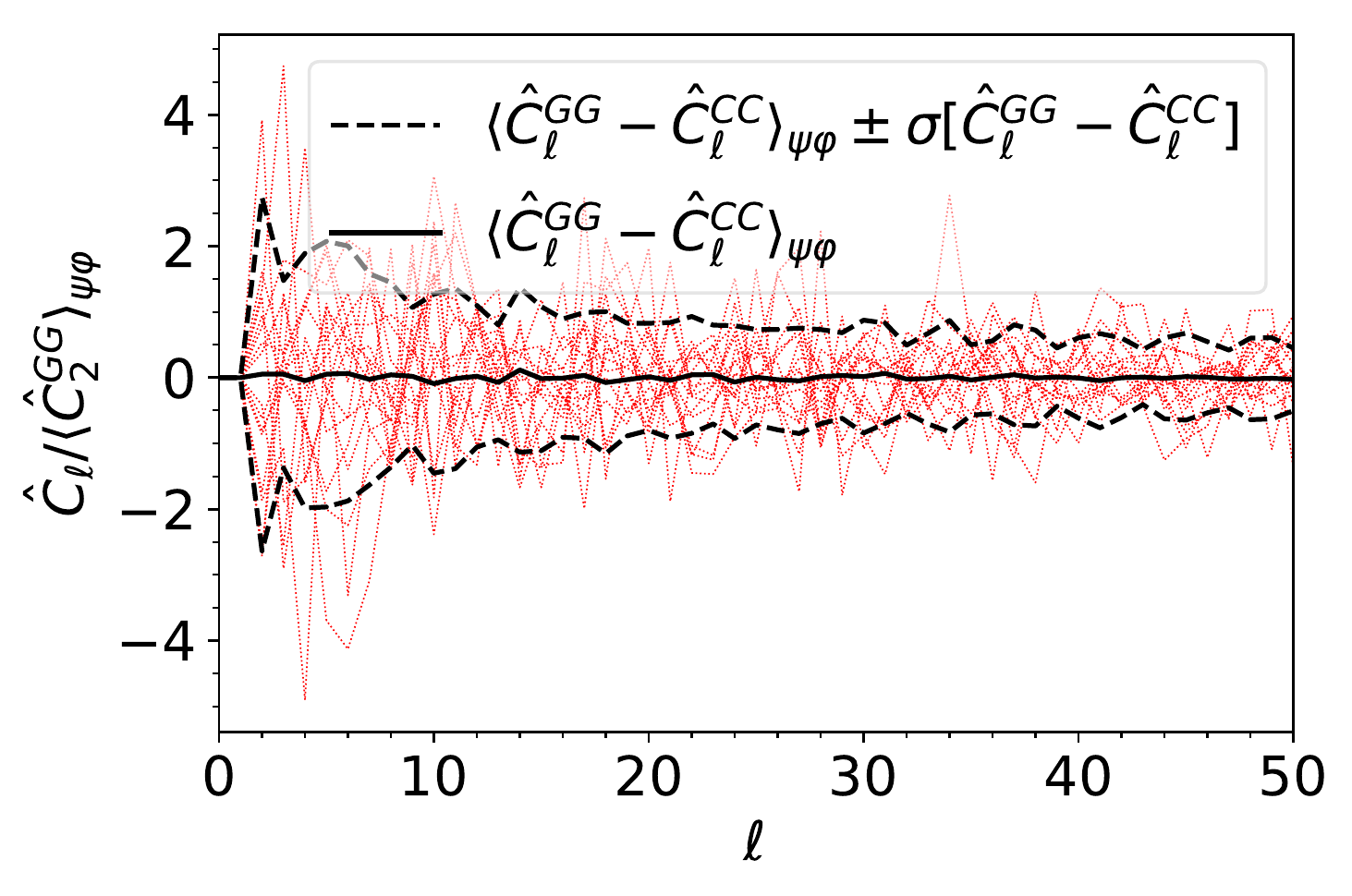}}
\subfigure[{$\mathbb{R}[C^{GC}_\ell]$}]{
\includegraphics[width = 0.32\textwidth]{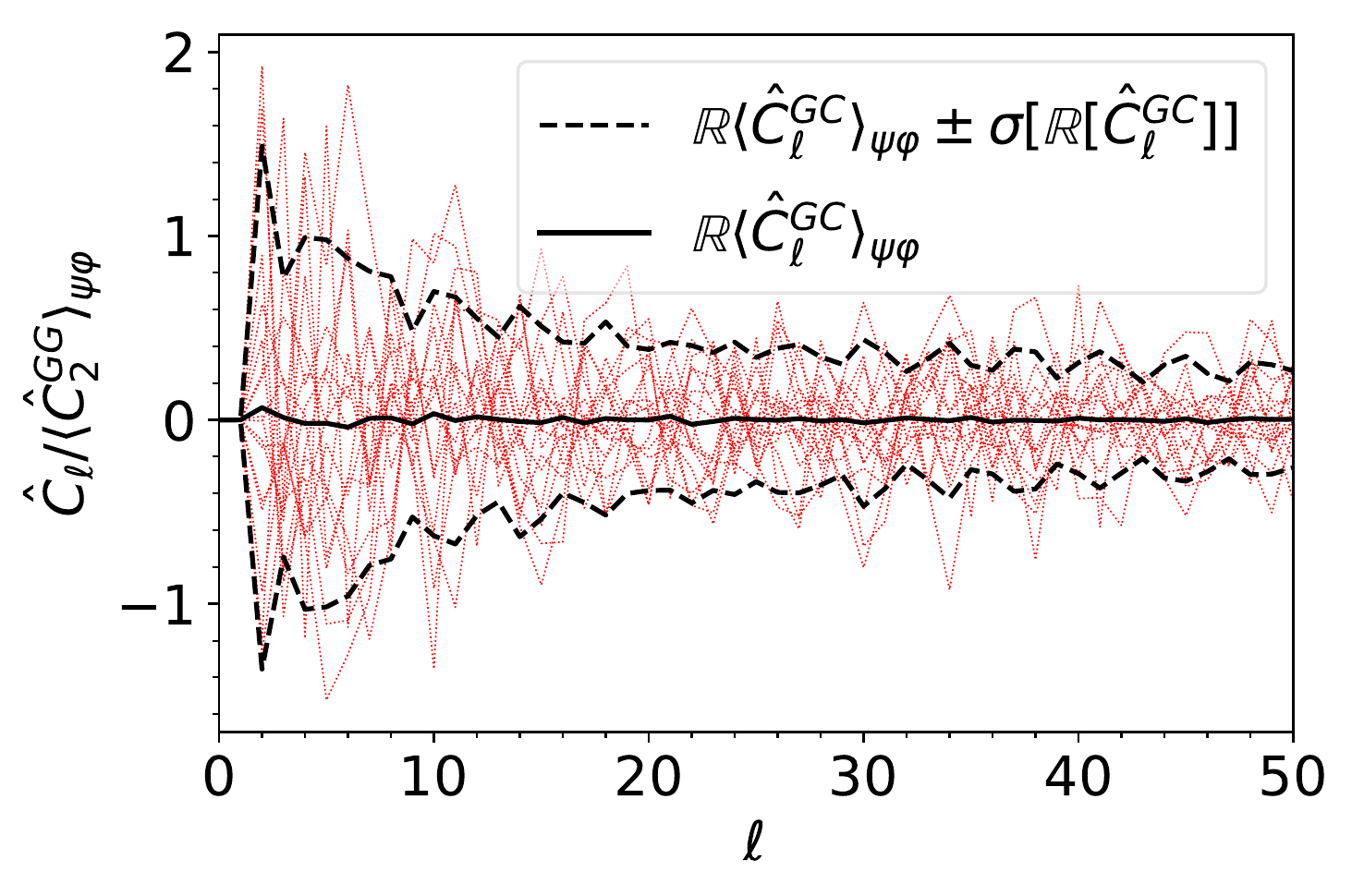}}
\subfigure[{$\mathbb{I}[C^{GC}_\ell]$}]{
\includegraphics[width = 0.32\textwidth]{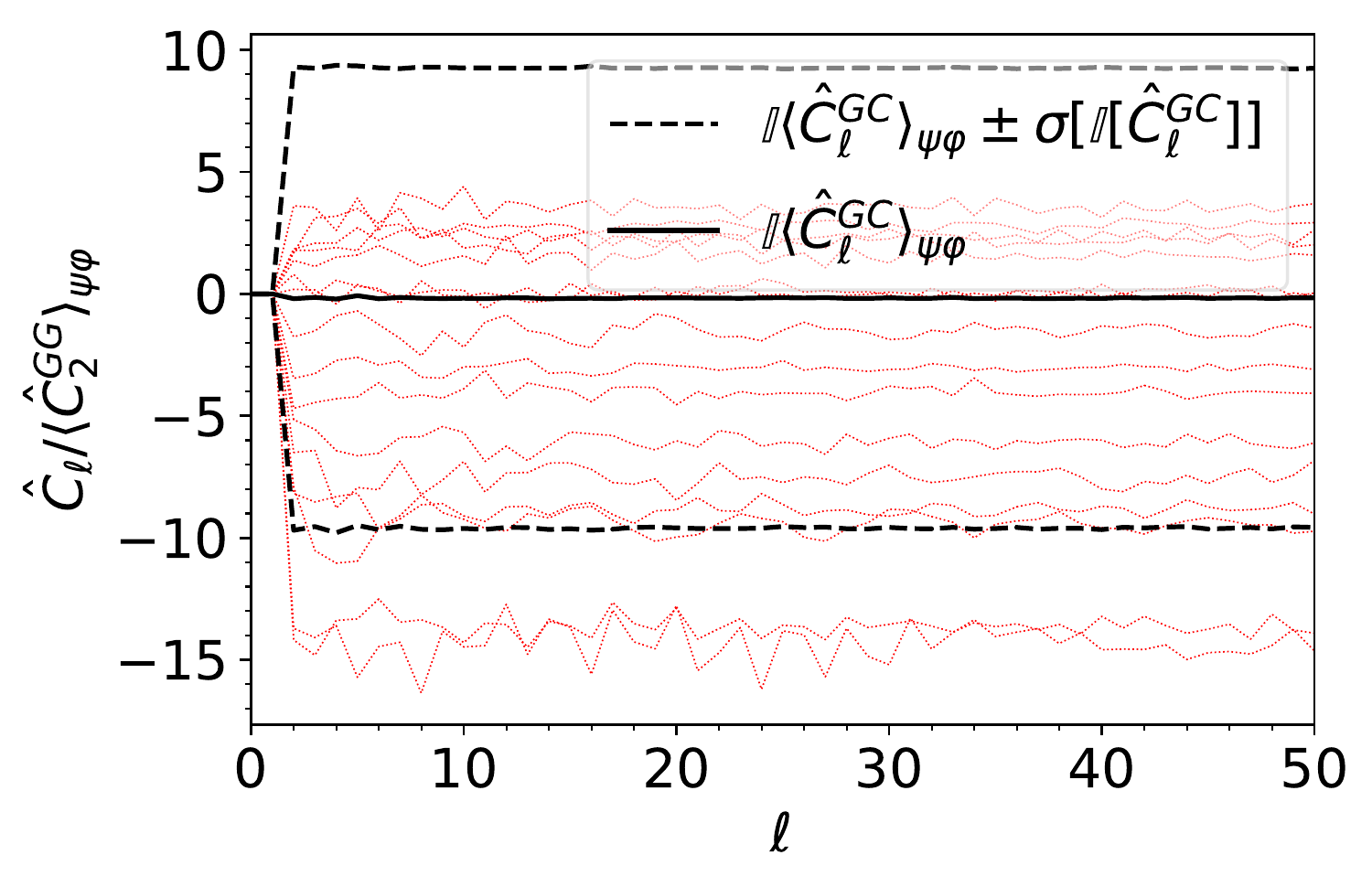}
\label{Fig:Multiple-All-k}}
\caption{As in Figure~\ref{Fig:Multiple-Noise}, but for the ``all-sky'' background of \citetalias{Mingarelli2017}, which contains a strong continuous gravitational-wave source.} \label{Fig:Multiple-All}
\end{figure*}

Given a particular realisation of a gravitational-wave background generated by local sources, we can compute the relevant power spectra. 
Using the same sample realisation as in \citetalias{Mingarelli2017} Fig.~2, this is done for two cases. In the ``all-sky'' background (Fig.~2d from \citetalias{Mingarelli2017}) there is one dominant source -- i.e. one loud enough to be detected by a PTA -- and a background from the unresolvable nearby sources. This dominant source is removed to produce an anisotropic, or  ``noise'', background.
The Mollweide projections of $\langle I \rangle_{\psi\varphi}$  for these two skies are given in Fig.~\ref{Fig:Moll-All-Noise}. We can see that the dominant source is many orders of magnitude larger than any other source, to the extent that it is the only one visible in the ``all-sky'' background. Note that plots of power have been smoothed to make the sources more visible.
Plotted in Figs~\ref{Fig:Multiple-Noise} and \ref{Fig:Multiple-All} are the power spectra for a sample of 16 from 512 simulations of different inclinations and principal polarisations of the source SMBHBs.  
For ease of comparison, particularly in terms of magnitudes, these are all normalised with respect to $\langle \hat{C}^{II}_0 \rangle_{\psi\varphi}$ and $\langle \hat{C}^{GG}_2 \rangle_{\psi\varphi}$ from the ``noise'' background for the Stokes parameters and amplitudes, respectively. 

The $C^{\langle I \rangle_{\psi\varphi} \langle I \rangle_{\psi\varphi} }_\ell$ spectra (i.e. those constructed by applying equations~\ref{Eq:CMB-T} and~\ref{Eq:SpectraDef} to $\langle I \rangle_{\psi\varphi}$) are those most closely related to those from \citetalias{Mingarelli2017}. There will be some slight variation in the shape of the spectra because of the factor of $f_i$ between $\langle I \rangle_{\psi\varphi}$ and $h^2_c$, but this is not a large effect. The factor of $2/\Delta f$ will not make a difference because it is constant for all sources and cancels out when the spectra are normalised. One key difference is in the change in shape for large $\ell$. The background field in \citetalias{Mingarelli2017} is Gaussian smoothed, whereas here it is not. Because of this, the power spectra here do not decay for large $\ell$ and are approximately white, though this will stop being the case when $\ell$ approaches the pixel scale. As mentioned, a real background will take into account detector effects and so will be more complicated. 

As in previous examples, the power spectra are approximately constant for all $\ell$ but there is still information in them. Firstly, while there is a single power spectrum for the unobservable $\langle I \rangle_{\psi\varphi}$ (or $h_c$) field, the 512 different instances have a wide variation of realised spectra. To see this define the average variation for auto-spectra
\begin{align}
\sigma^{F} =& \frac{1}{\ell_\text{max} - \ell_F}\sum_{\ell=\ell_F}^{\ell_\text{max}} \frac{\Delta C^{FF}_\ell}{\langle \hat{C}^{FF}_\ell \rangle_{\psi\varphi}} \, ,
\end{align}
where $\ell_F = 0$ for $F \in \{ I, V \}$, $\ell_F = 4$ for $F \in \{ E, B \}$ and $\ell_F = 2$ for $F \in \{ G, C \}$ is chosen so as avoid where the spectra are identically zero. The upper limit $\ell_\text{max}$ is set to 50, corresponding to an angular resolution of $\Delta\Omega \approx (180 \mathrm{deg})^2 / \ell_\text{max}^2 \approx 13 \mathrm{deg}^2$ -- a conservative estimate of the resolution of LISA compared to, for example, $\Delta\Omega \approx 0.3 \mathrm{deg}^2$ in \citet{Cutler1998}. This is similar to the result for PTAs from \citet{Sesana2010,TaylorEtAl:2015} where 2000 pulsars from the Square Kilometre Array and a signal-to-noise ratio of 10 would give an angular resolution of $\Delta \Omega \approx 8\mathrm{deg}^2$ -- corresponding to $\ell_\mathrm{max} \approx 63$. 
Using this, we can see that $\sigma^{I} \sim 0.7$ for the ``noise'' case and $\sim 1.2$ for the ``all-sky''. This is due to the fact that $I$ varies by a factor of 8 depending on the inclination -- as the binary can be edge- or face-on. This effect is also pronounced in the remaining auto-power spectra  
($\sigma^{V} \sim 0.8$, $\sigma^{E} \sim \sigma^{B} \sim 0.6$, $\sigma^{G} \sim \sigma^{C} \sim 0.3$ for the ``noise'' background $\sigma^{V} \sim 1.3$, $\sigma^{E} \sim \sigma^{B} \sim 0.9$, $\sigma^{G} \sim \sigma^{C} \sim 0.6$ for the ``all-sky'') particularly because it is theoretically possible to have any (but at most two) of $Q$, $U$ or $V$ as being zero given particular sets of orientations.

This is important for measurements. If a background were, by coincidence, made up of signals from face-on or close to face-on binaries then the overall signal would be a lot stronger than if they were all edge-on. Because of this, the power spectra for a given collection of binaries may be detectable or not depending on the collection of orientations. For a large number of binaries, this effect should be reduced as it is unlikely to have all of the binaries face-on. However, for a smaller number of sources or (as is the case in the ``all-sky'' background) a collection of sources with one dominant, it is plausible to have the alignments affecting the overall power in such a way. In such cases, the treatment of the signal as being Gaussian is perhaps inappropriate.

The ``all-sky'' spectra are noticeably whiter than the ``noise'' spectra, particularly for the $C^{II}_\ell$ spectra. This is related to the single-source spectra. If we consider the ``all-sky'' background to be a sum of the ``noise'' background and the single dominant source -- i.e. $I_{\text{all}} = I_{\text{noise}} + I_{\text{dominant}}$ -- then the power spectrum will be a sum of the two spectra in quadrature -- i.e. $\hat{C}^{II, \text{all}}_{\ell} = \hat{C}^{II, \text{noise}}_{\ell} + \hat{C}^{II, \text{dom}}_{\ell} + \hat{C}^{II, \text{cross}}_\ell$, where the $\hat{C}^{II, \text{dom}}_{\ell}$ will be white as in the single-source example.  As there is actually more power in the single dominant source than the rest of the background combined (and so significantly more in the $I^2$ terms relevant to the power spectra), the $\hat{C}^{II, \text{dom}}_{\ell}$ will be larger than the other terms and the power spectrum is white with small modulations. This also explains why the magnitude of the ``all-sky'' spectra are so much larger than the ``noise''.

A similar consequence of the dominant source can be seen in Fig.~\ref{Fig:Multiple-All-k}. Here it can be seen that, though on average it is zero, many of the simulated backgrounds have non-zero $\mathbb{I}[C^{GC}_\ell]$.
This is related to equations~\ref{Eq:GGCC-aniwn-b} and~\ref{Eq:DeltaFuncPred-b} and this spectrum measures the level of circular polarisation. In the ``noise'' case this is negligible but for the ``all-sky'' version the circular polarisation of the dominant source can lead to a non-zero sky-average value of $V$.

Another key observation is the relative levels of various power spectra. Using equations~\ref{Eq:IQUV-bin-a} and~\ref{Eq:IQUV-bin-d}, it can be shown that
\begin{subequations}
\begin{align}
\langle \hat{C}^{II}_{\ell} \rangle_{\psi\varphi} =& C^{\langle I \rangle_{\psi\varphi}\langle I \rangle_{\psi\varphi}}_{\ell} + \frac{1}{4\pi} \frac{103}{252} \sum_i \langle I_i \rangle_{\psi\varphi}^2  \, ,
\\
\langle \hat{C}^{VV}_{\ell} \rangle_{\psi\varphi} =& \frac{1}{4\pi} \frac{115}{84} \sum_i \langle I_i \rangle_{\psi\varphi}^2 \, ,
\end{align}
\end{subequations}
where the difference is primarily due to the fact that $\langle V \rangle_{\psi\varphi} = 0$ and so $C^{\langle V \rangle_{\psi\varphi}\langle V \rangle_{\psi\varphi}}_\ell = 0$. This explains why various spectra (including $\langle \hat{C}^{II}_{\ell} \rangle_{\psi\varphi}$, $\hat{C}^{\langle I \rangle_{\psi\varphi}\langle I \rangle_{\psi\varphi}}_{\ell}$, $\hat{C}^{\langle I \rangle_{\psi\varphi}\langle I \rangle_{\psi\varphi}}_{\ell} - \hat{C}^{\langle V \rangle_{\psi\varphi}\langle V \rangle_{\psi\varphi}}_{\ell}$ and $\langle \hat{C}^{II}_{\ell} \rangle_{\psi\varphi} - \langle \hat{C}^{VV}_{\ell} \rangle_{\psi\varphi}$) have the same shape, just shifted by a constant factor. 
Some of this shape, particularly for the ``noise'' example, will be due to an effective mask on the signal. As the electromagnetic signals (unlike gravitational waves) used to determine the positions and properties of the galaxies in the 2MASS survey are obscured by the galactic plane of the Milky Way, there is a region of the sky that is effectively masked in this simulation. To see this, compare the masking term in Fig.~\ref{Fig:Mask-vs-I-b} to Fig.~\ref{Fig:Multiple-Noise-a}.

As in many previous examples, the $\hat{C}^{EE}_{\ell}$ and $\hat{C}^{BB}_{\ell}$ have very similar shapes and are on average (over inclinations and principal polarisations) equal -- as can be seen in Figs~\ref{Fig:Multiple-Noise-f} and \ref{Fig:Multiple-All-f}. 
However, their difference for any given realisation can be of the order of $\langle C^{EE}_{\ell} \rangle_{\psi\varphi}$ for any given $\ell$ of a background. 
That they are on average equal can be explained using Appendix~\ref{App:Lemma}: it can be shown that $\langle Q(\hat{k}) Q(\hat{k}') \rangle_{\psi\varphi}  = \langle U(\hat{k}) U(\hat{k}') \rangle_{\psi\varphi} $ and $\langle Q(\hat{k}) U(\hat{k}') \rangle_{\psi\varphi}  = -\langle U(\hat{k}) Q(\hat{k}') \rangle_{\psi\varphi} = 0$, it follows that $\langle \hat{C}^{EE}_\ell \rangle_{\psi\varphi} = \langle \hat{C}^{BB}_\ell \rangle_{\psi\varphi}$.
The difference between the spectra in any particular realisation will be due to the finite number of sources in the background. This means that there can be a set of principal polarisations leading to a statistically significant difference in the values of $Q$ and $U$ across the whole sky. For a sky with a larger number of sources (as in the white dwarf binary example), this effect will be reduced.

\subsection {Anisotropy from large-scale structure}
\begin{figure}
\subfigure[$\log(I_s)$ for a simulation of the 5119 galaxies]{
\includegraphics[width = 0.48\textwidth]{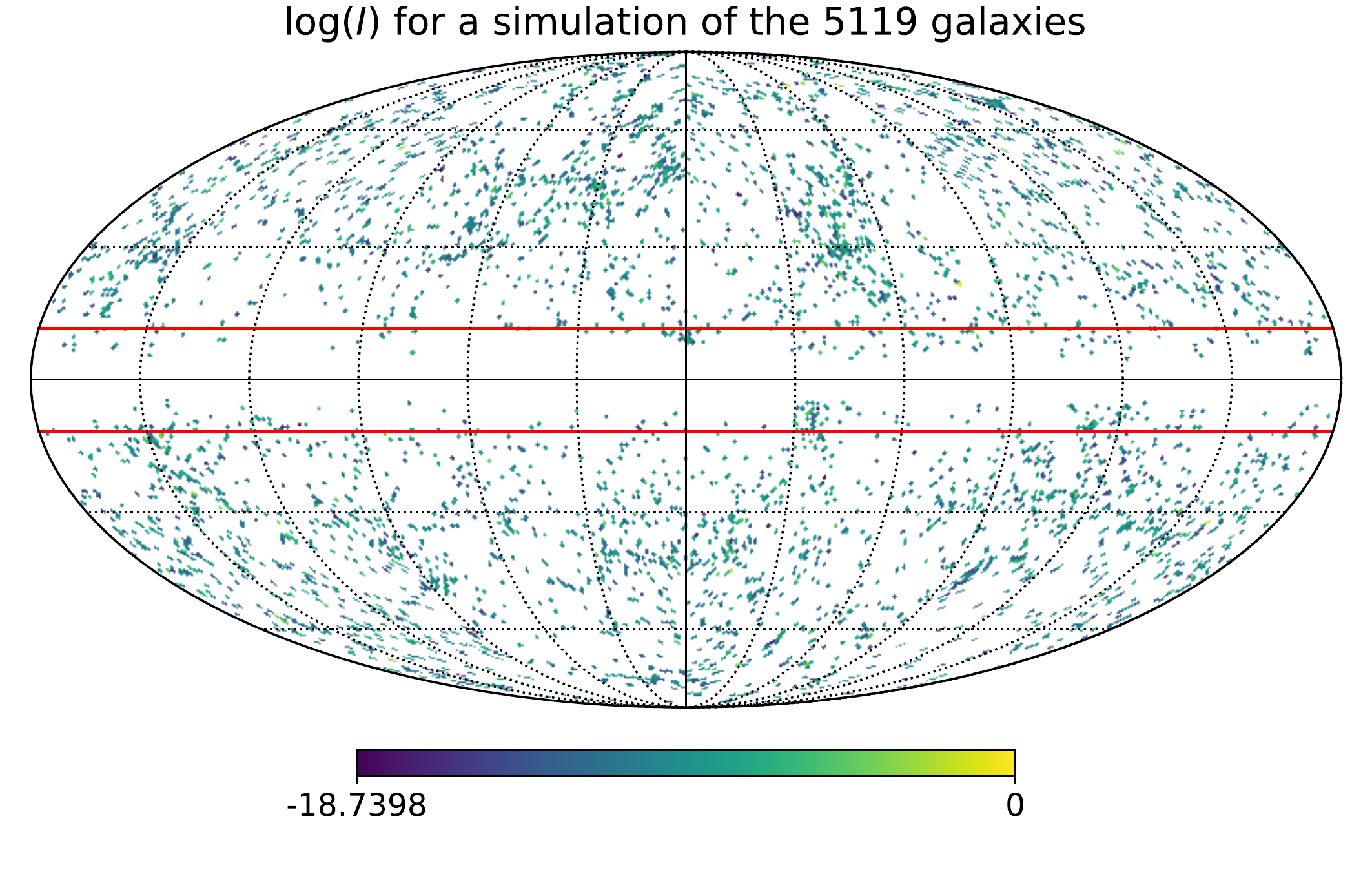}\label{Fig:5119-Loc-a}}
\subfigure[$\log(I_{si})$ for an isotropisation of 4803 galaxies]{
\includegraphics[width = 0.48\textwidth]{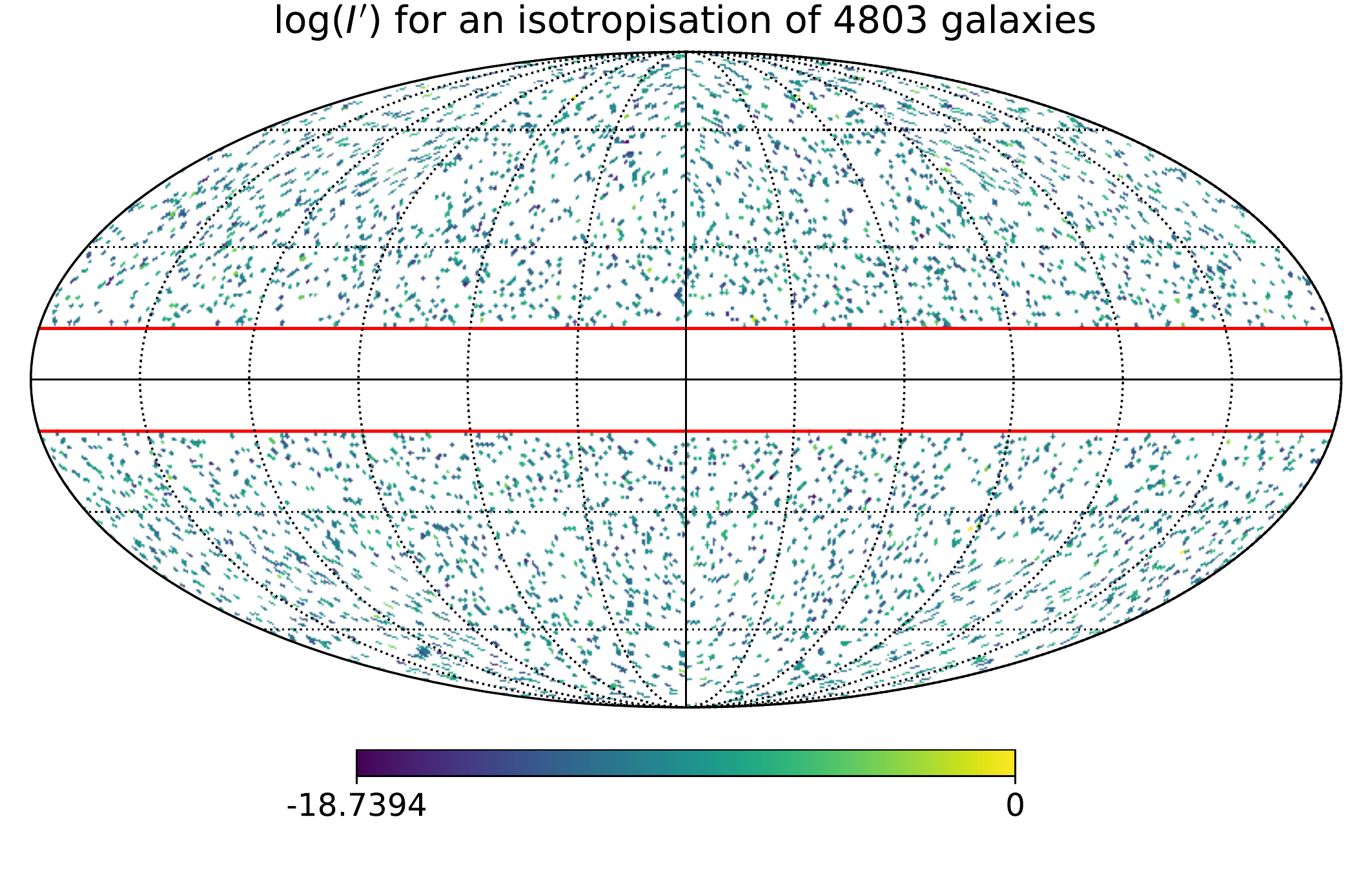}\label{Fig:5119-Loc-b}}
\caption{Locations and $\log(I/\max(I))$ for a single simulation of the 5119 galaxies considered by \citetalias{Mingarelli2017} including the 316 contained in the masked region, as indicated by the red lines at 0.2 radians from the galactic plane. Also plotted is an isotropisation of the 4803 galaxies outside of the band.} \label{Fig:5119-Loc}
\end{figure}

The assumption used thus far of independent power and sky-location for all of the signals is not the complete picture. While the presence of a SMBHB in a given galaxy will be due to many largely independent effects, the distribution of galaxies will be correlated due to large-scale structure. To see the effect of this we must consider a larger number of galaxies. Specifically we use the 5119 galaxies, out to 225 Mpc, from the 2MASS survey considered by \citetalias{Mingarelli2017}. We compare the power spectrum of the signal from the true distribution of the galaxies to those where the galaxies have been uniformly redistributed across the sky. 
To do this, we take the ``Nanohertz GW'' code \citep{nano_gw} and set the probability of a given galaxy hosting a binary to be 1. In real observations, we will be able to measure gravitational waves from the whole sky, but, as mentioned, this is not the case with the 2MASS survey and so it is not appropriate to simply randomly redistribute all 5119 galaxies.
Instead, we consider a mask over all positions in a band defined by being less than 0.2 radians from the galactic plane. This masked region contains 316 galaxies which are ignored for the rest of the analysis. The remaining 4803 galaxies are redistributed randomly with a uniform distribution for all positions outside of this band. The original distribution for a given simulation (indexed by $s$) and an example of the uniform distribution (indexed by $i$) for this simulation are given in Fig.~\ref{Fig:5119-Loc}. Using this, we can compute the power spectra for the two cases.

\begin{figure*}
\subfigure[]{
\includegraphics[width = 0.3\textwidth]{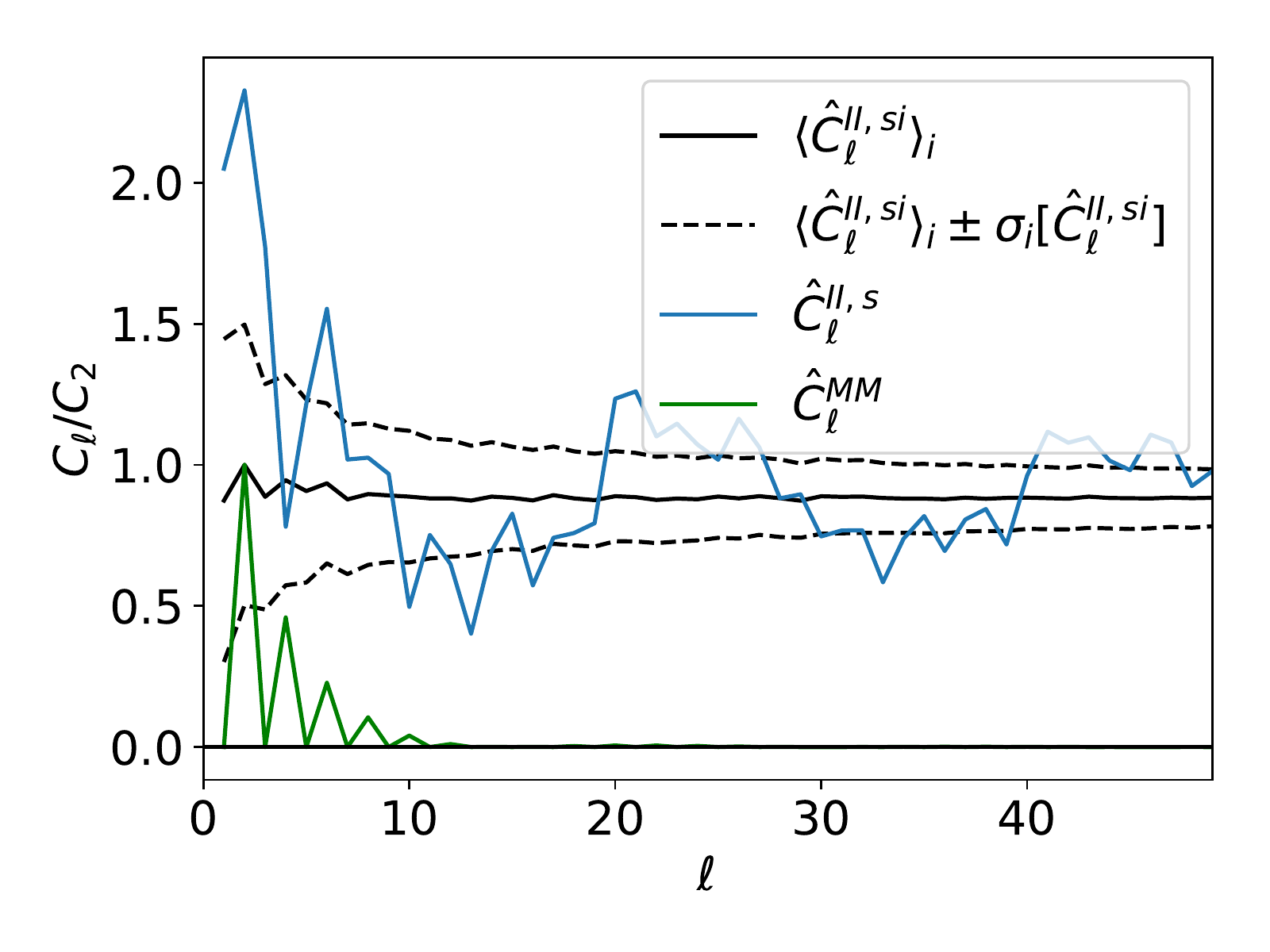}
\label{Fig:5119-PS-a}}
\subfigure[]{
\includegraphics[width = 0.3\textwidth]{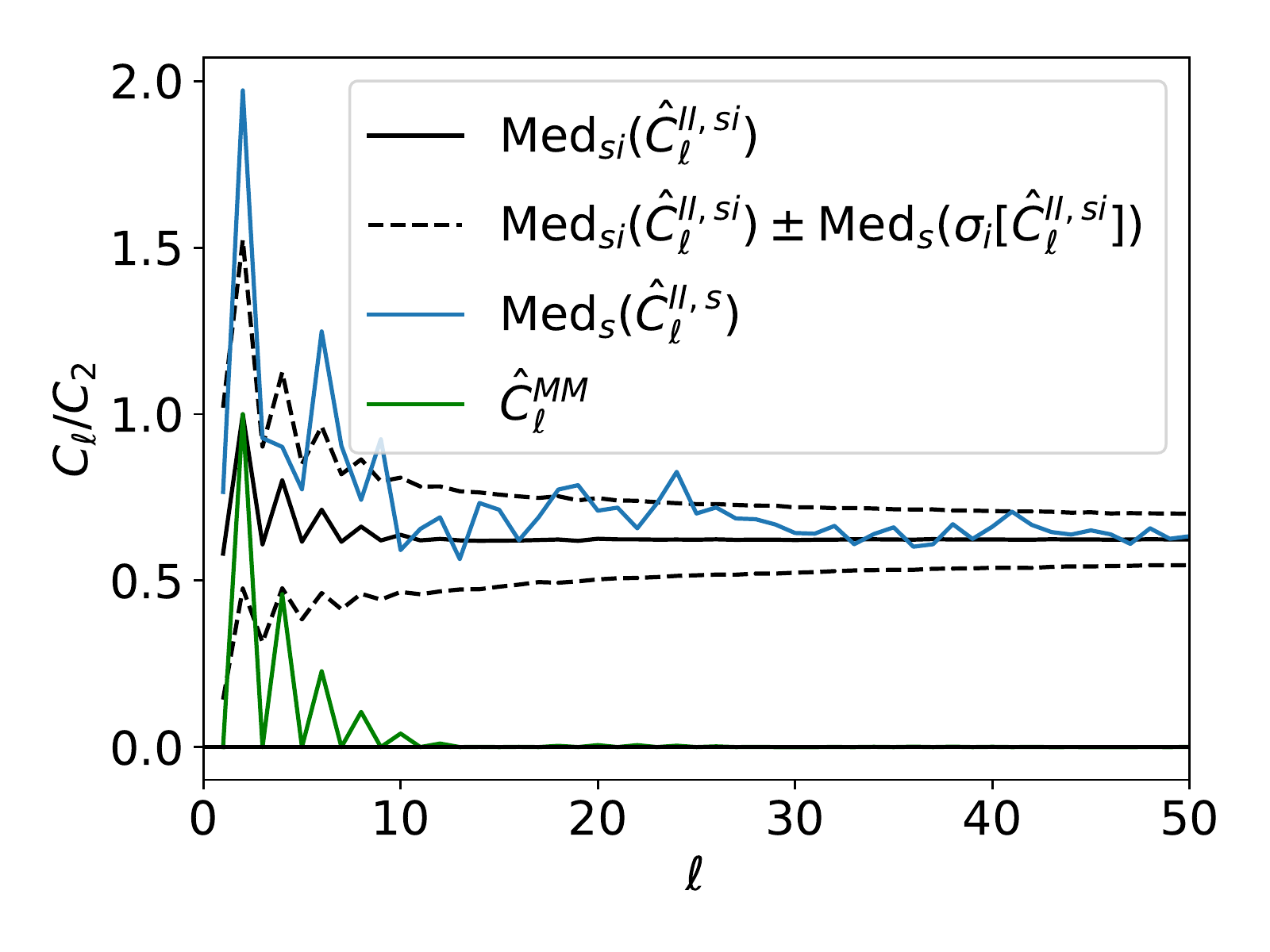}\label{Fig:5119-PS-b}}
\subfigure[]{
\includegraphics[width = 0.3\textwidth]{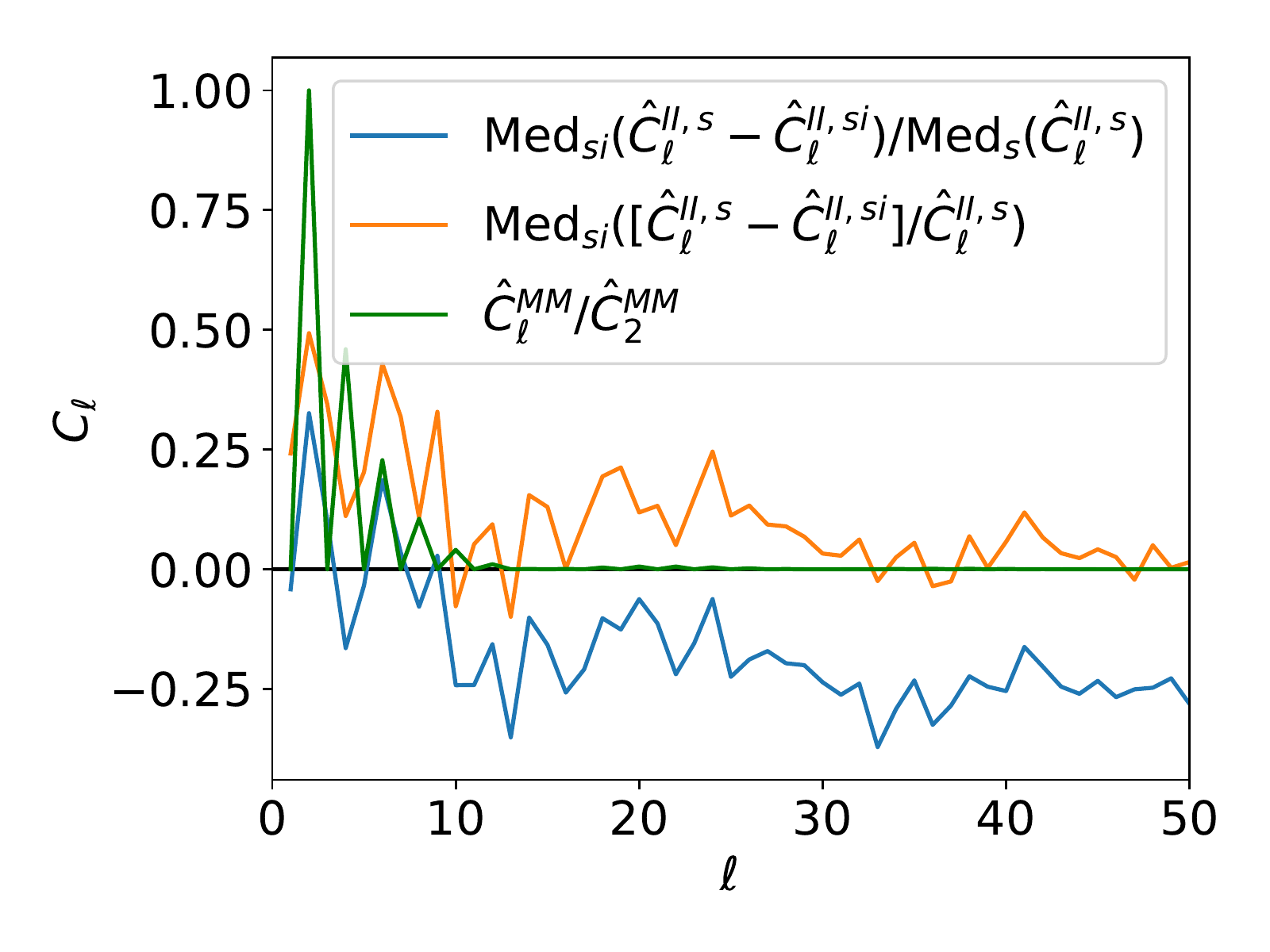}
\label{Fig:5119-PS-c}}
\caption{(a) Power spectra for a single simulation and 1024 isotropisations. The blue curve is the power spectrum for the masked true distribution ($C^{II,s}_\ell$) for the same simulation as in Fig.~\ref{Fig:5119-Loc-a}, the black curve is the mean power spectrum ($\langle \hat{C}^{II,si}_\ell \rangle_{i}$) over 1024 isotropised redistributions with the dashed lines representing one standard deviation. The green curve is the power spectrum of the mask. All plots are normalised with respect to the quadrupole of the averaged power spectrum, except for the mask distribution which is normalised with respect to its own quadrupole.
(b) Power spectra for $\mathrm{Med}_{s}(\hat{C}^{II,s}_\ell)$ and $\mathrm{Med}_{si}(\hat{C}^{II,si}_\ell)$ -- i.e. the median values of $\hat{C}^{II,s}_\ell$ and $\hat{C}^{II,si}_\ell$ over 128 simulations and $128 \times 1024$ isotropisations. The plotted standard deviation is the median value of $\sigma_i[\hat{C}^{II, si}_\ell ]$ over 128 simulations. All spectra are normalised with respect to the median value (over all simulations and isotropisations) of $\hat{C}^{II,si}_2$. The mask spectrum is defined as in (a).
(c) The median fractional difference between the initial simulations and their respective isotropisations. The mask spectrum is, again, defined as in (a).} 
\end{figure*}

The angular power spectra for this masked anisotropic simulation of the power and the average over 1024 isotropisations of it are plotted in Fig.~\ref{Fig:5119-PS-a} -- for a given simulation $s$ and set of isotropisations $i$, we define $\langle \hat{C}^{II,si}_\ell \rangle_{i}$ as the mean power spectrum and $\sigma_i[C^{II,si}_\ell]$ as the standard deviation\footnote{Note that these are both over isotropisations, $i$, so remain a function of the simulation, $s$.}. 
Using the metric of $\chi^2$ averaged over $\ell$,
\begin{align}
\chi^2 = \frac{1}{\ell_\mathrm{max}} \sum_{l = 1}^{\ell_\mathrm{max}} \frac{\left(\hat{C}^{II,s}_\ell - \langle \hat{C}^{II,si}_\ell \rangle_{i}\right)^2}{\left(\sigma_i[\hat{C}^{II,si}_\ell]\right)^2} \, , 
\end{align}
for $\ell_\mathrm{max} = 50$ gives $\chi^2 \approx 2.02$. This is typical\footnote{This example was chosen because of its typical $\chi^2$ value and because this deviation is comparatively easy to see.} 
as $\chi^2$ has, on average across different simulations, mean value $\approx 2.15$ and median $\approx 1.80$ (though over 128 simulations was as small as $\approx 0.662$ and as large as $\approx 13.2$) and is significantly larger than the case for different isotropic redistributions which is, by definition, $1$ on average. It is worth noting that, because of the large number of isotropisations (1024 different redistributions each for 128 simulations), there is a large range of values for the value of $\chi^2$ for a given $\hat{C}^{II, si}_\ell$ -- varying between $\sim 0.334 - 27.0$ in these examples. This is a small effect and only $\sim 2\%$ of the isotropised power spectra give $\chi^2 > 2$. 

Of course, a true measurement will not have this masking effect because gravitational waves are not obscured by the galaxy. It is, however, noticeable how different the spectrum of the true distribution is to the isotropised versions. As such, it appears that a measurement of the power spectrum of SMBHBs can, in principle, provide information on large-scale structure, though this may require more galaxies to be emitting in the nanohertz band than used by \citetalias{Mingarelli2017}.

A single simulation of the power emitted from the 5119 galaxies is plotted in Fig.~\ref{Fig:5119-PS-a} instead of a collection of them because of the issue of single dominant sources. As the overall power spectrum can be so affected by the presence of single, large source, as seen previously in this section, the overall strength of the power spectra for different simulations can vary massively and in a very non-Gaussian way.  
In principle such a source may be large enough to be individually observed and so removed from the analysis but one potential method of compensating for this is plotted in Fig.~\ref{Fig:5119-PS-b}. Here, instead of the spectra for a single simulation, or the mean over many simulations, we plot the median. That is, the median value of $\hat{C}^{II,s}_\ell$ and $\hat{C}^{II,si}_\ell$ for each $\ell$ harmonic over the 128 original simulations and $128 \times 1024$ isotropisations, respectively. The corresponding standard deviation term is the median over the standard deviations calculated for each initial simulation. Here it can be seen that, perhaps unsurprisingly, the median deviation is smaller -- corresponding to an equivalent $\chi^2$ of $\approx 0.616$. The (median) fractional difference between these is plotted in Fig.~\ref{Fig:5119-PS-c}. Here we can see that some of the features in Figs~\ref{Fig:5119-PS-a} and~\ref{Fig:5119-PS-b} are statistically significant -- for example, increased power in the original simulation at the $\ell = 2$ and $6$ harmonics over the isotropisations, in addition to reduced power at $\ell = 4$. Note that some of the features that may be statistically significant on average (i.e. in Figs~\ref{Fig:5119-PS-b} and~\ref{Fig:5119-PS-c}) such as an enhancement at $\ell = 8$ are not necessarily present in any given simulation.
A final note is that the majority of the structure of the spectra of the original simulations will, in general, be on larger scales -- i.e. for $\ell \lessapprox 10$.

One potentially informative metric related to this is cross-correlations of these spectra with the power spectrum of the large-scale structure. In theory, this should be similar to an autocorrelation of either spectra but any significant difference could inform us of properties the source -- such as preferential distribution of the emitting SMBHBs -- or of gravitational waves themselves -- such as a difference in the propagation of electromagnetic and gravitational waves, though current constraints on prorogation imply that this would require a very large degree of sensitivity.  
This will also have the benefit of a higher signal-to-noise ratio for this cross-spectrum than the auto-spectrum of the gravitational waves because the large-scale structure will itself have higher signal-to-noise than the gravitational wave field, even for the most sensitive future detectors.

\section{Conclusions}\label{Sec:conclusions}

We have compared the spin-2 spectra of the gravitational wave amplitude of a stochastic background to that of the spin-4 spectra of the gravitational Stokes parameters.
We can see that, in many examples, the amplitude analysis gives white (constant amplitude) spectra of statistically equal strength for the autocorrelations and zero for the cross-correlations. As such, we cannot infer much information directly from the power spectra in these cases, other than a measure of the overall signal strength and an indication of the expected parity symmetry. While it is possible to construct examples where the amplitude auto-spectra are not equal and the cross-spectrum is not zero (e.g. a background with a small number of sources or by breaking parity) and where they are not white (requiring a correlation in phase which is very difficult to set up), these simple constraints hold for a majority of cases considered.
However, this does mean that any detection of such non-standard power spectra indicates something unexpected about the background. 

The Stokes parameter analysis method, while still often giving white spectra, does show more information. Though the $C^{EE}_\ell$ and $C^{BB}_\ell$ spectra are usually equal, they differ from $C^{II}_\ell$ and $C^{VV}_\ell$. Further, in the Stokes case, it is easier to construct reasonable examples of backgrounds where the spectra are not white -- as can be seen in the white dwarf binary and the large-scale structure/SMBHB examples -- giving information on the distribution of sources. 

It should be further noted that the power spectra for Stokes parameters, other than $I$, share many of the same constraints as the amplitude spectra. That is, in the majority of examples they are white and $C^{EE}_\ell$ usually equals $C^{BB}_\ell$. The reasons for this are often similar: for example, in the case of binaries, the orientations of different binaries are unlikely to be correlated and so lead to white spectra in the same way as uncorrelated phases.

Because of these observations, it is clear that the Stokes parameter method will often provide more information than the amplitude (and in the majority of cases, the most valuable spectrum to consider is that of the $I$ Stokes parameter -- consistent with what has been studied in the past). As such it will, in general, be the more appropriate method to analyse the background. We will apply these techniques to simulations of white dwarf binaries (Brevik et al., 2018, in preparation) as both a method characterise the background and distinguish it from other sources of gravitational radiation.

For the application of these methods to real data, contaminated by noise and selection effects, much work will be necessary, going from pulsar-timing or gravitational-wave interferometer timestream data to maps of the background, e.g. \cite{RenziniContaldi2018}, and then to the spectra and related likelihood functions. Estimations of backgrounds made up of small numbers of sources, or examples with a single source that provides a significant part of the signal, will be subject to non-Gaussian statistics and it could be argued that the angular power spectra are not appropriate, or at least do not show the full picture. We expect that methods from the analysis of the CMB \citep[e.g.][]{BJK1998,Hivon2002,PlanckLikelihood2015}, large scale structure \cite[e.g.][]{BOSS2018}, and weak lensing \citep[e.g.][]{Alsing2017} will be particularly useful and we will pursue these techniques in future work.

\section*{Acknowledgements}

The authors would like to thank Arianna Renzini, Carlo Contaldi, Giulia Gubitosi, and Joao Magueijo for the their help. CC and AHJ were funded by the STFC in the UK. The Flatiron Institute is supported by the Simons Foundation. This research made heavy use of IPython~\citep{ipython} and its many open access libraries (\citealt{healpy, matplotlib, matplotlib2007, numpy, astropy, astropy2018, scipy}.), and C. Mingarelli's open access code ``Nanohertz GWs'', on github, \cite{nano_gw}.




\bibliographystyle{mnras}
\bibliography{references} 




\appendix
\section{Properties of spherical harmonics}
\subsection{Errors for power spectra of amplitude fields} \label{App:Errors}
The standard deviation for the power spectra given in  equation~\ref{Eq:SpectraDef-Measured} assumes that the fields considered (e.g. $I$, $Q$, $U$, $V$) are real as the calculation requires that the harmonic coefficients satisfy equation~\ref{Eq:m-mrelation}. 
This means that for power spectra generated from the complex amplitude fields, the standard deviation is more complicated. To see this, we split the fields into real and imaginary parts
\begin{align}
{}_{\pm 2} a_{\ell m}^r + i {}_{\pm 2} a_{\ell m}^i =& \int \text{d}^2\hat{k} ((h_+^r + i h_+^i) \pm i (h_\times^r + i h_\times^i)) {}_{\pm 2}Y_{\ell m}(\hat{k})
\notag\\
=& \int \text{d}^2\hat{k} (h_+^r \pm i h_\times^r ) {}_{\pm 2}Y_{\ell m}(\hat{k})
\notag\\
&+ i \int \text{d}^2\hat{k} (h_+^i \pm i h_\times^i ) {}_{\pm 2}Y_{\ell m}(\hat{k}) \, ,
\notag\\
a^{Gr/Gi}_{\ell m} =& +\frac{1}{\sqrt{2}}\left( {}_{+2}a_{\ell m}^{r/i} + {}_{-2}a_{\ell m}^{r/i} \right) \, ,
\notag\\
a^{Cr/Ci}_{\ell m} =& -\frac{i}{\sqrt{2}}\left( {}_{+2}a_{\ell m}^{r/i}- {}_{-2}a_{\ell m}^{r/i} \right)
\end{align}
which do satisfy equation~\ref{Eq:m-mrelation} and lead to power spectra
\begin{align}
C^{AA'}_\ell = \frac{1}{2\ell+1}\sum_m a^{A}_{\ell m} a^{A'*}_{\ell m}, \, A,A' \in \{Gr, Gi, Cr, Ci\} \, .
\end{align}
Using these, we can then show that the errors on the full power spectra are 
\begin{align} \label{Eq:Errors-for-Amplitude}
\Delta C^{MM}_\ell
=& \bigg[\frac{2}{(2\ell+1)f_\text{sky}} \big(
      C^{MrMr}_\ell C^{MrMr}_\ell  
\notag\\
&+ 2 C^{MrMi}_\ell C^{MrMi}_\ell +  C^{MirMi}_\ell C^{MiMi}_\ell 
\big) \bigg]^\frac{1}{2}, 
\notag\\
& M \in \{ G, C \} \, ,
\notag\\
\Delta[\mathbb{R}[C^{GC}_\ell]] 
=&   \bigg[\frac{1}{(2\ell+1)f_\text{sky}} \big(
C^{GrGr}_\ell C^{CrCr}_\ell  + C^{GrCr}_\ell C^{GrCr}_\ell
\notag\\
&+ C^{GrGr}_\ell C^{CiCi}_\ell  + C^{GiCi}_\ell C^{GiCi}_\ell 
\notag\\
&+ C^{GrGi}_\ell C^{CrCi}_\ell  + C^{GrCi}_\ell C^{GiCr}_\ell\big) \bigg]^\frac{1}{2} \, ,
\notag\\
\Delta[\mathbb{I}[C^{GC}_\ell]]
=& \bigg[\frac{1}{(2\ell+1)f_\text{sky}} \big(C^{GiGi}_\ell C^{CrCr}_\ell  + C^{GiCr}_\ell C^{GiCr}_\ell 
\notag\\
&+ C^{GrGr}_\ell C^{CiCi}_\ell  + C^{GrCi}_\ell C^{GrCi}_\ell 
\notag\\
&- C^{GrGi}_\ell C^{CrCi}_\ell  - C^{GiCi}_\ell C^{GrCr}_\ell \big) \bigg]^\frac{1}{2} \, . 
\end{align}
This can be significantly different from the result in equation~\ref{Eq:SpectraDef-Measured}.
For example in the case where $C^{GrGr}_\ell = C^{GiGi}_\ell$ and $C^{GrGi}_\ell = 0$, the value of $\Delta C^{GG}_\ell$ is a factor of $\sqrt{2}$ less for this method than if we had assumed that equation~\ref{Eq:SpectraDef-Measured} applied. 

\subsection{Sum of spherical harmonics}
Using the sign convention of \citet{Hu:1997hp,Goldberg1967} (rather than that of \citet{Gair2014}) -- specifically equations~2 of \citep{Hu:1997hp} and 3.1 of \citep{Goldberg1967} -- the power spectra of an uncorrelated field (either spin weighted or spin-0) will contain terms like the following:
\begin{align} \label{Eq:Sum-sYlm1}
\frac{1}{2\ell +1 }\sum_{m = \ell}^\ell & \int \text{d}^2\hat{k} g(f, \hat{k}) {}_{s_1}Y_{\ell m}^*(\theta, \phi) {}_{s_2}Y_{\ell m}(\theta, \phi)
\notag
\\\nopagebreak
=& \frac{1}{2\ell+1} \int \text{d}^2\hat{k} g(f, \hat{k}) \sqrt{\frac{2\ell + 1}{4\pi}} {}_{s_2}Y_{\ell, -s_1}(0,0) \, ,
\end{align}
where the relevant pairs of spins are
\begin{align} \label{Eq:Sum-sYlm2}
{}_{0}Y_{\ell 0}(0, 0) =& \sqrt{\frac{2\ell + 1}{4\pi}} \, ,
\notag\\
{}_{\pm s}Y_{\ell, \mp s}(0,0) =& \sqrt{\frac{2\ell + 1} {4\pi}} \, ,
\notag\\
{}_{\pm s}Y_{\ell, \pm s}(0,0) =& 0 \, .
\end{align} 

\subsection{Correlation of Stokes parameters} \label{App:CorrSP}
Assuming that the phases, $\phi_+$ and $\phi_\times$, for each direction are independent of each other and $|h_+|$ and $|h_\times|$ and are uniformly distributed, $\phi_{+, \times} \sim \text{U} (0, 2\pi)$, means that we can say certain things about the expected  (i.e. ensemble averaged) correlations of the Stokes parameters. First note that the expectation value of functions of the phases only is equivalent to an integral over the phases
\begin{align}
\langle f(\phi_+, \phi_\times) \rangle =& \int \text{d}\phi_+ \int \text{d}\phi_\times f(\phi_+, \phi_\times) P(\phi_+) P(\phi_\times) 
\notag\\
=&  \left(\frac{1}{2\pi}\right)^2 \int_0^{2\pi} \text{d}\phi_+ \int_0^{2\pi} \text{d}\phi_\times f(\phi_+, \phi_\times) \, .
\end{align}
Specifically consider
\begin{align}
\langle & \sin(\phi_+ - \phi_\times) \rangle 
\notag\\
&=
\left(\frac{1}{2\pi}\right)^2\int_0^{2\pi}  \int_0^{2\pi} \sin(\phi_+ - \phi_\times) \text{d}\phi_+  \text{d}\phi_\times 
= 0 \, ,
\notag\\
\langle & \cos(\phi_+ - \phi_\times) \rangle 
\notag\\
&= 
\left(\frac{1}{2\pi}\right)^2\int_0^{2\pi}  \int_0^{2\pi} \cos(\phi_+ - \phi_\times) \text{d}\phi_+  \text{d}\phi_\times = 0 \, ,
\notag\\
\langle & \sin(\phi_+ - \phi_\times) \cos(\phi_+' - \phi_\times')\rangle 
\notag\\\notag
=&
\begin{cases}
\langle \sin(\phi_+ - \phi_\times) \rangle \langle \cos(\phi_+' - \phi_\times')\rangle, & \text{if} \, \hat{k} \neq \hat{k}
\notag\\
\langle \sin(\phi_+ - \phi_\times)  \cos(\phi_+' - \phi_\times')\rangle, & \text{if} \, \hat{k} = \hat{k}
\end{cases}
\\\notag
=&
\begin{cases}
0, & \text{if}\,  \hat{k} \neq \hat{k}
\\
\left(\frac{1}{2\pi}\right)^2\int_0^{2\pi}  \int_0^{2\pi} \sin(\phi_+ - \phi_\times) \cos(\phi_+ - \phi_\times) \text{d}\phi_+  \text{d}\phi_\times, & \text{if} \, \hat{k} = \hat{k}
\end{cases}
\\\notag
=&
\begin{cases}
0, & \text{if} \, \hat{k} \neq \hat{k}
\\
\frac{1}{8\pi^2}\int_0^{2\pi} \int_0^{2\pi} \sin(2\phi_+ - 2\phi_\times) \text{d}\phi_+  \text{d}\phi_\times, & \text{if} \, \hat{k} = \hat{k}
\end{cases}
\\\notag
=&
\begin{cases}
0, &\text{if} \, \hat{k} \neq \hat{k}
\\
0, & \text{if} \, \hat{k} = \hat{k}
\end{cases}
\\ 
=& 0 ,
\end{align}
which means that
\begin{align}
\langle I(f, \hat{k}) U(f' \hat{k}')\rangle =& \langle 2 (|h_+(f, \hat{k})|^2 + |h_\times(f, \hat{k})|^2) 
\notag\\
&\times |h_+(f', \hat{k}')||h_\times(f', \hat{k}')| \sin(\phi_+' -\phi_\times')  \rangle \label{Eq:IU-corr}
\notag\\
=& 2\langle (|h_+(f, \hat{k})|^2 + |h_\times(f, \hat{k})|^2)  
\notag\\
&\times|h_+(f', \hat{k}')||h_\times(f', \hat{k}')|  \rangle \langle \sin(\phi_+' -\phi_\times')\rangle 
\notag\\
=& 0 \, ,
\notag\\
\langle V(f, \hat{k}) Q(f' \hat{k}')\rangle =& \langle 2 |h_+(f, \hat{k})||h_\times(f, \hat{k})| \sin(\phi_+ -\phi_\times) 
\notag\\
&\times (|h_+(f', \hat{k}')|^2 - |h_\times(f', \hat{k}')|^2)\rangle 
\notag\\
=& 2\langle (|h_+(f', \hat{k}')|^2 - |h_\times(f', \hat{k}')|^2)
\notag\\
&\times |h_+(f, \hat{k})||h_\times(f, \hat{k})| \rangle \langle \sin(\phi_+ -\phi_\times)\rangle 
\notag\\
=& 0 \, ,
\notag\\
\langle V(f, \hat{k}) U(f' \hat{k}')\rangle =& \langle 2 |h_+(f, \hat{k})||h_\times(f, \hat{k})| \sin(\phi_+ -\phi_\times) 
\notag\\
&\times 2 |h_+(f', \hat{k}')||h_\times(f', \hat{k}')| \cos(\phi_+' -\phi_\times')\rangle 
\notag\\
=&
4\langle |h_+(f, \hat{k})||h_\times(f, \hat{k})| |h_+(f', \hat{k}')||h_\times(f', \hat{k}')| \rangle
\notag\\
& \times \langle \sin(\phi_+ -\phi_\times)\cos(\phi_+' -\phi_\times')\rangle 
\notag\\
=&
0 \, .
\end{align}
The expected power spectrum for $C^{VB}_\ell$ is then
\begin{align} \label{Eq:VB-corr}
& C^{VB}_\ell(f)
\notag \\
=& \frac{1}{2i} \frac{1}{2\ell + 1} \sum_m \left(\langle b^V(f)_{\ell m} {}_{+4}b_{\ell m}^*(f)\rangle - \langle b^V(f)_{\ell m} {}_{-4}b_{\ell m}^*(f)\rangle \right)
\notag\\
=&
\frac{1}{2i} \frac{1}{2\ell + 1} \sum_m \iint \text{d}^2\hat{k}  \text{d}^2\hat{k}
\notag\\
&\times\bigg(
\left[\langle V(f, \hat{k}) Q(f', \hat{k}') \rangle - i\langle V(f, \hat{k}) U(f', \hat{k}') \rangle \right]{}_{+4}Y_{\ell m}^*(\hat{k}) Y_{\ell m}(\hat{k}')
\notag\\
&+ \left[ \langle V(f, \hat{k}) Q(f', \hat{k}') \rangle + i \langle V(f, \hat{k}) U(f', \hat{k}') \rangle \right] {}_{-4}Y_{\ell m}^*(\hat{k}) Y_{\ell m}(\hat{k}') \bigg)
\notag\\
=&0 \, .
\end{align}

For $\langle V(f, \hat{k}) V(f', \hat{k}')) \rangle$ consider
\begin{align}
&\langle \sin(\phi_+ - \phi_\times) \sin(\phi_+' - \phi_\times')\rangle \notag\\
&=
\begin{cases}
\langle \sin(\phi_+ - \phi_\times) \rangle \langle \sin(\phi_+' - \phi_\times')\rangle, & \text{if} \, \hat{k} \neq \hat{k}
\\
\langle \sin(\phi_+ - \phi_\times)  \sin(\phi_+' - \phi_\times')\rangle, & \text{if} \, \hat{k} = \hat{k}
\end{cases}
\notag\\
&=
\begin{cases}
0, & \text{if} \, \hat{k} \neq \hat{k}
\\
\frac{1}{4\pi^2}\int_0^{2\pi} \left[ \int_0^{2\pi} \sin^2(\phi_+ - \phi_\times) \text{d}\phi_+ \right] \text{d}\phi_\times, & \text{if} \, \hat{k} = \hat{k}
\end{cases}
\notag\\
&=
\begin{cases}
0, & \text{if} \, \hat{k} \neq \hat{k}
\\
\frac{1}{2}, & \text{if} \, \hat{k} = \hat{k}
\end{cases}
\notag \\
&= \frac{1}{2}\delta^2(\hat{k}, \hat{k}') \, .
\end{align}
Applying this to $V$ gives
\begin{align} \label{Eq:WN-VVcorr}
\langle V(f, \hat{k}) V(f' \hat{k}')\rangle =& \langle 2 |h_+(f, \hat{k})||h_\times(f, \hat{k})| \sin(\phi_+ -\phi_\times) 
\notag\\
&\times 2 |h_+(f', \hat{k}')||h_\times(f', \hat{k}')| \sin(\phi_+' -\phi_\times')\rangle 
\notag\\
=&
4\langle |h_+(f, \hat{k})||h_\times(f, \hat{k})| |h_+(f', \hat{k}')||h_\times(f', \hat{k}')| \rangle
\notag\\
&\times\langle \sin(\phi_+ -\phi_\times)\sin(\phi_+' -\phi_\times')\rangle 
\notag\\
=& 2\langle |h_+(f, \hat{k})||h_\times(f, \hat{k})| |h_+(f, \hat{k})||h_\times(f, \hat{k})| \rangle 
\notag\\
&\times\delta^2(\hat{k}, \hat{k}') = g(f, \hat{k}) \delta^2(\hat{k}, \hat{k}') \, .
\end{align}
Using the same methods as equations~\ref{Eq:GGCC-aniwn}, this can be seen to lead to a white, but not necessarily zero, $C^{VV}_\ell$ spectrum.

Making the further assumption that $h_+$ and $h_\times$ are statistically identical gives
\begin{align}
&\langle I(f, \hat{k}) Q(f' \hat{k}')\rangle 
\notag\\
&= \langle 2 (|h_+(f, \hat{k})|^2 + |h_\times(f, \hat{k})|^2) 
\cdot (|h_+(f', \hat{k}')|^2 - |h_\times(f', \hat{k}')|^2) \rangle 
\notag\\
&= \langle |h_+(f,\hat{k})|^2 |h_+(f',\hat{k}')|^2 - |h_\times(f,\hat{k})|^2|h_\times(f',\hat{k}')|^2 \rangle 
\notag\\
&\phantom{=}+ \langle |h_\times(f,\hat{k})|^2 |h_+(f',\hat{k}')|^2 - |h_+(f,\hat{k})|^2|h_\times(f',\hat{k}')|^2 \rangle 
\notag\\
&= 0 \, .
\end{align}
This, combined with equation~\ref{Eq:IU-corr} gives $C^{IE}_\ell = 0$ in the same way as equation~\ref{Eq:VB-corr}.

\section{Correlations of harmonics of a spin-weighted field} \label{App:Lemma}
Given any spin-$s$ background $(F + iG)(\hat{k})$ on the sphere with harmonics ${}_{\pm 2}a_{\ell m}$ we can define (pseudo-)scalar harmonics $a^E_{\ell m}$ and $a^B_{\ell m}$. Then 
$\langle F(\hat{k}) F(\hat{k}') \rangle  = \langle G(\hat{k}) G(\hat{k}') \rangle $ and $\langle F(\hat{k}) G(\hat{k}') \rangle  = -\langle G(\hat{k}) F(\hat{k}') \rangle$
if and only if
$\langle a^E_{\ell m} a^{E*}_{\ell' m'} \rangle = \langle a^B_{\ell m} a^{B*}_{\ell' m'} \rangle$ and $\langle a^E_{\ell m} a^{B*}_{\ell' m'} \rangle = -\langle a^B_{\ell m} a^{E*}_{\ell' m'} \rangle$.

For example, this means that for any pair of field that are statistically identical (so $\langle F(\hat{k}) F(\hat{k}') \rangle  = \langle G(\hat{k}) G(\hat{k}') \rangle $) and are independent with mean zero ($\langle F(\hat{k}) G(\hat{k}') \rangle = \langle F(\hat{k}) \rangle \langle G(\hat{k}') \rangle = 0 = -\langle G(\hat{k}) F(\hat{k}') \rangle $) then  we immediately get that $C^{EE}_\ell = C^{BB}_\ell$ and $C^{EB}_\ell = -C^{BE}_\ell$. It can be further shown that, with these same assumptions, $C^{EB}_\ell = C^{BE}_\ell = 0$ and parity is satisfied.


\bsp	
\label{lastpage}
\end{document}